\documentclass[prb,aps,twocolumn,superscriptaddress]{revtex4-2}
\usepackage{amsmath}
\usepackage{subfigure}
\usepackage{color}
\usepackage{bbm}
\usepackage{amssymb}
\usepackage{epsfig}
\usepackage{multirow}
\usepackage{amsbsy}
\usepackage{array}
\usepackage{diagbox}
\usepackage{bm}
\usepackage{extarrows}
\usepackage{graphicx}
\usepackage{appendix}
\usepackage{txfonts}
\usepackage{lipsum} 
\usepackage{tablefootnote}
\usepackage{soul}
\graphicspath{{figures/}}
\allowdisplaybreaks[4]

\usepackage[colorlinks=true,linkcolor=blue,citecolor=blue,urlcolor=blue,bookmarks=false]{hyperref}

\begin{document}
	
\title{Interplay of Zeeman field, Rashba spin-orbit interaction, and superconductivity: Transition temperature and quasiparticle excitations}

\author{Chen Pang}
\affiliation {Institute of Physics, Chinese Academy of Sciences, Beijing 100190, China}
\affiliation{School of Physical Sciences, University of Chinese Academy of Sciences, Beijing 100190, China}	

\author{Yi Zhou}
\email{yizhou@iphy.ac.cn}
\affiliation {Institute of Physics, Chinese Academy of Sciences, Beijing 100190, China}

\date{\today}
	
\begin{abstract}
This paper provides a theoretical analysis on the effects of an external Zeeman field and Rashba spin-orbit interactions on superconductivity. We have extensively studied their influence on the superconducting transition temperature $T_c$ and the quasiparticle excitation energy. Our investigation includes a detailed examination of both the $s$-wave and $p$-wave pairing states. Implications for the recently discovered family of superconductors, A$_2$Cr$_3$As$_3$ (A = Na, K, Rb and Cs), as well as the validation of our theory have been discussed.
\end{abstract}
	
\maketitle

\section{Introduction}

Superconductivity, acknowledged as a unique phenomenon of macroscopic quantum coherence that exhibits zero electrical resistance and the Meissner-Ochsenfeld effect (perfect diamagnetism), has remained a central topic in modern physics since its discovery by H. Kamerlingh Onnes in 1911~\cite{tinkham}.
Two key features of superconductivity are the transition temperature $T_c$ and the pairing symmetry of the Cooper pairs.
These characteristics can be experimentally examined by observing responses to external electric and magnetic fields.

As an example, the Knight shift and the spin-lattice relaxation rate $1/T_1$ in nuclear magnetic resonance (NMR) are crucial to assessing spin susceptibility in superconducting states, which in turn reveals the pairing symmetry through spin channels~\cite{BCS1957,Leggett1975,AndersonandMorel1961,AndersonandBrinkman1973,Balian1963}.
In the majority of theoretical investigations, an electric or magnetic field imposed on a quantum many-particle system is considered as a perturbation, resulting in linear response theory. However, in NMR measurements, a substantial signal is detectable only with a strong external magnetic field. The energy scale of such a magnetic field can match or surpass the superconducting energy gap; thus, the applied magnetic field should not be considered a mere perturbation. Instead, one must determine the ground state of the many-particle system under the influence of this external magnetic field. Consequently, the system's responses to the external magnetic field should be assessed based on this revised ground state.

Alongside the applied magnetic field, the spin-orbit interaction modifies the superconducting ground state as well. In a crystalline solid, this interaction arises from a broken inversion symmetry, leading to the splitting of energy bands that would otherwise be degenerate. These effects are commonly referred to as the Dresselhaus effect~\cite{Dresselhaus} or the Rashba effect~\cite{Rashiba1960}. The primary difference originates from either the inherent asymmetry within the uniaxial crystals or the irregularity of the interface or surface.  

The spin-orbit coupling (SOC) of electrons arises from relativistic effects, as elaborated in Ref.~\cite{dresselhaus2007group}. 
In the context of symmetry-based energy band theory, notably through $\mathbf{k}\cdot\mathbf{p}$ perturbation theory, the spin-orbit interaction for an individual electron is conventionally represented as:
$H_{\text{SO}} = \mathbf{g}_{\mathbf{k}}\cdot\hat{\sigma}$,
where $\hat{\sigma}$ is a vector consisting of the three Pauli matrices, and $\mathbf{g}_{\mathbf{k}}$ includes only odd powers of the linear momentum $\mathbf{k}$. Furthermore, this paper examines a simplified form of SOC, defined by
\begin{equation*} 
H_{\text{SO}} =
g\mathbf{k}\cdot\hat{\sigma},
\end{equation*}
with the particular simplification $\mathbf{g}_{\mathbf{k}}=g\mathbf{k}$. 
It is noted that this form is straightforward, preserving time reversal symmetry while breaking both spin rotational and spatial inversion symmetry. Although it does not exactly align with the specifics of a typical non-centrosymmetric superconductor, as a leading term, it aptly captures the fundamental physics and is commonly used in theoretical studies~\cite{bauer2012non,Yip14,Samohkin15,Smidman17}. 

In the late 1950s, NMR experiments conducted on BCS ($s$-wave pairing) superconductors showed that the Knight shift remains nearly unchanged below $T_c$ for heavy elements such as Hg~\cite{Reif1957} and Sn~\cite{AndroesKnight1959}. Following these findings, several theoretical papers with similar titles were published by Ferrell~\cite{Ferrell1959}, Martin and Kadanoff~\cite{Kadanoff1959}, Schrieffer~\cite{Schrieffer1959}, and Anderson~\cite{AndersonKnight1959}. Anderson~\cite{AndersonKnight1959} proposed that the slight changes observed in the NMR Knight shift are due to disturbances in spin conservation, which stem from significant SOC in elements with high atomic numbers and tiny superconducting grains. He presented this idea through the notation of time-reversal pairing states~\cite{Andersontheoryofdirty1959}. These pioneering investigations have attracted significant attention and stimulated numerous theoretical studies~\cite{Appel1965,shiba1976effect,zhogolev1972magnetic}. Recently, there has been considerable research on superconducting systems with broken spatial inversion symmetry that host considerable Rashba SOC in the bulk ~\cite{Frigeri2004,Frigeri_2004,Samokhin2007,Edelstein2008,bauer2012non,Yip14,Samohkin15,Smidman17}.

In this study, we comprehensively examined the effects of a Zeeman magnetic field and Rashba SOC on both $s$-wave and p-wave pairing superconductors. The paper is organized as follows. Section II introduces the model Hamiltonian and presents a brief discussion of our results and their analysis. Section III investigates the influence of a Zeeman field and/or Rashba SOC on an s-wave pairing superconductor. Next, Section IV examines various $p$-wave pairing superconductors, emphasizing their markedly different responses to the Zeeman field and/or Rashba SOC in contrast to the s-wave superconductor. Finally, Section V provides a summary and our conclusions. For the reader's convenience, Table~\ref{tab:notation} lists some commonly used notations throughout the paper.

\begin{table}[tb]
\renewcommand\arraystretch{2.0}
\setlength{\tabcolsep}{1.0ex}
\centering
\caption{Common notations.}\label{tab:notation}
\begin{tabular}{c|l}
\hline\hline
Symbols & Meanings\\ \hline
$g$ & Strength of Rashba SOC \\ \hline
$\mathbf{H}$ & External Zeeman field \\ \hline
$N(0)$ & Density of states at the Fermi level in the normal state \\ \hline
$\Delta$ & Superconducting gap \\ \hline
$T_c$ & Superconducting transition temperature \\ \hline
$\Delta_0$ & $\Delta(T=0,g=0,\mathbf{H}=0)$ \\ \hline
$E_c$ & Condensation energy, $E_c=F_{n}(T=0)-F_{s}(T=0)$ \\ \hline
$E_{c0}$ & $E_{c}(g=0,\mathbf{H}=0)$ \\ \hline
$H_0$ & Defined by $E_{c0}=N(0)\mu_B^2{}H_{0}^2$\footnote{For the $s$-wave pairing state, $H_0=\Delta_0/\sqrt{2}\mu_B$; for the pairing state $k_z\hat{z}$, $H_0=\Delta_0/\sqrt{6}\mu_B$; and for the other $p$-wave pairing states considered in this paper, $H_0=\Delta_0/\sqrt{3}\mu_B$.} \\ \hline
$H_P$ & Pauli limit of $s$-wave pairing state, $H_P\approx H_0=\Delta_0/\sqrt{2}\mu_B$ \\ \hline
$T_{c0}$ & $T_{c}(g=0,\mathbf{H}=0)$\\ \hline
$T_{c}^{\pm}$ & $T_c\pm{}0^{+}$\\ \hline
\end{tabular}
\end{table} 

\section{Theoretical model and formulation}
	
{\bf{}Mean field Hamiltonian:} Generally, a standard mean field Hamiltonian that characterizes a superconducting system includes two parts: the unpaired component $H_0$ and the paired component $H_{\text{SC}}$, and can be expressed as follows:
\begin{equation}\label{hamilt} \begin{aligned}
H&=H_0+H_{\text{SC}},\\
H_0&=\sum_{\mathbf{k},\alpha,\beta}c_{\mathbf{k},\alpha}^{\dagger}H^{\alpha\beta}_0(\mathbf{k})c_{\mathbf{k},\beta},\\
H_{\text{SC}}&=\frac{1}{2}\sum_{\mathbf{k},\alpha,\beta}[c_{\mathbf{k},\alpha}^{\dagger}\Delta_{\alpha\beta}(\mathbf{k})c_{-\mathbf{k},\beta}^{\dagger}+h. c. ],
\end{aligned}\end{equation} 
with $\mathbf{k}$ representing the wave vector, $\alpha,\beta=\uparrow, \downarrow$ denoting spin indices, and $c_{\mathbf{k},\alpha}^{\dagger}$ ($c_{\mathbf{k},\alpha}$) indicating the creation (annihilation) of an electron having wave vector $\mathbf{k}$ and spin $\alpha$.

{\bf{}Normal state:} First, we shall consider the normal state of the system, which is governed by the unpaired part $H_0$. In the presence of an external Zeeman field $\mathbf{H}$ and Rashba SOC with coupling strength $g$, the unpaired part $H_0$ is given by
\begin{equation} \begin{aligned}
H_0(\mathbf{k})=\xi_{\mathbf{k}}\sigma_0+\mu_B\mathbf{H}\cdot\hat{\sigma}+g\mathbf{k}\cdot\hat{\sigma},
\label{hamilnormal}
\end{aligned}\end{equation}
where $\xi_{\mathbf{k}}$ is the energy of an electron with a wave vector $\mathbf{k}$ and measured relative to the chemical potential $\mu$, $\mu_B$ is the Bohr magneton. $\sigma_0$ is the identity matrix, and $\hat{\sigma}$ comprises a three-component vector of Pauli matrices.

\begin{figure}[tb]
\centering
\subfigure[]{
\label{Fig:FS}
\includegraphics[width=0.9\linewidth]{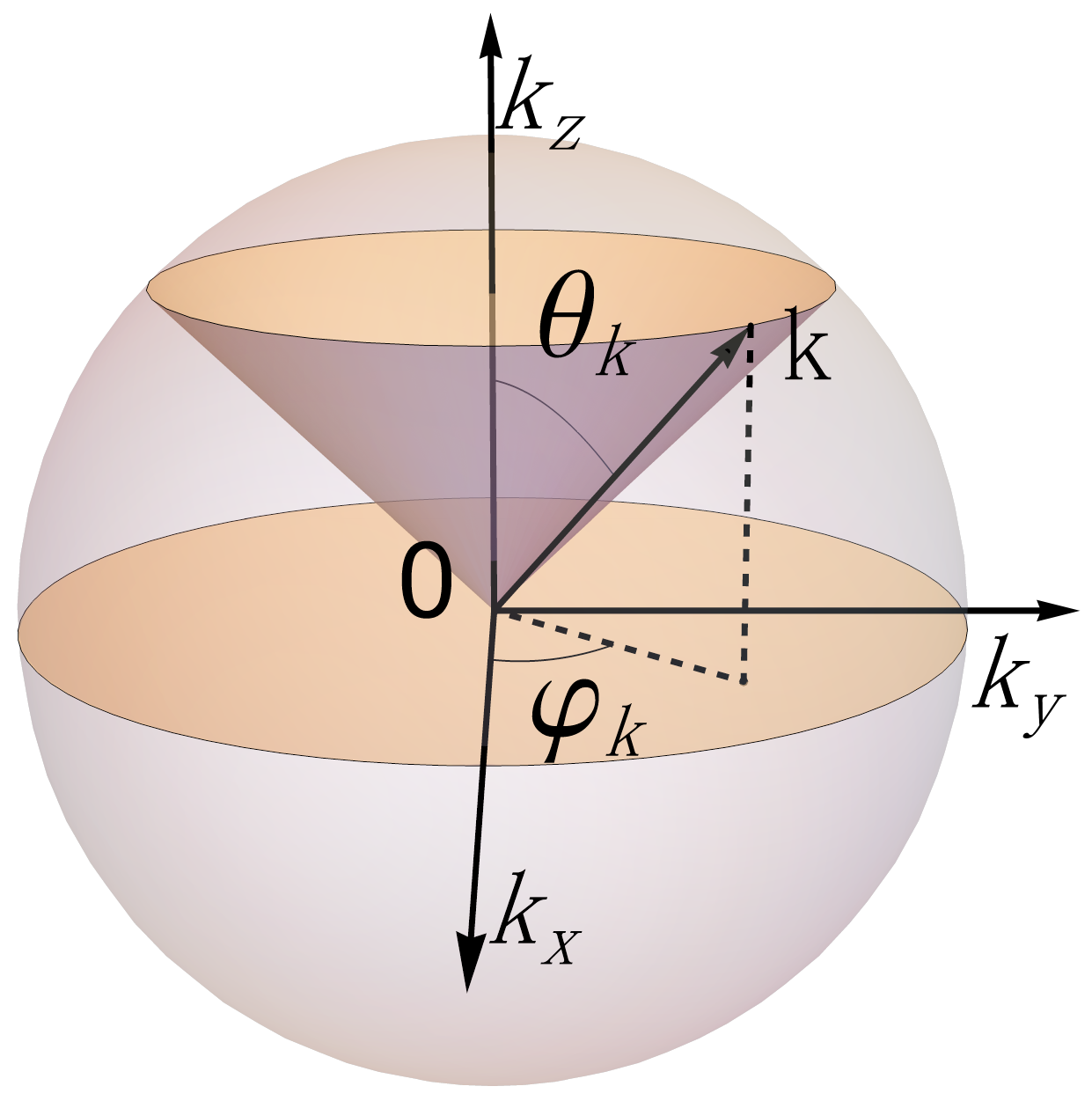} }
\subfigure[]{
\label{Fig:dispersion}
\includegraphics[width=0.96\linewidth]{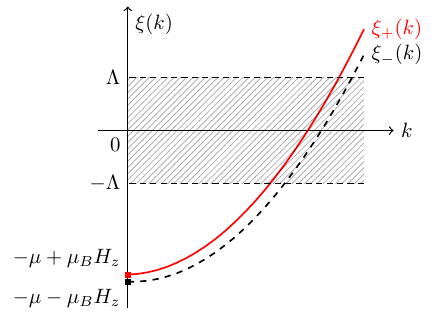}}
\caption{ (a) A spherical Fermi surface and (b) the energy dispersion $\xi_{\pm}(k)$ of the normal state Hamiltonian $H_0$ as given in Eq.~\eqref{eq:xi0}. Here $k=|\mathbf{k}|$, $\mu$ is chemical potential, and $\Lambda$ is the energy cutoff of the Cooper pairing. For an $s$-wave superconductor, $\Lambda=\hbar\omega_D$, where $\omega_D$ is the Debye frequency.
}
\end{figure}

Therefore, the normal state electrons can be considered as moving within an effective field $\mu_B\mathbf{H}+g\mathbf{k}$ in the direction indicated by the unit vector:
\begin{subequations}\label{eq:nk}
\begin{equation}
\hat{n}_{\mathbf{k}}(\mathbf{H}) = \frac{\mu_B\mathbf{H}+g\mathbf{k}}{\lvert\mu_B\mathbf{H}+g\mathbf{k}\rvert}, 
\end{equation}
which are characterized by two angles, $\Theta_\mathbf{k}$ and $\Phi_\mathbf{k}$, as follows,
\begin{equation}
\hat{n}_{\mathbf{k}} = (\sin\Theta_\mathbf{k}\cos\Phi_\mathbf{k},\sin\Theta_\mathbf{k}\sin\Phi_\mathbf{k},\cos\Theta_\mathbf{k}).
\end{equation}
\end{subequations}
Explicitly, $\Theta_\mathbf{k}$ and $\Phi_\mathbf{k}$ can be formulated as: \begin{equation}\label{eq:Thetak}
\begin{split}
\Theta_\mathbf{k} &= \arccos\frac{\mu_B{}H_z+g|\mathbf{k}|\cos\theta_\mathbf{k}}{\lvert\mu_B\mathbf{H}+g\mathbf{k}\rvert},\\
\tan\Phi_\mathbf{k} &= \frac{\mu_B{}H_y+g|\mathbf{k}|\sin{\theta_\mathbf{k}}\sin\varphi_\mathbf{k}}{\mu_BH_x+g|\mathbf{k}|\sin{\theta_\mathbf{k}}\cos\varphi_\mathbf{k}},
\end{split}
\end{equation} 
where $\theta_{\mathbf{k}}$ and $\varphi_{\mathbf{k}}$ denote the polar and azimuthal angles of the wave vector $\mathbf{k}$, respectively, as illustrated in Fig.~\ref{Fig:FS}. Note that $\Theta_{\mathbf{k}}$ is specified within the region $[0,\pi]$. Notably, when the external Zeeman field aligns with the $z$ direction, i.e., $\mathbf{H}=H_{z}\hat{z}$, $\Phi_{\mathbf{k}}$ reduces to: $\Phi_\mathbf{k} = \varphi_\mathbf{k}$.

The eigenvalues of $H_0(\mathbf{k})$ are calculated as follows,
\begin{subequations}\label{eq:eigenH}
\begin{equation}\label{eq:xi0}
\xi_{\mathbf{k}\pm} = \xi_{\mathbf{k}}\pm \lvert\mu_B\mathbf{H}+g\mathbf{k}\rvert,
\end{equation}
and the corresponding eigenstates are found to be
\begin{equation}\label{eigennewp}
\lvert\mathbf{k},+\rangle = \lvert\mathbf{k}\rangle \otimes\begin{pmatrix}
\cos\frac{\Theta_{\mathbf{k}}}{2}\\
\sin\frac{\Theta_{\mathbf{k}}}{2}\mathrm{e}^{i\Phi_{\mathbf{k}}}
\end{pmatrix},\,
\lvert\mathbf{k}, -\rangle = \lvert \mathbf{k}\rangle \otimes
\begin{pmatrix}
\sin\frac{\Theta_{\mathbf{k}}}{2}\\
-\cos\frac{\Theta_{\mathbf{k}}}{2}\mathrm{e}^{i\Phi_{\mathbf{k}}}
\end{pmatrix}.
\end{equation}
\end{subequations}

{\bf{}Pairing interaction and gap equation:} In general, Cooper pairs arise from an effective attraction, which can be expressed in a four-fermion form as follows:
\begin{equation}\label{eq:Hint}
H_{int} = -\frac{1}{2}\sum_{\mathbf{k},\mathbf{k}^{\prime},\alpha,\beta,\alpha^{\prime},\beta^{\prime}} V_{\beta\alpha\alpha^{\prime}\beta^{\prime}}(\mathbf{k},\mathbf{k}^{\prime}) c_{\mathbf{k},\alpha}^\dagger c_{-\mathbf{k},\beta}^\dagger c_{\mathbf{k}^{\prime},\alpha^{\prime}} c_{-\mathbf{k}^{\prime},\beta^{\prime}},    
\end{equation}
where $V_{\beta\alpha\alpha^{\prime}\beta^{\prime}}(\mathbf{k},\mathbf{k}^{\prime})$ represents the pairing potential with spin indices $\alpha,\beta,\alpha^{\prime},\beta^{\prime}=\uparrow,\downarrow$.
The mean field decomposition of $H_{int}$ in the pairing channel leads to the Hamiltonian $H_{\text{SC}}$ in Eq.~\eqref{hamilt} and the gap equation as follows~\cite{Sigrist1991},
\begin{equation}\label{gapeqwhole}    
\Delta_{\alpha\beta}(\mathbf{k}) = -\sum_{\mathbf{k}^{\prime},\alpha^{\prime},\beta^{\prime}}V_{\beta\alpha\alpha^{\prime}\beta^{\prime}}(\mathbf{k},\mathbf{k}^{\prime})\langle c_{\mathbf{k}^{\prime},\alpha^{\prime}} c_{-\mathbf{k}^{\prime},\beta^{\prime}}\rangle.
\end{equation}
A generalized Bogoliubov transformation (see Appendix~\ref{app:BTM}) \begin{equation}\label{eq:BT}
c_{\mathbf{k},\alpha}=\sum_{s=\pm}u^{\alpha{}s}_{\mathbf{k}}\psi_{\mathbf{k},s} + v_{\mathbf{k}}^{\alpha{}s}\psi_{-\mathbf{k},s}^{\dagger}
\end{equation} 
allows us to diagonalize the mean field Hamiltonian given in Eq.~\eqref{hamilt} and subsequently solve the gap equation \eqref{gapeqwhole} in a self-consistent method. Here, $\psi_{\mathbf{k},s}$ and $\psi_{\mathbf{k},s}^{\dagger}$ refer to the Bogoliubov quasiparticle operators, and $s=\pm$ denotes the quasiparticle band index, which should be differentiated from that of the normal state given in Eq.~\eqref{eigennewp}. Furthermore, we will represent the quasiparticle energy for a superconducting state as $E_{\mathbf{k}\pm}$ hereafter, highlighting its distinction from Eq.~\eqref{eq:xi0}.

{\bf{}Pairing channels:} Assuming a spherical Fermi surface depicted in Fig.~\ref{Fig:FS} and a rotationally symmetric pairing interaction, the matrix element $V(\mathbf{k}, \mathbf{k}')$ can be written as follows~\cite{Leggett1975,Sigrist1991}, \begin{equation}\label{eq:Vl}
V(\mathbf{k},\mathbf{k}')=\sum_{l=0}^{\infty}(2l+1)V_lP_l\left(\cos\theta_{\mathbf{k}, \mathbf{k}'}\right).
\end{equation}
Here, $\theta_{\mathbf{k},\mathbf{k}'}$ is the angle between $\mathbf{k}$ and $\mathbf{k}'$, and $P_l(x)$ are Legendre polynomials. In particular, the $l=0$ channel with $V_0>0$ leads to $s$-wave pairing, while the primary channel for $p$-wave pairing involves $l=1$ with $V_{1}>0$. 

Our theoretical model is characterized by $V(\mathbf{k},\mathbf{k}')$ along with the pairing channel parameters $V_l$ outlined in Eq.~\eqref{eq:Vl}. Consequently, the corresponding gap equation~\eqref{gapeqwhole} can be derived and solved through the Bogoliubov transformation.

{\bf{}Pairing functions:} In the paired part $H_{\text{SC}}$, the pairing function $\Delta(\mathbf{k})$ takes the form of a $2\times{}2$ matrix: 
\begin{equation} \begin{aligned}\label{eq:generic-pair}
\Delta(\mathbf{k})=\begin{pmatrix}
\Delta_{\uparrow\uparrow}(\mathbf{k})&\Delta_{\uparrow\downarrow}(\mathbf{k})\\
\Delta_{\downarrow\uparrow}(\mathbf{k})&\Delta_{\downarrow\downarrow}(\mathbf{k})
\end{pmatrix}.
\end{aligned}\end{equation}
Following Balian and Werthamer~\cite{Balian1963}, we obtain
\begin{equation}
\Delta(\mathbf{k})=i\Delta\sigma_y
\label{dels}
\end{equation}
for an $s$-wave and spin-singlet pairing superconductor, and
\begin{equation}
\Delta(\mathbf{k})=i\left[\mathbf{d}(\mathbf{k})\cdot \hat{\sigma}\right]\sigma_y
\end{equation}
for a $p$-wave and spin-triplet pairing superconductor. In this context, $\Delta$ represents the isotropic pairing function and $\mathbf{d}(\mathbf{k})$ is a complex vector, dubbed ``d-vector".

{\bf{}Pairing functions in the pseudo-spin basis:} To facilitate further analysis, the pairing Hamiltonian $H_{\text{SC}}$ can alternatively be written in the pseudo-spin basis as follows, 
\begin{equation}
H_{\text{SC}}=\frac{1}{2}\sum_{\mathbf{k},s,s^{\prime}}[c_{\mathbf{k},s}^{\dagger}\Delta_{ss^{\prime}}(\mathbf{k})c_{-\mathbf{k},s^{\prime}}^{\dagger}+h. c. ],
\end{equation}
where $s,s^{\prime}=\pm$ label the two eigenstates specified in Eq.~\eqref{eq:eigenH}, and $\Delta_{ss^{\prime}}(\mathbf{k})$ are defined by
\begin{widetext}
\begin{equation}\label{eq:pair-pspin}
\begin{split}
\Delta_{++} = & \cos\frac{\Theta_\mathbf{k}}{2}\left(\Delta_{\uparrow\uparrow}\cos\frac{\Theta_{-\mathbf{k}}}{2}+\mathrm{e}^{-i\Phi_{-\mathbf{k}}}\Delta_{\uparrow\downarrow}\sin\frac{\Theta_{-\mathbf{k}}}{2}\right) + \mathrm{e}^{-i\Phi_\mathbf{k}}\sin\frac{\Theta_\mathbf{k}}{2}\left(\Delta_{\downarrow\uparrow}\cos\frac{\Theta_{-\mathbf{k}}}{2}+\mathrm{e}^{-i\Phi_{-\mathbf{k}}}\Delta_{\downarrow\downarrow}\sin\frac{\Theta_{-\mathbf{k}}}{2}\right),\\
\Delta_{--} = & \sin\frac{\Theta_\mathbf{k}}{2}\left(\Delta_{\uparrow\uparrow}\sin\frac{\Theta_{-\mathbf{k}}}{2}-\mathrm{e}^{-i\Phi_{-\mathbf{k}}}\Delta_{\uparrow\downarrow}\cos\frac{\Theta_{-\mathbf{k}}}{2}\right) -\mathrm{e}^{-i\Phi_\mathbf{k}}\cos\frac{\Theta_\mathbf{k}}{2}\left(\Delta_{\downarrow\uparrow}\sin\frac{\Theta_{-\mathbf{k}}}{2}-\mathrm{e}^{-i\Phi_{-\mathbf{k}}}\Delta_{\downarrow\downarrow}\cos\frac{\Theta_{-\mathbf{k}}}{2}\right),\\
\Delta_{+-} = & \cos\frac{\Theta_\mathbf{k}}{2}\left(\Delta_{\uparrow\uparrow}\sin\frac{\Theta_{-\mathbf{k}}}{2}-\mathrm{e}^{-i\Phi_{-\mathbf{k}}}\Delta_{\uparrow\downarrow}\cos\frac{\Theta_{-\mathbf{k}}}{2}\right) + \mathrm{e}^{-i\Phi_\mathbf{k}}\sin\frac{\Theta_\mathbf{k}}{2}\left(\Delta_{\downarrow\uparrow}\sin\frac{\Theta_{-\mathbf{k}}}{2}-\mathrm{e}^{-i\Phi_{-\mathbf{k}}}\Delta_{\downarrow\downarrow}\cos\frac{\Theta_{-\mathbf{k}}}{2}\right),\\
\Delta_{-+} = & \sin\frac{\Theta_\mathbf{k}}{2}\left(\Delta_{\uparrow\uparrow}\cos\frac{\Theta_{-\mathbf{k}}}{2}+\mathrm{e}^{-i\Phi_{-\mathbf{k}}}\Delta_{\uparrow\downarrow}\sin\frac{\Theta_{-\mathbf{k}}}{2}\right) - \mathrm{e}^{-i\Phi_\mathbf{k}}\cos\frac{\Theta_\mathbf{k}}{2}\left(\Delta_{\downarrow\uparrow}\cos\frac{\Theta_{-\mathbf{k}}}{2}+\mathrm{e}^{-i\Phi_{-\mathbf{k}}}\Delta_{\downarrow\downarrow}\sin\frac{\Theta_{-\mathbf{k}}}{2}\right).
\end{split}
\end{equation}
\end{widetext}
Note that $\mbox{tr}\left[\Delta(\mathbf{k})\Delta(\mathbf{k})^{\dagger}\right]$ does not depend on the spin basis, and hence we can state the following relation:
$|\Delta_{++}|^2+|\Delta_{+-}|^2+|\Delta_{-+}|^2+|\Delta_{--}|^2=|\Delta_{\uparrow\uparrow}|^2+|\Delta_{\uparrow\downarrow}|^2+|\Delta_{\downarrow\uparrow}|^2+|\Delta_{\downarrow\downarrow}|^2$.
More information on the intraband and interband pairing functions can be found in Appendix~\ref{app:Theta}.

\section{$s$-wave pairing state}

The most simplest model for an $s$-wave pairing superconducting state can be achieved by setting 
\begin{equation*}
V_0>0,\,\,\,\mbox{and}\,\,\, V_{l}=0\,\,\,\mbox{for}\,\,\, l\geq{}1    
\end{equation*}
in the pairing potential as described in Eq.~\eqref{eq:Vl}. 

In the remaining part of this section, we will explore the impact of the Zeeman field and Rashba SOC on the superconducting transition temperature $T_c$ and the quasiparticle excitations in an $s$-wave pairing superconductor.

\subsection{Superconducting transition temperature}

\begin{figure*}[tb]
\centering
\subfigure[]{
\includegraphics[width=0.49\linewidth]{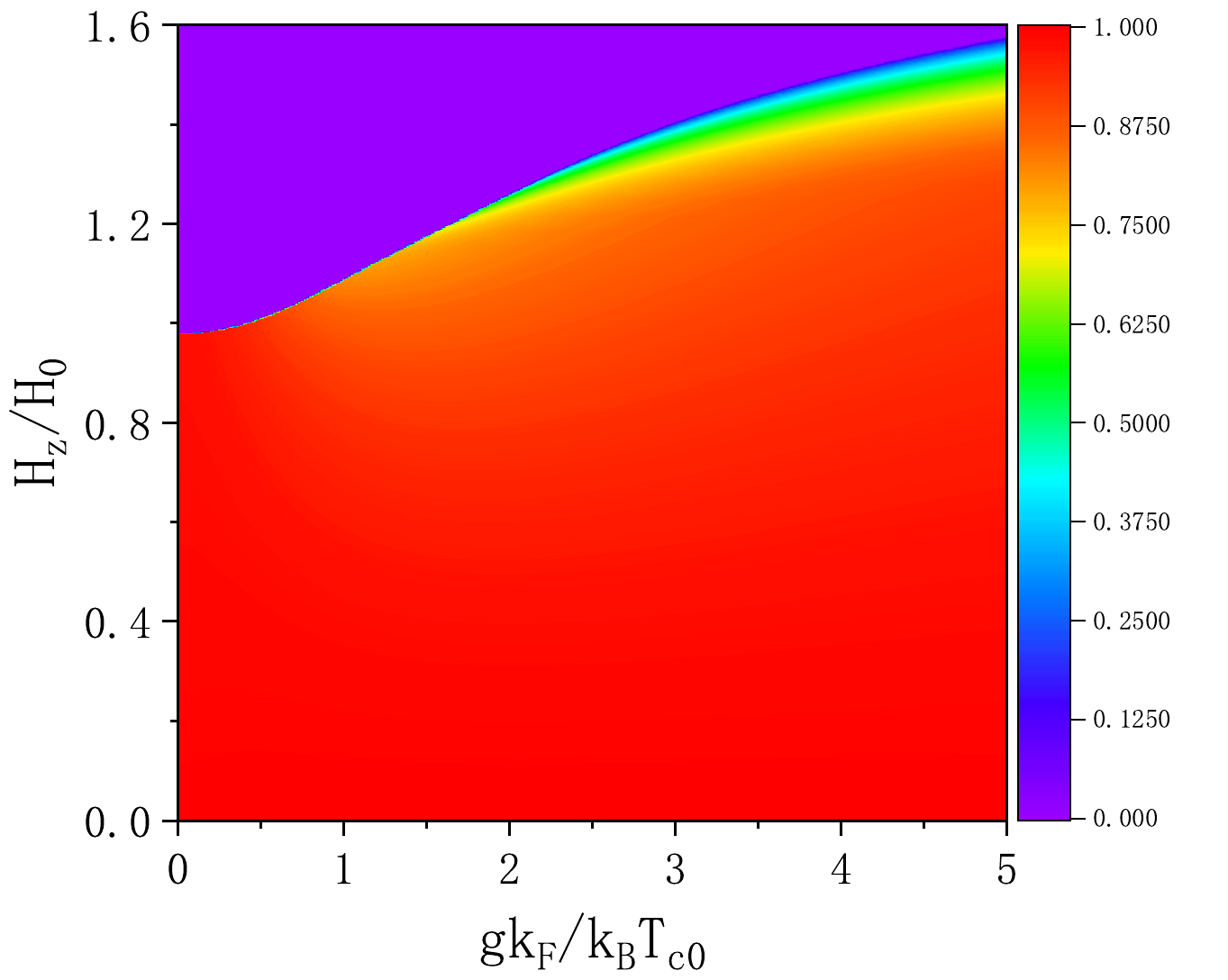}}
\subfigure[]{
\includegraphics[width=0.49\linewidth]{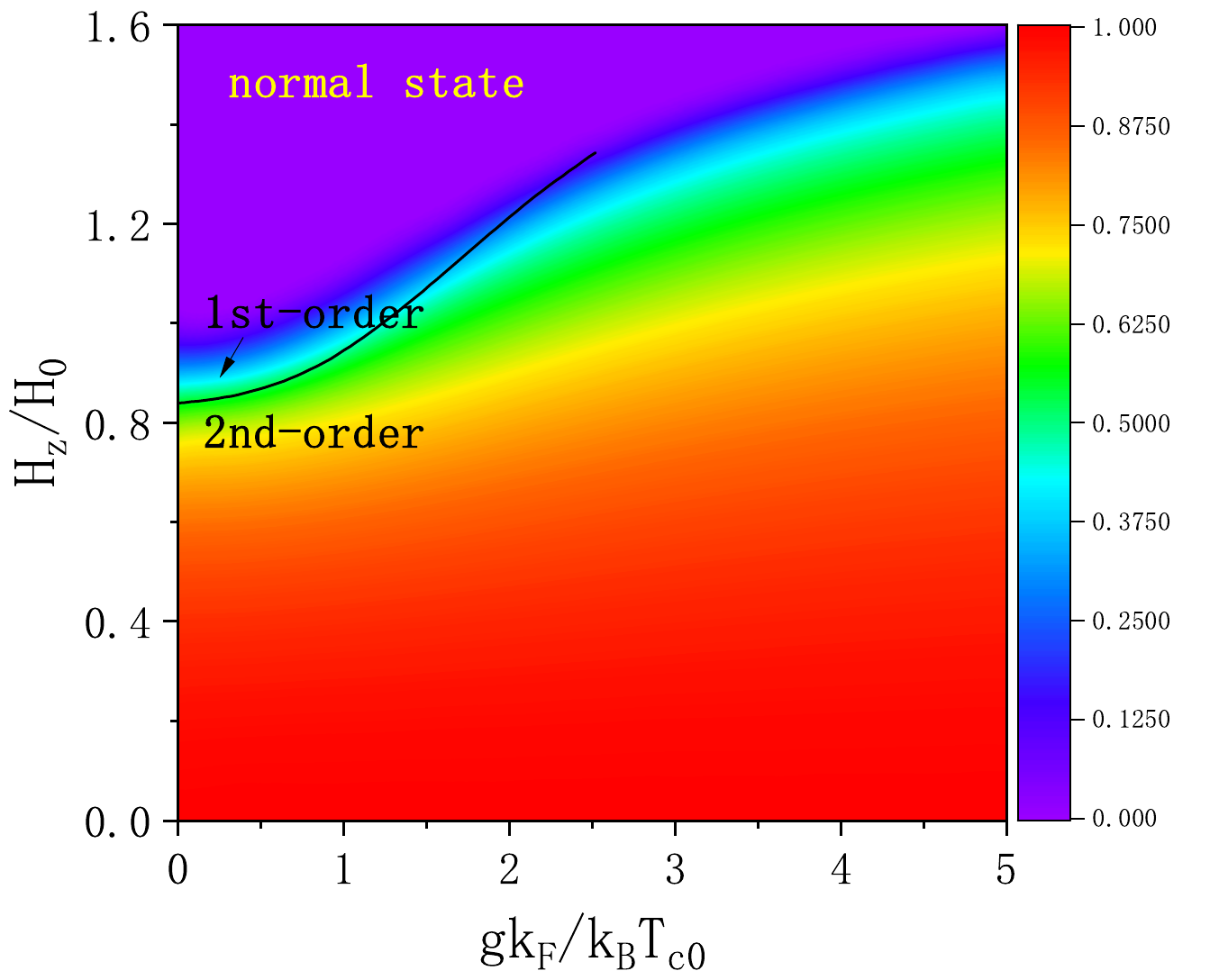}}
\caption{ The $s$-wave pairing state: (a) The superconducting gap at absolute zero temperature $\Delta(T=0)$ measured in $\Delta_0$, and (b) the superconducting transition temperature $T_c$ measured in $T_{c0}$ as functions of the Rashba SOC $g$ and the Zeeman field $H_z$, determined self-consistently from Eq.~\eqref{eq:sgap}. The solid line in (b) represents the boundary between the first- and second-order phase transitions across $T_c$.}
\label{fig:gap_magSOC-s}
\end{figure*}

By iterative solving Eq.~\eqref{eq:sgap}, one can determine the superconducting gap $\Delta(T)$ and the superconducting transition temperature $T_c$. Fig.~\ref{fig:gap_magSOC-s} illustrates how the zero temperature superconducting gap $\Delta(T=0)$ and the superconducting transition temperature $T_c$ vary with the Rashba SOC and the Zeeman field. In what follows, we will examine the effects of the Zeeman field and Rashba SOC individually, before analyzing their combined influence.

{\bf{}Zeeman field effect:} Assuming without loss of generality, we consider a Zeeman field $\mathbf{H}=H_{z}\hat{z}$. The Zeeman field aligns the electron spin, which hinders the formation of a spin-singlet Cooper pair and reduces $T_c$. When $H_z=0$, the superconducting $T_c$ reaches its maximum value of $T_{c0}=T_{c}(H_z=0)$, but when the Zeeman field $H_z$ exceeds the Pauli limit~\cite{Clogston1962,Maki1964}, the superconducting $T_c$ vanishes. According to Maki and Tsuneto~\cite{Maki1964}, there exists a critical threshold of $T_0=0.556T_{c0}$, which dictates whether the superconductor to normal state phase transition is second or first order. Specifically, the phase transition is  second order when $T_c>T_0$, and becomes of first order when $T_c<T_0$.

{\bf{}Effect of Rashba SOC:} Without the Zeeman field, Rashba SOC affects neither the superconducting gap $\Delta(T)$ nor the superconducting transition temperature $T_c$. The unchanged $\Delta(T)$ and $T_c$ are attributed to time-reversal symmetry and the time-reversal pairing mechanism proposed by Anderson~\cite{Andersontheoryofdirty1959}.

\begin{figure}[tb]
\centering
\includegraphics[width=1.0\linewidth]{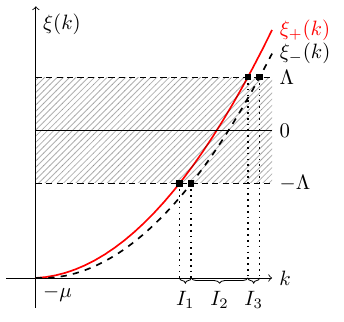} 
\caption{Normal state energy dispersion $\xi_{\pm}(\mathbf{k})$ in the presence of a finite Rashba SOC and vanishing Zeeman field. The shadow area indicates the energy window where Cooper pairs are formed. Here $\Lambda$ is the energy cutoff, and $\mu$ is the chemical potential.}
\label{soc}
\end{figure}

In particular, the normal state Hamiltonian $H_{0}$ can be represented in the pseudo-spin basis of $\{\lvert\mathbf{k}, \pm\rangle\}$ as characterized in Eq.~\eqref{eigennewp}, and corresponds to
\begin{equation}\label{hamilsoc}
H_0(\mathbf{k})=\xi_{\mathbf{k}}\sigma_0+g\lvert\mathbf{k}\lvert \sigma_z,
\end{equation}
leading to the normal state energy dispersion $\xi_{\mathbf{k}\pm}=\xi_{\mathbf{k}}\pm{}g|\mathbf{k}|$ illustrated in Fig.~\ref{soc}. In the $\mathbf{k}$-space, the Cooper pairing region restricted by the energy cutoff $\Lambda$ can be divided into three regions: $I_1$, $I_2$, and $I_3$. In area $I_{1(3)}$, Cooper pairs are limited to the $\xi_{+(-)}(\mathbf{k})$ branch, whereas in area $I_2$ they appear in both branches.
Consequently, the $s$-wave pairing function $\Delta(\mathbf{k})$ given in Eq.~\eqref{dels} can be explicitly expressed in the pseudo-spin basis:
\begin{equation}
\Delta(\mathbf{k})=\left\{\begin{aligned}
&\,-\frac{1}{2}\Delta \mathrm{e}^{-i\varphi_{\mathbf{k}}}(\sigma_0+\sigma_z), &\mathbf{k}\in I_1,\\
&\,-\Delta \mathrm{e}^{-i\varphi_{\mathbf{k}}}\sigma_z,  &\mathbf{k}\in I_2,\\
&\,\frac{1}{2}\Delta \mathrm{e}^{-i\varphi_{\mathbf{k}}}(\sigma_0-\sigma_z), &\mathbf{k}\in I_3.
\end{aligned}
\right.
\end{equation}
This diagonal form without interband pairing is anticipated in the basis formed by the Kramers pair and is owing to the rotational symmetry of the $s$-wave pairing. Furthermore, we can conclude that the diagonal form of the pairing function $\Delta(\mathbf{k})$ results in a gap equation independent of the strength of Rashba SOC (see Appendix~\ref{app:BT} for more details). Therefore, both the gap function $\Delta(T)$ and the superconducting $T_c$ remain unchanged, as confirmed by numerical verification.

\begin{figure}[tb]
\centering
\includegraphics[width=1.0\linewidth]{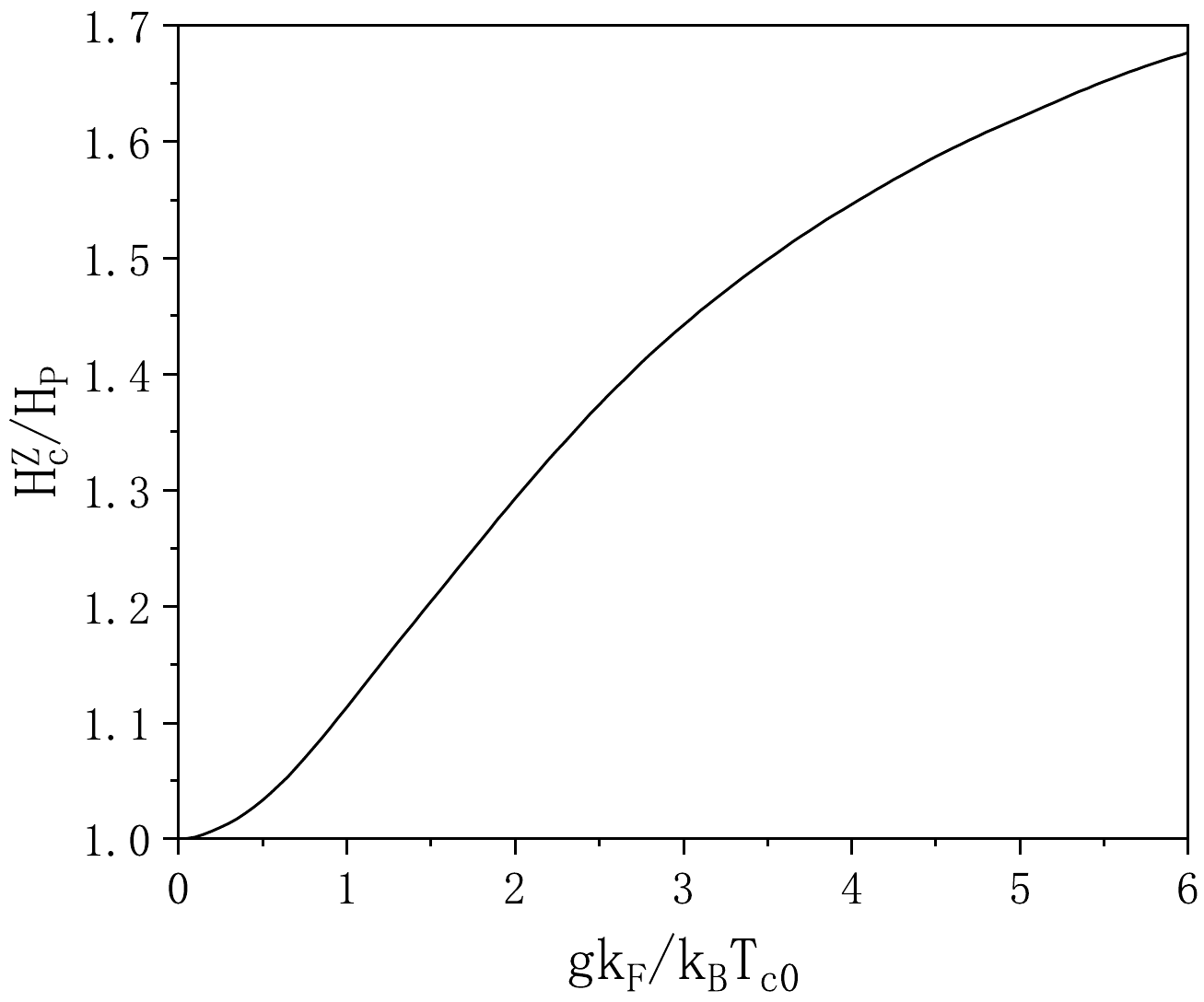}
\caption{The $s$-wave pairing state: The threshold Zeeman field $H_c^{Z}$ plotted in $H_P$ versus $gk_F/k_B{}T_{c0}$.
}
\label{crizeeman}
\end{figure}

{\bf{}Combination of Zeeman field and Rashba SOC:} When both the Zeeman field and Rashba SOC are present, as illustrated in Fig.~\ref{fig:gap_magSOC-s}, they give rise to contrasting effects on $T_c$:
\begin{itemize}
\item{} The Zeeman field generally reduces $T_c$. Regardless of the presence of Rashba SOC, $T_c$ decreases as $H_z$ increases and it is entirely suppressed once $H_z$ reaches a threshold $H_c^{Z}$. As illustrated in Fig.~\ref{crizeeman}, the critical value $H_c^{Z}$ increases monotonically with respect to $g$.
\item{} Furthermore, when $g$ is finite, the critical Zeeman field $H_c^{Z}$ exceeds the Pauli limit $H_P$, as can be seen in Fig.~\ref{crizeeman}. 
\item{} When $0<H_z<H_{c}^{Z}$, Rashba SOC enhances $T_c$ and tends to return it to the intrinsic value $T_{c0}$ in the limit $g{}k_F/\mu_B{}H_z\to\infty$.
\end{itemize}

The relationship between superconducting $T_c$ and both $H_z$ and $g$ can be understood as follows: By expressing the $s$-wave pairing state in the pseudo-spin basis defined in Eqs.~\eqref{eq:pair-pspin}, we can derive the intraband and interband pairing functions:
\begin{equation}\label{eq:s-pseudo-spin}
\begin{split}
&\Delta_{++}=-\Delta_{--}=-\Delta{}\mathrm{e}^{-i\varphi_{\mathbf{k}}} \sin{\frac{\Theta_{\mathbf{k}}+\Theta_{-\mathbf{k}}}{2}},\\
&\Delta_{+-}=\Delta_{-+}=\Delta{}\mathrm{e}^{-i\varphi_{\mathbf{k}}} \cos{\frac{\Theta_{\mathbf{k}}+\Theta_{-\mathbf{k}}}{2}}.
\end{split}
\end{equation} 
As the ratio of $g{}k_{F}/\mu_{B}H_z$ increases from $0$ to $1$ and subsequently to $\infty$, the angle $(\Theta_{\mathbf{k}}+\Theta_{\mathbf{-k}})/2$ increases consistently from $0$ to $\pi/4$ and then to $\pi/2$ (see Appendix~\ref{app:Theta}). Hence, it can be inferred that:

\emph{For an $s$-wave pairing superconductor, an external Zeeman field favors the interband pairing with $\Delta_{+-}=\Delta_{-+}$, while Rashba SOC promotes the intraband pairing with $\Delta_{++}=-\Delta_{--}$.}

When $H_z=0$, the interband pairing is completely eliminated, which means $\Delta_{+-}=\Delta_{-+}=0$, but the intraband pairing stays the same. In this scenario, Rashba SOC has no effect on either $\Delta(T)$ or $T_c$. Meanwhile, in the situation where $H_z\neq{}0$, increasing $g$ enhances the intraband pairing while reducing the interband pairing, leading to a rise $T_c$ that almost reaches $T_{c0}$ as $g{}k_F\gg{}\mu_B{}H_z$.

\subsection{Quasiparticles: Bogoliubov Fermi surface}

This subsection examines the quasiparticle excitations influenced by the Zeeman field and/or the Rashba SOC.

{\bf{}Zeeman field effect:} For an $s$-wave pairing state in a single Zeeman field, $\mathbf{H}=H_{z}\hat{z}$, the Hamiltonian described in Eq.~\eqref{hamilt} can be diagonalized through the Bogoliubov transformation, yielding the quasiparticle energy dispersion as follows: 
\begin{equation}\label{eq:Ek-SH}
E_{\mathbf{k}\pm}=E_{\mathbf{k}}\pm\mu_BH_z=\sqrt{\xi_{\mathbf{k}}^2+\Delta^2}\pm\mu_B{}H_z.
\end{equation}
With an increase in $H_z$, the superconducting gap $\Delta$ decreases slowly, on the order of $(\mu_BH_{z}/\hbar\omega_{D})\Delta$, while the quasiparticle energy splitting becomes larger, until $H_z=H_P$ when the quasiparticle excitations turn unstable.

{\bf{}Effect of Rashba SOC:} Next, we activate the Rashba SOC and deactivate the Zeeman field, concentrating on the case where the Rashba SOC is much weaker than the energy cutoff, as indicated in $gk_F\ll\Lambda$. In this context, illustrated in Fig.~\ref{soc}, the primary contribution originates from the $I_2$ region, where the quasiparticle energy spectrum is 
\begin{equation}
E_{\mathbf{k}\pm}=\sqrt{\xi_{\mathbf{k}\pm}^2+ \Delta^2}=\sqrt{(\xi_{\mathbf{k}}\pm g\lvert \mathbf{k}\lvert )^2+ \Delta^2}.
\end{equation}
Therefore, the Rashba SOC merely adjusts the chemical potential $\mu$ by $\pm{}gk_{F}$, leaving the quasiparticle energy gap unaffected.

{\bf{}Combination of Zeeman field and Rashba SOC:}
We continue our investigation of the $s$-wave superconducting state under the influence of both a Zeeman field and Rashba SOC. For simplicity, the Zeeman field is assumed to be aligned along the $z$-axis. As a result, $H_{0}(\mathbf{k})$ and $\Delta(\mathbf{k})$ in Eq.~\eqref{hamilt} can be explicitly expressed as follows:
\begin{equation}
\begin{split}
H_0(\mathbf{k})&=\xi_{\mathbf{k}}\sigma_0+\mu_BH_z\sigma_{z}+g\mathbf{k}\cdot\sigma,\\
\Delta(\mathbf{k})&=i\Delta\sigma_y.
\end{split}
\end{equation}
The energy spectrum of Bogoliubov quasiparticles can be found by diagonalizing the $4\times4$ matrix (see Appendix~\ref{app:BT}). The explicit form can be derived by specifying the polar angle $\theta_{\mathbf{k}}=0,\pi/2,\pi$:
\begin{equation}
\begin{split}
E_{\mathbf{k}\pm}|_{\theta_{\mathbf{k}}=0}=&\sqrt{(\xi_{\mathbf{k}}\pm g|\mathbf{k}|)^2+\Delta^2}\pm \mu_BH_z,\\
E_{\mathbf{k}\pm}|_{\theta_{\mathbf{k}}=\frac{\pi}{2}}=&\left\{\xi_{\mathbf{k}}^2+(g|\mathbf{k}|)^2+(\mu_BH_z)^2+\Delta^2\right.\\
&\left.\pm 2\left[\xi_{\mathbf{k}}^2(g|\mathbf{k}|)^2+\xi_{\mathbf{k}}^2(\mu_BH_z)^2+(\mu_BH_z)^2\Delta^2\right]^{\frac{1}{2}}\right\}^{\frac{1}{2}}, \\
E_{\mathbf{k}\pm}|_{\theta_{\mathbf{k}}=\pi}=&\sqrt{(\xi_{\mathbf{k}}\pm g|\mathbf{k}|)^2+\Delta^2}\mp \mu_BH_z.
\end{split}
\label{Eq:eigen_mag_soc}
\end{equation}

\begin{figure}[tb]
\centering
\subfigure[]{\includegraphics[width=0.36\linewidth]{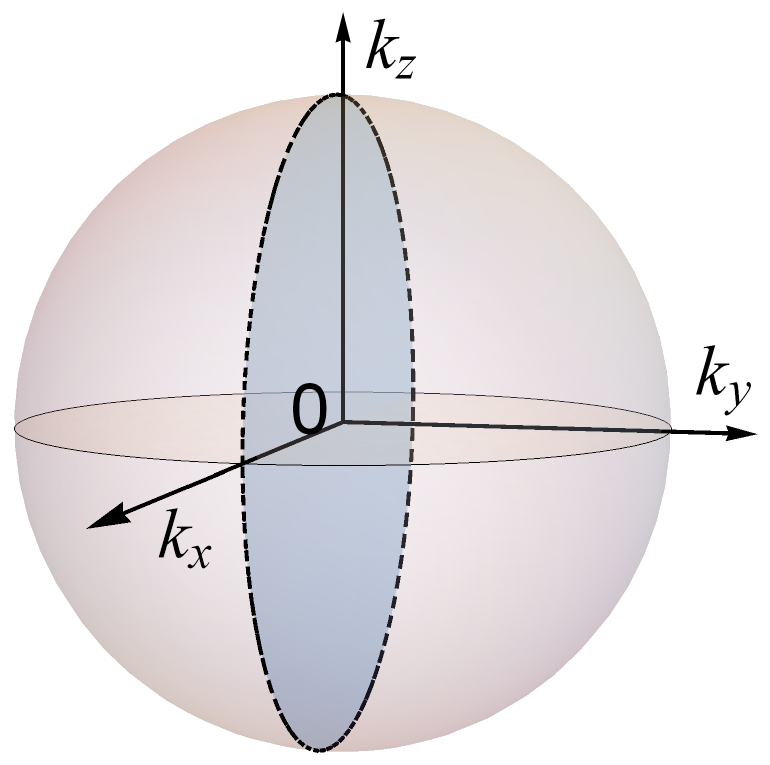} }
\subfigure[]{\includegraphics[width=0.6\linewidth]{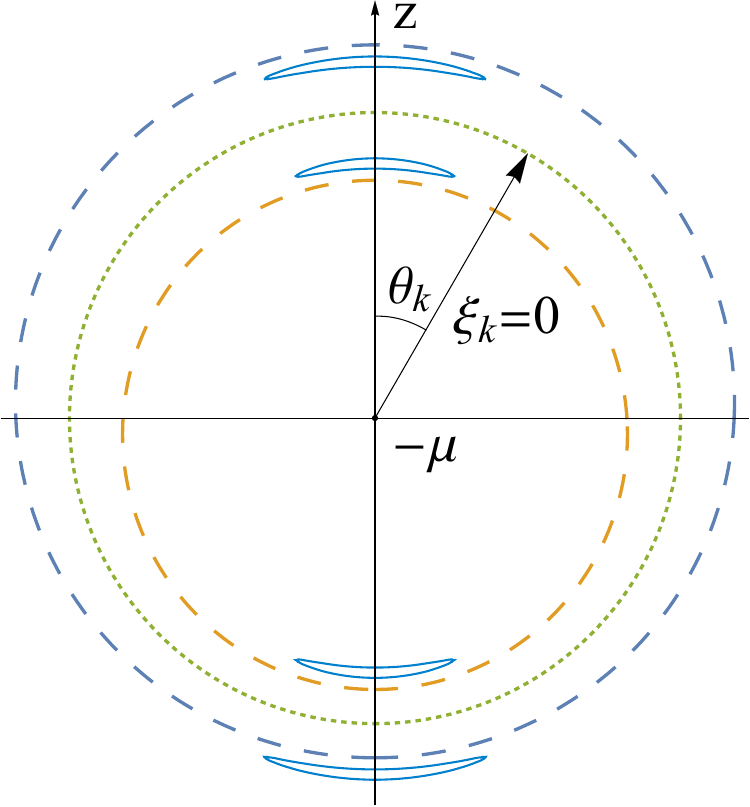}}
\subfigure[]{\includegraphics[width=0.96\linewidth]{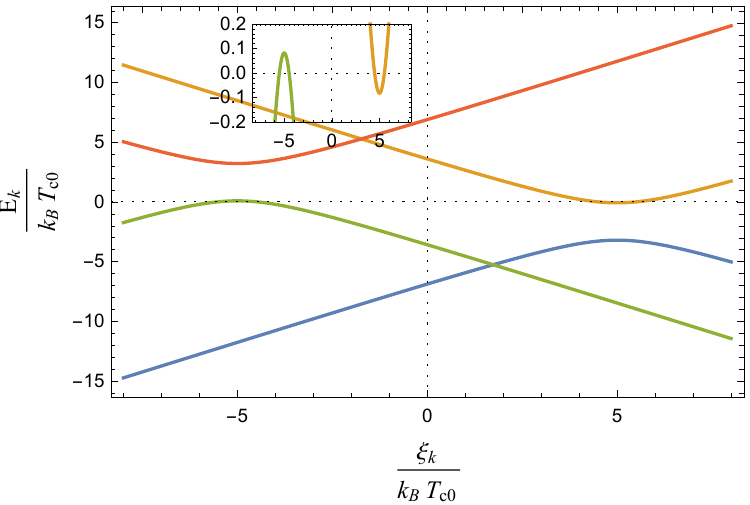}}  
\caption{The $s$-wave pairing state: Bogoliubov Fermi surface for quasiparticles and quasiholes, which maintains rotational symmetry along the $z$-axis (Zeeman field direction). (a) Fermi sphere. (b) Vertical cross-section of the Fermi sphere: The dotted line matches the dashed circle in (a), the dashed lines represent the Fermi surfaces in the normal state, i.e., $\xi_{\mathbf{k}\pm}=0$, and the solid lines depict the intersection of the Bogoliubov Fermi surfaces in the superconducting state with the vertical cross-section. (c) Bogoliubov quasiparticle (quasihole) energy spectra as a function of $\xi_{\mathbf{k}}\approx \hbar v_F(|\mathbf{k}|-k_F)$, with $\mathbf{k}$ aligned along the $z$-axis as shown in (b). Inset in (c): A close-up view of the zero energy level crossing. Here, ${gk_F}/{k_BT_{c0}}=5$ and ${H_z}/{H_0}=1.32$ are chosen.} 
\label{FS:quasi}
\end{figure}

Interestingly, there exist a critical Rashba SOC value, $g_c$, and a minimum Zeeman field value, $H_{c2}^Z$, both crucial for the emergence of the Bogoliubov Fermi surface. The combined breaking of time-reversal and spatial inversion symmetry, owing to Rashba SOC and the Zeeman field, lifts the degeneracy in the quasiparticle energy spectrum. Consequently, the energy dispersion branches of the quasiparticles (or quasiholes) shift and divide within the energy-momentum space, as depicted in Fig.~\ref{FS:quasi}. As $g$ and $H_z$ become larger, the asymmetry in energy splitting and dispersion becomes more pronounced. These energy branches can cross the zero energy level when $g>g_{c}$ and $H_{c2}^{Z}<H_{z}<H_{c}^{Z}$, creating Fermi pockets for quasiparticles and quasiholes. In contrast, if $g<g_c$, the energy dispersion never intersects zero energy, even when $H=H_c^Z-0^+$.

The critical Rashba SOC $g_{c}$ can be identified by the appearance of the Bogoliubov Fermi surface as described in the following. As depicted in Fig.~\ref{FS:quasi}(b) and as is evident from \eqref{Eq:eigen_mag_soc}, the Bogoliubov Fermi surface first emerges at polar angles $\theta_{\mathbf{k}}=0,\pi$, without extending to the equator $\theta_{\mathbf{k}}=\pi/2$. With increasing $g$, the quasiparticle energy $E_{\mathbf{k}\pm}|_{\theta_{\mathbf{k}=0,\pi}}$ intersects the zero energy level if its minimum value is negative, which is given by
\begin{equation*}
\min\{E_{\mathbf{k}\pm}|_{\theta_{\mathbf{k}=0,\pi}}\}=\Delta-\mu_B{}H_z<0.
\end{equation*}
Therefore, the critical value $g_c$ is defined as
\begin{equation}
\Delta(T=0,g_c,H_z)-\mu_B{}H_z=0,
\end{equation}
for $g<g_c$ and $0<H_z<H_{c}^{Z}$, $\Delta(T=0,g,H_z)$ decreases monotonically as $H_z$ increases and is bounded below by $\Delta(T=0,g,H_z=H_{c}^{Z}-0^{+})>\mu_{B}H_{c}^{Z}$. Hence, the quasiparticle energy does not cross the zero energy level. In contrast, for $g>g_c$, $\Delta(T=0,g,H_z=H_{c}^{Z}-0^{+})<\mu_{B}H_{c}^{Z}$, indicating that the energy of the quasiparticle intersects the zero energy level.

\subsection{Finite total momentum pairing: FFLO state}

In the end of this section, we would like to explore the feasibility of a finite total momentum Cooper pairing state, such as the prominent Fulde-Ferrell (FF)~\cite{FF_state} and Larkin-Ovchinnikov (LO)~\cite{LO_state} states, together referred to as FFLO, or pair density waves as discussed in recent studies (see Ref.~\cite{PDW} for a comprehensive review). Specifically, for an $s$-wave pairing superconducting state where each Cooper pair has a total momentum $\mathbf{q}$, one can derive a similar self-consistent gap equation for the pairing function $\Delta_{\mathbf{q}}$, solve it and compute the condensation energy $E_{c}$, defined as the energy difference between the normal and superconducting ground states, as a function of $q$. Thus, an FFLO ground state is identified when the condensation energy $E_c$ reaches a maximum at a finite $q=|\mathbf{q}|$. The technical details can be found in Appendix~\ref{free_ff}

\begin{figure}[tb]
\centering
\includegraphics[width=1.0\linewidth]{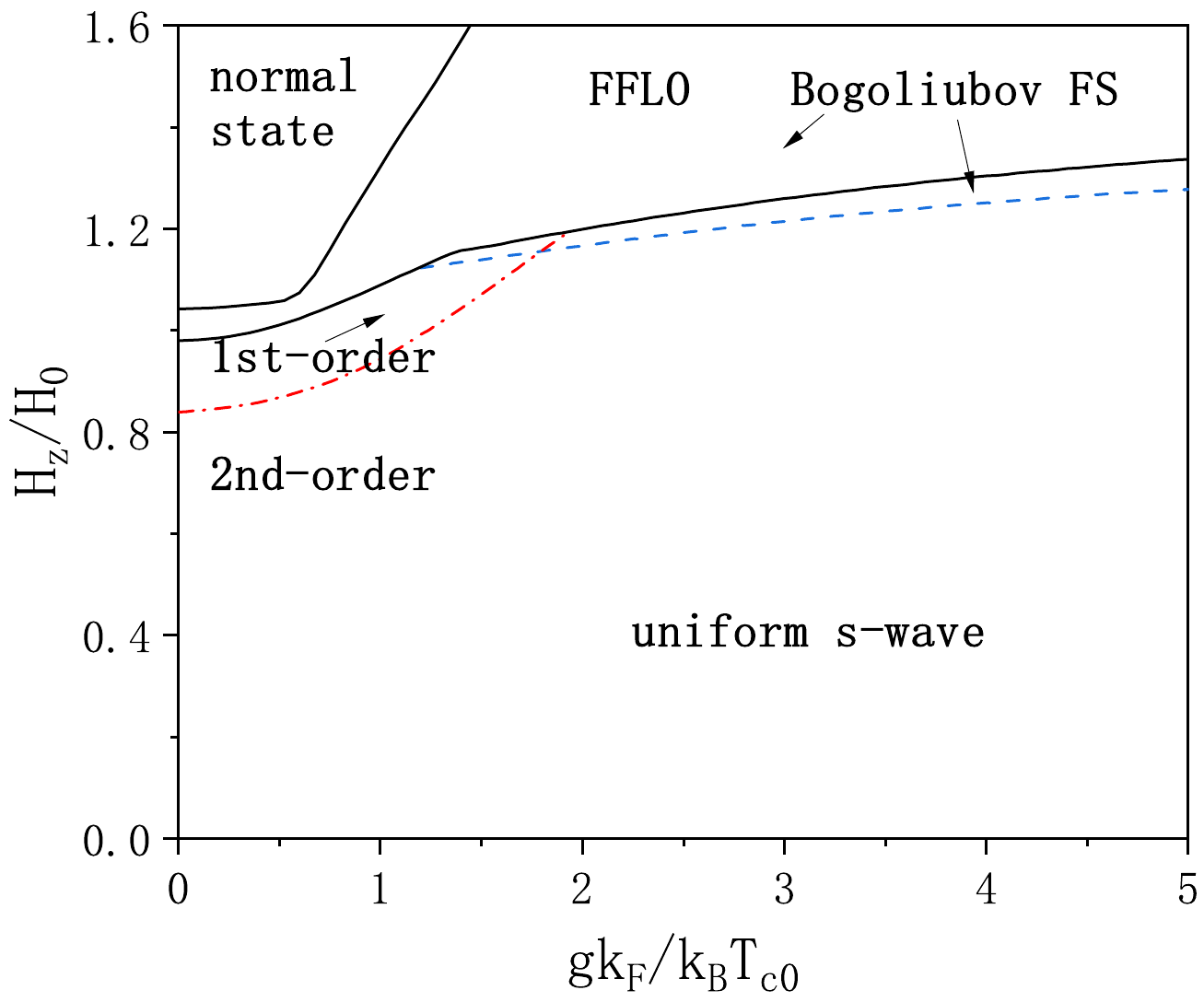}
\caption{Phase diagram of $s$-wave pairing states. The solid lines differentiate the FFLO phase from both the uniform $s$-wave pairing phase and the normal phase. The dashed line indicates the boundary between areas with and without Bogoliubov Fermi surfaces. The dashed-dotted line represents the distinction between first- and second-order phase transitions across $T_c$ within the uniform $s$-wave pairing phase. }
\label{fig:swave_phasediagram}
\end{figure}

A phase diagram including the FFLO state is illustrated in Fig.~\ref{fig:swave_phasediagram}. It shows that a sufficiently strong Zeeman field leads to an FFLO ground state. Although the region of the FFLO state is limited without Rashba SOC, it will be enlarged by sufficiently large Rashba SOC, i.e. $gk_F/k_BT_{c0}\gtrsim{}1$. This means that the Rashba SOC can stabilize the FFL0 state in a Zeeman field. 
On the other hand, by combining the previous analysis, we can identify the boundary that separates superconducting states with Bogoliubov Fermi surfaces from those without them. The appearance of Bogoliubov Fermi surfaces is noted in the FFLO phase as well as in adjacent uniform superconducting states, given that both the Zeeman field and Rashba SOC are considerably strong.

\section{$p$-wave pairing states}
\begin{table*}[tb]
\caption{Examples of $p$-wave pairing states. Here $Y_{1,\pm1}(\hat{\mathbf{k}})=\sqrt{\frac{3}{8\pi}}\sin{\theta_{\mathbf{k}}}\mathrm{e}^{\pm i\varphi_{\mathbf{k}}}$ and $Y_{1,0}(\hat{\mathbf{k}})=\sqrt{\frac{3}{4\pi}}\cos{\theta_{\mathbf{k}}}$ are spherical harmonics.}
\label{pwave}
\setlength{\tabcolsep}{1.9ex}
\renewcommand\arraystretch{2.0}
\begin{tabular}{c|c|c|c|c|c|c}
\hline\hline
& \multicolumn{2}{c|}{Opposite-spin pairing} & \multicolumn{4}{c}{Equal-spin pairing}\\
\hline
Notation & $(k_x+ik_y)\hat{z}$ & $k_z\hat{z}$ & $k_x\hat{x}+k_y\hat{y}$ & $k_y\hat{x}-k_x\hat{y}$ & $k_x\hat{x}-k_y\hat{y}$ & $k_y\hat{x}+k_x\hat{y}$\\
\hline

\multirow{2}{*}{$\mathbf{d}(\mathbf{k})$} & \multirow{2}{*}{$\Delta\sin\theta_{\mathbf{k}}\mathrm{e}^{i\varphi_{\mathbf{k}}}\hat{z}$} & \multirow{2}{*}{$\Delta\cos\theta_{\mathbf{k}}\hat{z}$} & \multicolumn{4}{c}{$\Delta\sin\theta_{\mathbf{k}}\left(\cos\phi_{\mathbf{k}}\hat{x}+\sin\phi_{\mathbf{k}}\hat{y}\right)$}\\
\cline{4-7}

& & & $\phi_{\mathbf{k}}=\varphi_{\mathbf{k}}$ & $\phi_{\mathbf{k}}=\varphi_{\mathbf{k}}-\pi/2$ & $\phi_{\mathbf{k}}=-\varphi_{\mathbf{k}}$ & $\phi_{\mathbf{k}}=\pi/2-\varphi_{\mathbf{k}}$\\
\hline

$\Delta(\mathbf{k})$ & \multicolumn{2}{c|}{$\begin{pmatrix}
0 & \Delta_0(\mathbf{k})\\
\Delta_0(\mathbf{k}) & 0
\end{pmatrix}$} & \multicolumn{4}{c}{$\begin{pmatrix}
\Delta_{\uparrow\uparrow}(\mathbf{k}) & 0\\
0 & \Delta_{\downarrow\downarrow}(\mathbf{k})
\end{pmatrix}$}\\
\hline
$\Delta_0(\mathbf{k})$ & $\sqrt{8\pi/3}\Delta Y_{1, 1}(\hat{\mathbf{k}})$ & $\sqrt{4\pi/3}\Delta Y_{1, 0}(\hat{\mathbf{k}})$ & \multicolumn{4}{c}{0}\\
\hline

$\Delta_{\uparrow\uparrow}(\mathbf{k})$ & \multicolumn{2}{c|}{0} & $-\sqrt{8\pi/3}\Delta Y_{1,-1}(\hat{\mathbf{k}})$ & $-i\sqrt{8\pi/3}\Delta Y_{1,-1}(\hat{\mathbf{k}})$ & $-\sqrt{8\pi/3}\Delta Y_{1,1}(\hat{\mathbf{k}})$ & $i\sqrt{8\pi/3} \Delta Y_{1,1}(\hat{\mathbf{k}})$\\
\hline

$\Delta_{\downarrow\downarrow}(\mathbf{k})$ & \multicolumn{2}{c|}{0} &$\sqrt{8\pi/3} \Delta Y_{1,1}(\hat{\mathbf{k}})$ & $-i\sqrt{8\pi/3} \Delta Y_{1,1}(\hat{\mathbf{k}})$ & $\sqrt{8\pi/3} \Delta Y_{1,-1}(\hat{\mathbf{k}})$ & $i\sqrt{8\pi/3} \Delta Y_{1,-1}(\hat{\mathbf{k}})$\\
\hline\hline
\end{tabular}
\end{table*}

For $p$-wave pairing superconductors, the primary pairing channel is defined by $l=1$. We focus on the simplest $p$-wave pairing model, given by a reduced form of the matrix element from Eq.~\eqref{eq:Vl}, as follows:
\begin{equation}\label{eq:Vp}
\begin{split}
V(\mathbf{k}, \mathbf{k}')&=3V_1\cos \theta_{\mathbf{k}, \mathbf{k}'}\\\
&=4\pi V_1\sum_{m=0, \pm1}Y_{1, m}(\theta_{\mathbf{k}}, \varphi_{\mathbf{k}})Y_{1, m}^{*}(\theta_{\mathbf{k}'}, \varphi_{\mathbf{k}'}),
\end{split}
\end{equation}
where $Y_{l,m}(\theta,\varphi)$ are spherical harmonics. In this study, we restrict ourselves to unitary states in which $\Delta^{\dagger}(\mathbf{k})\Delta(\mathbf{k})$ is proportional to the unit matrix, giving the quasiparticle energy dispersion as:
\begin{equation} \begin{aligned}
E_{\mathbf{k}}&=\sqrt{\xi_{\mathbf{k}}^2+\lvert\mathbf{d}(\mathbf{k})\lvert^2}.
\label{eq:Ek-us}
\end{aligned}
\end{equation} 
For a unitary state, the condition $\mathbf{d}(\mathbf{k})\times\mathbf{d}^{*}(\mathbf{k})=0$ holds true. Moreover, $\mathbf{d}(\mathbf{k})$ is a real vector encompassing three components except for an overall phase factor~\cite{Sigrist1991}.

The $p$-wave and spin-triplet pairing states are classified into opposite-spin pairing (OSP) and equal-spin pairing (ESP) categories~\cite{Leggett1975}. Table~\ref{pwave} provides examples of these $p$-wave pairing states. It is important to note that this classification depends on the chosen spin quantization axis and can change as a result. For instance, by taking the spin quantization
axis along z and x axes, we obtain the following relation:
\begin{equation}\label{eq:xz}
\left|\uparrow_{z}\downarrow_{z}\right\rangle+\left|\downarrow_{z}\uparrow_{z}\right\rangle = \left|\uparrow_{x}\uparrow_{x}\right\rangle -\left|\downarrow_{x}\downarrow_{x}\right\rangle,   
\end{equation}
where $\left|\uparrow_{\alpha}\right\rangle$ and $\left|\downarrow_{\alpha}\right\rangle$ are defined as $\sigma_{\alpha}\left|\uparrow_{\alpha}\right\rangle=+\left|\uparrow_{\alpha}\right\rangle$ and $\sigma_{\alpha}\left|\downarrow_{\alpha}\right\rangle=-\left|\downarrow_{\alpha}\right\rangle$ respectively, with $\alpha=x,z$. (For further information, refer to Appendix~\ref{app:spin-quantization-axis}.) To discuss OSP and ESP states, the spin quantization axis must be established. Throughout this paper, unless specified differently, we will align the spin quantization axis along the $z$-axis, especially when there is no external Zeeman field, or along the direction of the imposed Zeeman field.

It is widely accepted that the superconducting transition temperature $T_c$ of an ESP state is not affected by an external Zeeman field aligned with the spin quantization axis. This occurs because the superconducting condensate interacts with the Zeeman field and retains its Zeeman energy. In contrast, in an $s$-wave pairing state and an OSP state, the superconducting condensate typically does not interact with the Zeeman field, leading to a loss of Zeeman energy and a subsequent decrease in $T_c$.

 In the presence of both the Zeeman field and Rashba SOC, deriving a general analytical expression for the quasiparticle energy dispersion is challenging. However, when either the Zeeman field or Rashba SOC is present alone, such an expression can be obtained.
\begin{itemize}
\item{} For a system subjected solely to a Zeeman field $\mathbf{H}$, the energy spectrum of the Bogoliubov quasiparticles is altered and given by
\begin{equation}\label{eq:EtripltH}
E_{\mathbf{k}\pm}=\sqrt{
\left(\sqrt{\xi_{\mathbf{k}}^2+\left|\mathbf{d}_{\mathbf{k}}\cdot\hat{\mathbf{H}}\right|^2}\pm\mu_{B}|\mathbf{H}|\right)^2+\left|\mathbf{d}_{\mathbf{k}}\times\hat{\mathbf{H}}\right|^2},
\end{equation}
where $\hat{\mathbf{H}}=\mathbf{H}/|\mathbf{H}|$.

\item{} Conversely, with only a finite Rashba SOC $g\mathbf{k}\cdot\hat{\sigma}$, the Bogoliubov quasiparticle energy is
\begin{equation}\label{eq:EtripltG}
E_{\mathbf{k}\pm}=\sqrt{
\left(\sqrt{\xi_{\mathbf{k}}^2+\left|\mathbf{d}_{\mathbf{k}}\times\hat{\mathbf{k}}\right|^2}\pm{}g\left|\mathbf{k}\right|\right)^2+\left|\mathbf{d}_{\mathbf{k}}\cdot\hat{\mathbf{k}}\right|^2}.
\end{equation}

\item{} In scenarios where both the Zeeman field and Rashba SOC are present, an analytical form of the Bogoliubov quasiparticle energy spectrum is generally not available but can be computed numerically, as in the $s$-wave pairing state case.
\end{itemize}

These findings allow for the analysis of the effects of the Zeeman field and/or Rashba SOC on arbitrary $p$-wave pairing states.
In the rest of this section, we will examine the OSP and ESP states listed in Table~\ref{pwave} respectively.

\subsection{Opposite-spin pairing state}

Let's consider, without loss of generality, an OSP state characterized by a \( d \)-vector, specified as:
\begin{equation}\label{eq:dvec-OSP}
\mathbf{d}(\mathbf{k})=d_{z}(\mathbf{k})\hat{z}.
\end{equation}
In particular, for the two OSP states shown in Table~\ref{pwave}, the expressions are:
\begin{subequations}\label{eq:dvec1}
\begin{equation}
\mathbf{d}(\mathbf{k}) = \Delta\sin\theta_{\mathbf{k}}\mathrm{e}^{i\varphi_{\mathbf{k}}}\hat{z},
\end{equation}
which corresponds to the $(k_x+ik_y)\hat{z}$ pairing state, and
\begin{equation}
\mathbf{d}(\mathbf{k}) = \Delta\cos\theta_{\mathbf{k}}\hat{z},
\end{equation}
\end{subequations}
which corresponds to the $k_z\hat{z}$ pairing state.

\subsubsection{Superconducting transition temperature}
\begin{figure*}[htb]
\subfigure[]{
\includegraphics[width=0.49\linewidth]{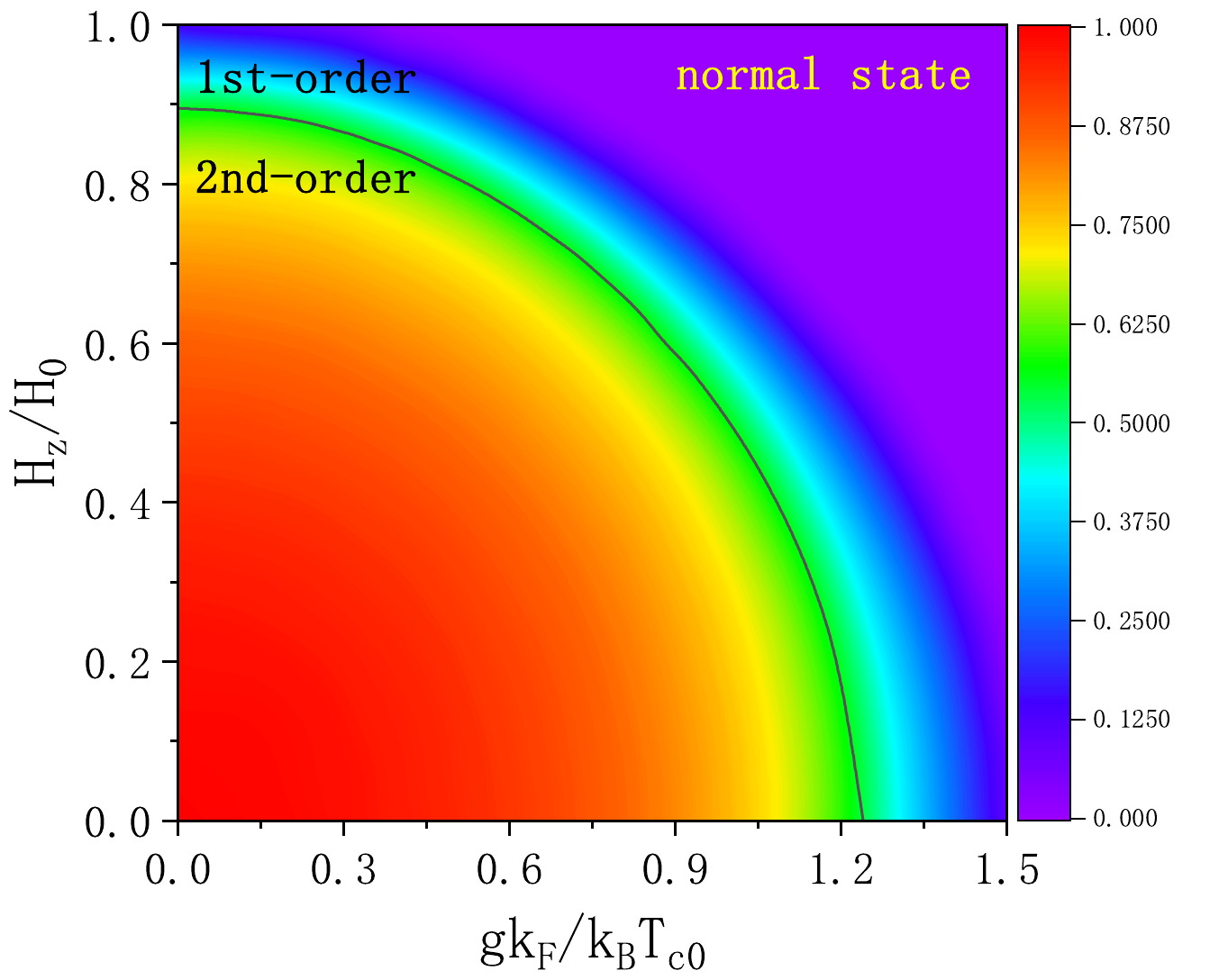}}
\subfigure[]{
\includegraphics[width=0.49\linewidth]{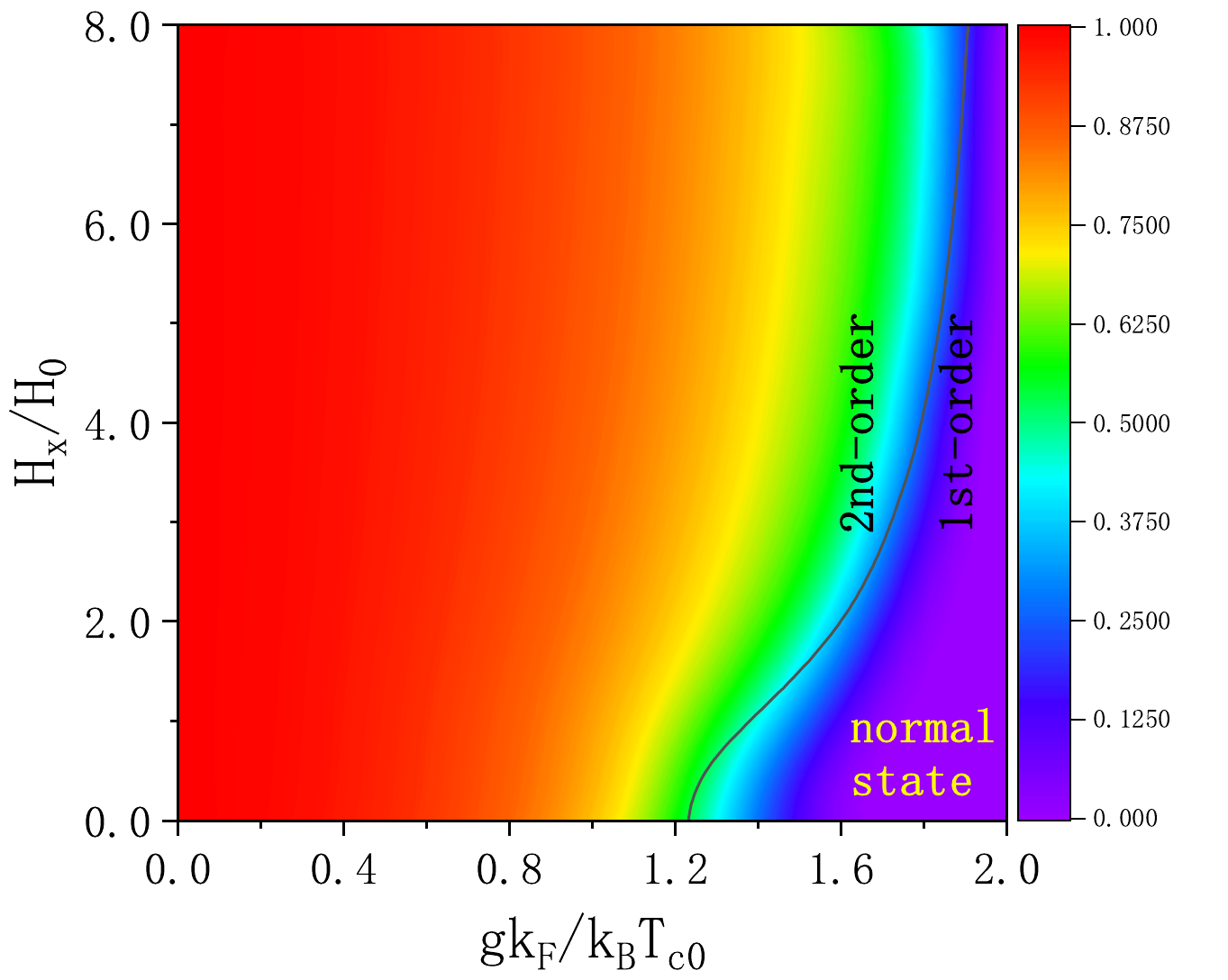}}
\subfigure[]{
\includegraphics[width=0.49\linewidth]{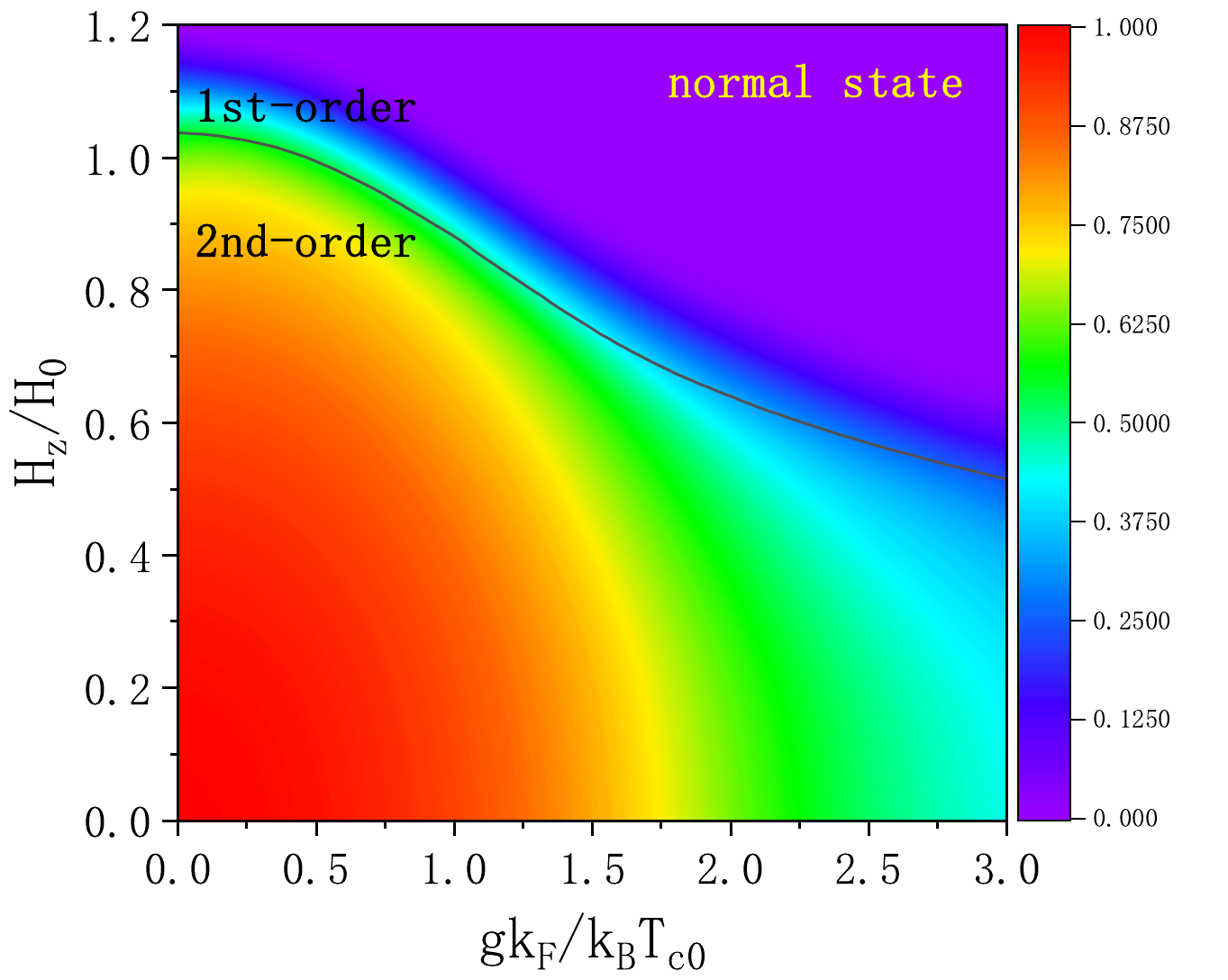}}
\subfigure[]{
\includegraphics[width=0.49\linewidth]{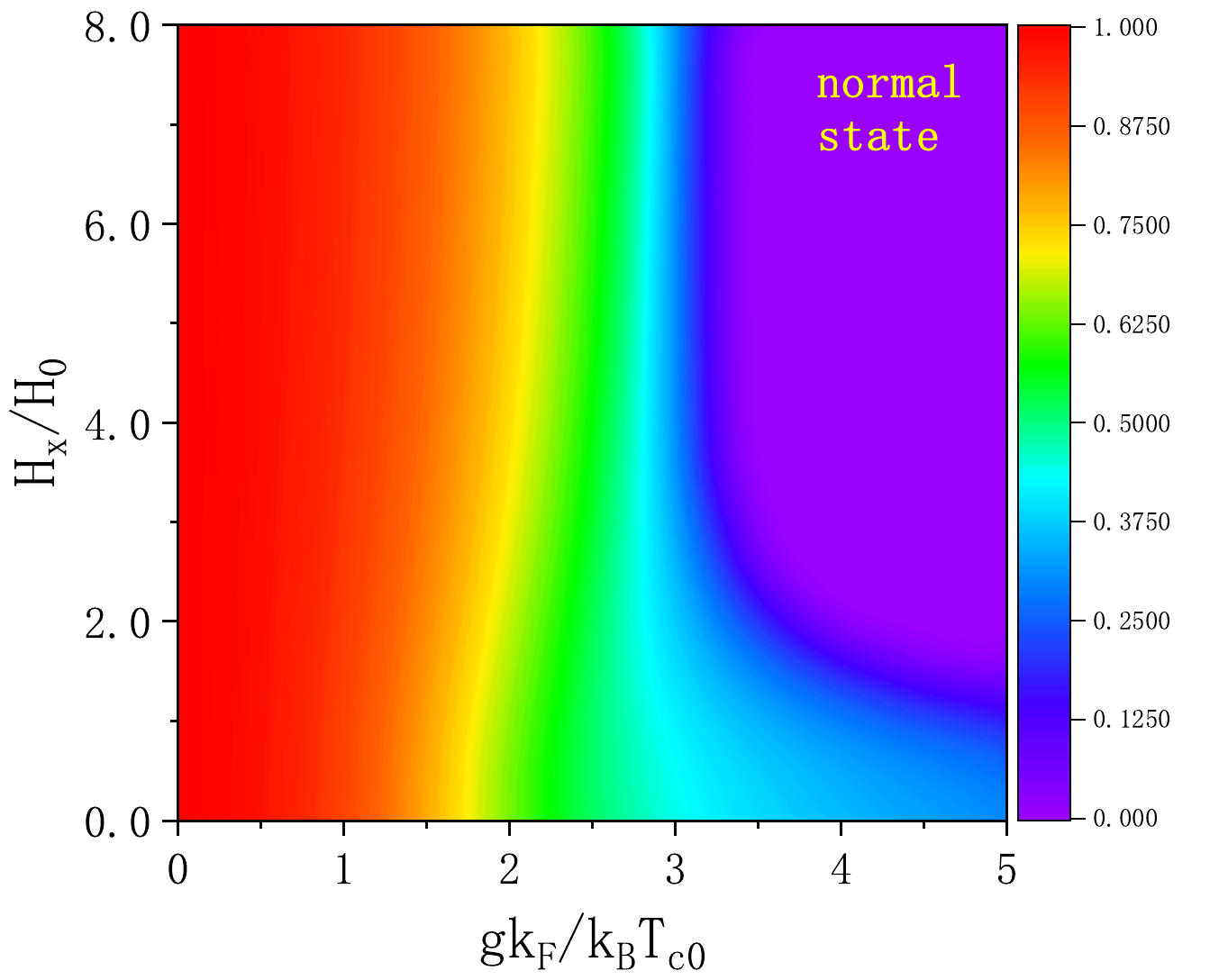}}
\caption{ Superconducting transition temperature $T_c$ in units of $T_{c0}$ as a function of the Rashba SOC and the Zeeman field $\mathbf{H}$. (a) Pairing state $(k_x+ik_y)\hat{z}$ in a Zeeman field $\mathbf{H}=H_z\hat{z}$ and (b) when subjected to a Zeeman field $\mathbf{H}=H_x\hat{x}$. (c) Pairing state $k_z\hat{z}$ in a Zeeman field $\mathbf{H}=H_z\hat{z}$ and (d) when in a Zeeman field $\mathbf{H}=H_x\hat{x}$. Solid lines represent the boundaries between the first and second order phase transitions across $T_c$.
}
\label{fig:tc_magsoc-dz}
\end{figure*}

Using the pairing model defined in Eq.~\eqref{eq:Vp} and the d-vector given in Eq.~\eqref{eq:dvec1}, the associated gap equation, as detailed in \eqref{gapeqwhole}, can be solved self-consistently. The numerical solution of the superconducting transition temperature $T_c$, which is a function of the Rashba SOC and the Zeeman field, is illustrated in Fig.~\ref{fig:tc_magsoc-dz}. Some features are observed and analyzed below.

{\bf{}Effects of Zeeman field:} A parallel Zeeman field $\mathbf{H}=H_{z}\hat{z}$ reduces $T_c$ in both OSP states, regardless of whether Rashba SOC is present. In contrast, a perpendicular Zeeman field $\mathbf{H}=H_{\perp}\left(\cos\phi\hat{x}+\sin\phi\hat{y}\right)$ leaves $T_c$ unaffected if Rashba SOC is absent. In this scenario, the spin quantization axis can be aligned along $\cos\phi\,\hat{x}+\sin\phi\,\hat{y}$, allowing the pairing state $d_z(\mathbf{k})\hat{z}$ to be regarded as an ESP state, consistent with Eq.~\eqref{eq:xz}.

{\bf{}Rashba SOC effects:} As the Rashba SOC strength $g$ increases, the superconducting transition temperature $T_c$ consistently decreases. However, the nature of this decrease differs significantly between the two OSP states $(k_x+ik_y)\hat{z}$ and $k_z\hat{z}$. Without the Zeeman field, superconductivity in the pairing state $(k_x+ik_y)\hat{z}$ is completely suppressed at $gk_F/k_B{}T_{c0}=1.52$, and the phase transition from normal to superconducting becomes first-order if $T_c$ is sufficiently low. In contrast, the superconducting phase transition across $T_c$ remains continuous in the pairing state $k_z\hat{z}$.

In the following, we examine the continuous phase transitions in more detail. When $\mathbf{H}=0$ and $g$ is finite, the $T_c$ equation for a continuous superconducting phase transition is given by 
\begin{equation}\label{eq:Tc-p-H0}
\ln\left(\frac{T_c}{T_{c0}}\right)=\alpha\left[\psi\left(\frac{1}{2}\right)-\mbox{Re}\,\psi\left(\frac{1}{2}+\frac{igk_F}{2\pi k_BT_{c}}\right)\right],
\end{equation}
where $\alpha$ assumes the values $4/5$ and $2/5$ for the two OSP states $(k_x+ik_y)\hat{z}$ and $k_z\hat{z}$ respectively, and $$\psi(z)=-\gamma+\sum_{n=0}^\infty\frac{z-1}{(n+1)(n+z)}$$
which represents the digamma function, with the Euler constant $\gamma\sim{}0.5772$.

Considering the pairing state $k_z\hat{z}$ with $\alpha=2/5$ in Eq.~\eqref{eq:Tc-p-H0}, when the superconducting phase transition at $T_c$ remains continuous solely under the influence of Rashba SOC, the $T_c$ equation provides the solution:
\begin{equation}\label{eq:Tc_kz-H0}
\frac{T_c}{T_{c0}}=\left[\frac{\mathrm{e}^{-\gamma}}{4}\cdot \frac{2\pi k_BT_{c0}}{gk_F}\right]^{\frac{2}{3}},
\end{equation}
under the condition $k_BT_{c}\ll gk_F/2\pi$. Refer to Appendix~\ref{app:tc_2nd} for a comprehensive set of calculations. On the other hand, in the case of the pairing state $(k_{x}+ik_{y})\hat{z}$, this continuous phase transition turns into a first-order transition when $gk_{F}\gtrsim{}1.23{}k_BT_{c0}$.

\subsubsection{Quasiparticle excitation energy}

\begin{figure*}[tb]
\centering
{\includegraphics[width=0.96\linewidth]{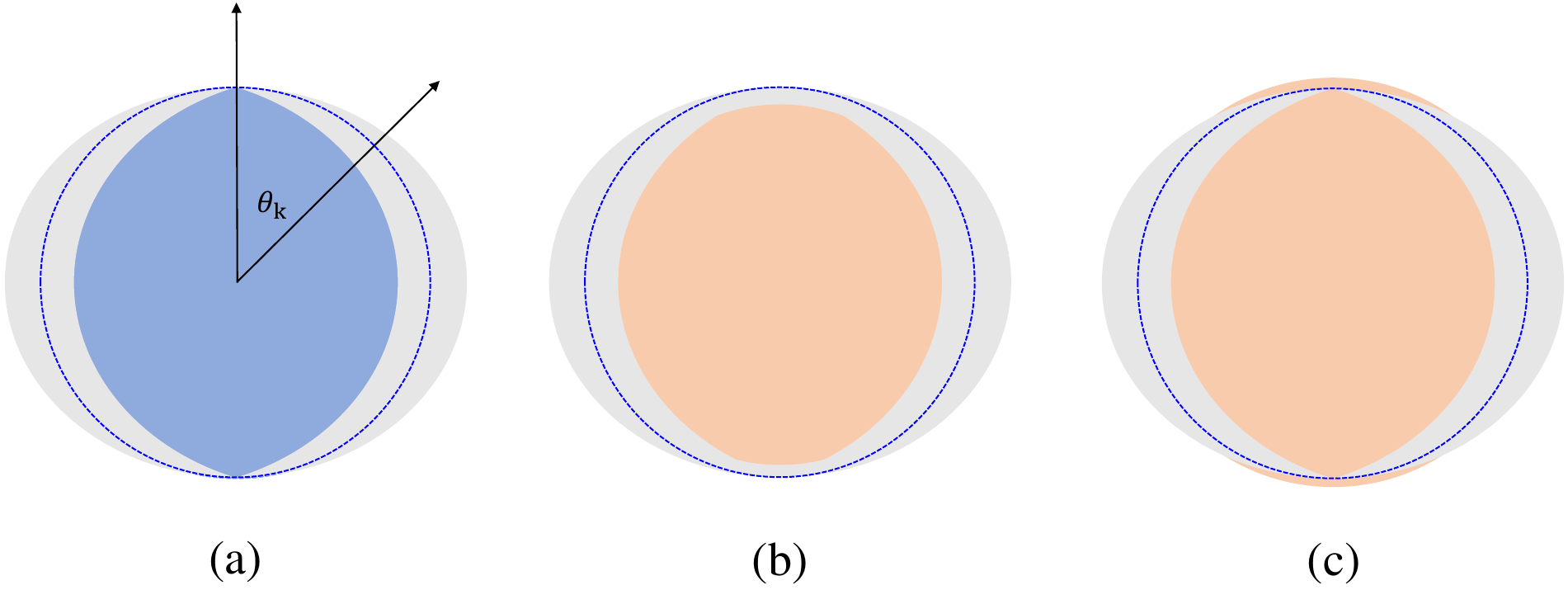}}  
\caption{The pairing state $(k_x+ik_y)\hat{z}$ in a parallel Zeeman field $\mathbf{H}=H_z\hat{z}$: Dashed circles represent the Fermi surface when the Zeeman field is absent. The gray areas indicate the unoccupied (gapped) states within the superconducting ground state. The gap configuration (a) without the Zeeman field and for the quasiparticle branches (b) $E_{\mathbf{k}+}$ and (c) $E_{\mathbf{k}-}$ under the parallel Zeeman field.} 
\label{fig:nodalS1}
\end{figure*}

Without the presence of a Zeeman field and Rashba SOC, the quasiparticle energy dispersion of an OSP state is simplified as follows:
\begin{equation}
E_{\mathbf{k}}=\sqrt{\xi_{\mathbf{k}}^2+|d_{z}(\mathbf{k})|^2}.
\end{equation}
As a result, the pairing state $(k_{x}+ik_{y})\hat{z}$ has two nodal points located at $\theta_{\mathbf{k}}=0$ and $\pi$ on the Fermi surface. Meanwhile, the pairing state $k_{z}\hat{z}$ features a nodal line at $\theta_{\mathbf{k}}=\pi/2$.

In what follows, we explore how the Zeeman field and Rashba SOC affect the quasiparticle excitation energy, with particular attention to their energy gap configuration.

{\bf{}Zeeman field effect:} We first consider a parallel Zeeman field $\mathbf{H}=H_z\hat{z}$. With the d-vector aligned along the $\hat{z}$-axis and the Cooper pair's total spin confined to the $xy$-plane, the parallel Zeeman field splits the quasiparticle energy dispersion into two distinct branches. Consequently, a simplified form of Eq.~\eqref{eq:EtripltH} reads:
\begin{equation}\label{eq:ospEkHz}
E_{\mathbf{k}\pm} =\left|\sqrt{\xi_{\mathbf{k}}^2+|d_z(\mathbf{k})|^2}\pm\mu_B H_z\right|.  
\end{equation}
Next, without loss of generality, we analyze a perpendicular Zeeman field $\mathbf{H}=H_{\perp}\left(\cos\phi\hat{x}+\sin\phi\hat{y}\right)$. Thus, the quasiparticle energy dispersion given by Eq.~\eqref{eq:EtripltH} can be written as
\begin{equation}\label{eq:ospEkHx}
E_{\mathbf{k}\pm}=\sqrt{(\xi_{\mathbf{k}} \pm \mu_B H_{\perp})^2 + |d_z(\mathbf{k})|^2}.
\end{equation}
Here, $d_{z}(\mathbf{k})=\Delta\sin\theta_\mathbf{k}\mathrm{e}^{i\varphi_{\mathbf{k}}}$ corresponds to the pairing state $(k_x+ik_y)\hat{z}$, while $d_{z}(\mathbf{k})=\Delta\cos\theta_\mathbf{k}$ describes the $k_z\hat{z}$ pairing state, as outlined in Eq.~\eqref{eq:dvec1}.
From Eqs.~\eqref{eq:ospEkHz} and \eqref{eq:ospEkHx}, it can be seen that:
\begin{itemize}
\item{} A perpendicular Zeeman field only shifts the chemical potential $\mu$ without affecting the gap structure. In contrast, a parallel Zeeman field can significantly alter the energy gap structure in an OSP state.
\item{} The pairing state $(k_x+ik_y)\hat{z}$ features two nodal points at $\theta_{\mathbf{k}}=0$ or $\pi$ within the energy gap, when the Zeeman field is absent. However, the parallel Zeeman field converts these nodal points into nodal surfaces defined by $\Delta\sin\theta_{\mathbf{k}}\leq\mu_{B}|H_z|$. Refer to Fig.~\ref{fig:nodalS1} for an illustration.

\item{} For the pairing state $k_z\hat{z}$, the parallel Zeeman field transforms the nodal line at $\theta_{\mathbf{k}=\pi/2}$ into a nodal surface characterized by $\Delta\cos\theta_{\mathbf{k}}\leq\mu_{B}|H_z|$.
\end{itemize}

{\bf{}Effect of Rashba SOC:} We now examine the case of a finite Rashba SOC and an absent magnetic field. The Rashba SOC splits the normal state band into two separate branches, resulting in the creation of two Fermi surfaces. In the superconducting phase, the quasiparticle energy dispersion in Eq.~\eqref{eq:EtripltG} is given by:
\begin{equation}
E_{\mathbf{k}\pm}=\sqrt{
\left(\sqrt{\xi_{\mathbf{k}}^2+\left|d_{z}(\mathbf{k})\right|^2\sin^2\theta_{\mathbf{k}}}\pm{}g\left|\mathbf{k}\right|\right)^2+\left|d_{z}(\mathbf{k})\right|^2\cos^2\theta_{\mathbf{k}}}.
\end{equation}

For the $(k_x + ik_y)\hat{z}$ pairing state, the angular-dependent quasiparticle energy gap, which is defined as the minimum of $E_{\mathbf{k}-}$ in the $\hat{\mathbf{k}}$ direction, is given by:
\begin{equation*}
\Delta(\theta_{\mathbf{k}}) =  
\begin{cases} 
\Delta \sin\theta_{\mathbf{k}} \left|\cos\theta_{\mathbf{k}}\right|, & \text{if}\ \sin^2\theta_{\mathbf{k}} < \frac{g k_F}{\Delta}, \\
\Delta \sqrt{\sin^2 \theta_{\mathbf{k}} \left(1-\frac{2g k_F}{\Delta} \right) + \left(\frac{gk_F}{\Delta}\right)^2}, & \text{if}\ \sin^2\theta_{\mathbf{k}} > \frac{g k_F}{\Delta}.
\end{cases}
\end{equation*}
Thus, a small $g (<\Delta/k_F)$ does not alter the two nodal points located at $\theta_{\mathbf{k}} = 0$ and $\pi$ on each of the separate Fermi surfaces. However, a sufficiently large $g (> \Delta/k_F)$ introduces an extra nodal line at $\theta_{\mathbf{k}} = \pi/2$ on both separate Fermi surfaces.

In contrast, for the pairing state $k_z\hat{z}$, the angular dependence of the quasiparticle energy gap is given by
\begin{widetext}
\begin{equation*}
\Delta(\theta_{\mathbf{k}}) = \begin{cases} 
\Delta \cos^2\theta_{\mathbf{k}}, & \text{if} \ \sin\theta_{\mathbf{k}}|\cos\theta_{\mathbf{k}}| < \frac{g k_F}{\Delta}, \\
\Delta \sqrt{\left( \sin\theta_{\mathbf{k}}|\cos\theta_{\mathbf{k}}| - \frac{g k_F}{\Delta} \right)^2 + \cos^4 \theta_{\mathbf{k}}}, & \text{if} \ \sin\theta_{\mathbf{k}}|\cos\theta_{\mathbf{k}}| > \frac{g k_F}{\Delta}.
\end{cases}
\end{equation*}
\end{widetext}
Therefore, the Rashba SOC maintains the nodal line at $\theta_{\mathbf{k}} = \pi/2$ on each Fermi surface intact.
 
\subsubsection{Discussions}

Firstly, we would like to delve further into the nodal surfaces evident in OSP states, which contrast with the Bogoliubov Fermi surface observed in the $s$-wave pairing state when both the Zeeman field and Rashba SOC are present. In the latter scenario, breaking of both time-reversal and spatial-inversion symmetries eliminates the degeneracy $\mathbf{k}$ to $-\mathbf{k}$ in the quasiparticle spectrum. This leads to the elevation of one quasihole branch and the descent of the corresponding quasiparticle branch, with both intersecting the zero-energy level, thus forming a Bogoliubov Fermi surface for both quasiparticles and quasiholes. However, in the presence of either a parallel Zeeman field or Rashba SOC within an OSP state, the quasiparticle dispersion splits into two branches, $E_{\mathbf{k}+}$ and $E_{\mathbf{k}-}$. The degeneracy condition $E_{\mathbf{k}\pm}=E_{-\mathbf{k}\pm}$ remains valid. Consequently, the quasiparticles and quasiholes in one branch elevate, while in the other branch they descend. This results in a quasiparticle Fermi surface in one branch and a complete gap in the other branch. An illustration of such a quasiparticle Fermi surface can be seen in Fig.~\ref{fig:nodalS1}.

Next, we will explore how the Zeeman field and Rashba SOC affect the superconducting transition temperature $T_c$ in more detail.
\begin{itemize}
\item{}As illustrated by Eqs.~\eqref{eq:ospEkHz} and \eqref{eq:ospEkHx}, the anisotropic pairing state $d_{z}(\mathbf{k})\hat{z}$ exhibits distinct responses to magnetic fields aligned parallel to the spin quantization axis ($\mathbf{H}=H_z{}\hat{z}$) and perpendicular to it [$\mathbf{H}=H_{\perp}\left(\cos\phi\hat{x}+\sin\phi\hat{y}\right)$]. In particular, the energy dispersion $E_{\mathbf{k}\pm}$ depends on $H_z$ and $H_{\perp}$ in different ways. 

\item{}Furthermore, according to Eq.~\eqref{eq:ospEkHx}, the self-consistent gap equation and its solution $\Delta(T)$ are not affected by $H_{\perp}$ because the perpendicular Zeeman field only shifts the chemical potential $\mu$ in this scenario. In contrast, both the gap equation and $\Delta(T)$ show dependence on $H_z$. Consequently, the superconducting transition temperature $T_c$ remains constant as $H_{\perp}$ increases but decreases with increasing $H_z$.

\item{}In the absence of the Zeeman field, the pairing state $k_z\hat{z}$ demonstrates greater stability against Rashba SOC compared to the pairing state $(k_x+ik_y)\hat{z}$. The transition from superconducting to normal state pairing state $k_z\hat{z}$ remains continuous, with the critical temperature $T_c$ decreasing according to a power law when $k_BT_{c}\ll gk_F/2\pi$, as shown in Eq.~\eqref{eq:Tc_kz-H0}.
\end{itemize}

\subsection{Equal-spin pairing states}
\begin{figure*}[htb]
\centering
\includegraphics[width=1.0\textwidth]{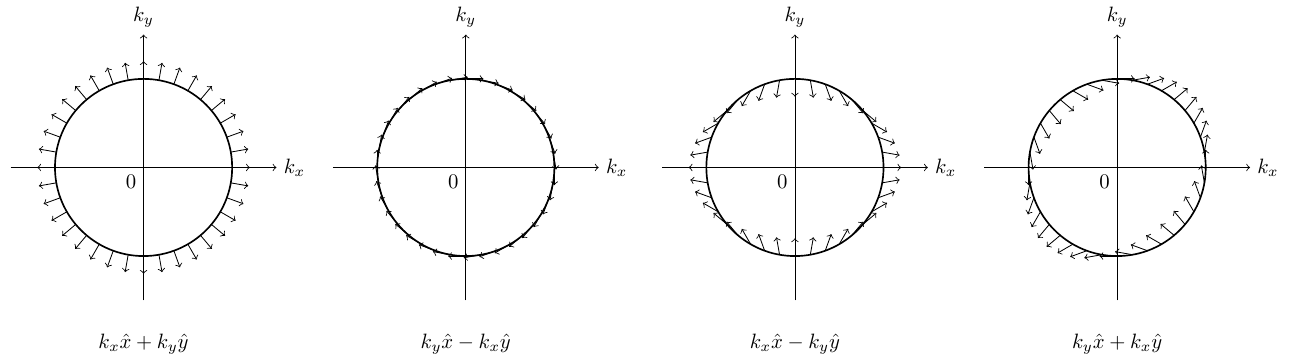}
\caption{The four ESP states listed in Table~\ref{pwave} describe the $\mathbf{d}(\mathbf{k})$ configuration on the Fermi surface as it intersects the $xy$ plane. The states $k_x\hat{x}+k_y\hat{y}$ and $k_y\hat{x}-k_x\hat{y}$ remain unaffected by any rotation about the $z$ axis. In contrast, under a rotation $C_8$, the state $k_x\hat{x}-k_y\hat{y}$ transforms into the state $k_y\hat{x}+k_x\hat{y}$ and changes by a minus sign under rotation $C_4$. These transformations are consistent with Eq.~\eqref{eq:Dn}. } 
\label{fig:4ESP_states}
\end{figure*}

In this subsection, we examine the four ESP states listed in Table~\ref{pwave}. Their associated d-vectors can be unified as follows:
\begin{equation}\label{eq:dvec4}
\mathbf{d}(\mathbf{k}) = \Delta\sin\theta_{\mathbf{k}}\left(\cos\phi_{\mathbf{k}}\hat{x}+\sin\phi_{\mathbf{k}}\hat{y}\right),
\end{equation}
where the angle $\phi_{\mathbf{k}}$ is specified in Table~\ref{pwave}. The d-vector configurations on the Fermi surface intersecting the $xy$ plane for these four ESP states are illustrated in Fig.~\ref{fig:4ESP_states}.

The final two ESP states presented in Table~\ref{pwave}, specifically $k_x\hat{x}-k_y\hat{y}$ and $k_y\hat{x}+k_x\hat{y}$, serve as the basis for a two-dimensional irreducible representation ($E$ or $E_2$) of the dihedral group $D_n$, where $n=3$ or $n\ge{}5$. To demonstrate this, we analyze the two generators of the $D_n$ group: the $n$-fold rotation $C_{n}$ about the $z$-axis:
\begin{subequations}
\begin{equation}
C_{n} = \exp\left(-i\frac{2\pi}{n}J_z\right) = \exp\left[-i\frac{2\pi}{n}\left(L_z+\frac{1}{2}\sigma_z\right)\right],
\end{equation}
and a two-fold rotation $C_2^{\prime}$ around an axis in the $xy$-plane, e.g.,
\begin{equation}
C_{2}^{\prime} = \exp\left(-i\pi{}J_x\right) = \exp\left[-i\pi\left(L_x+\frac{1}{2}\sigma_x\right)\right],
\end{equation}
\end{subequations}
where $J_{z(x)}$, $L_{z(x)}$, and $\sigma_{z(x)}/2$ are the $z(x)$-components of the total, orbital, and spin angular momentum, respectively. Generally, a group element $g \in SO(3)$ operates on the d-vector as follows,
\begin{equation}
g\mathbf{d}(\mathbf{k}):D(g)\mathbf{d}[D^{-1}(g)\mathbf{k}].    
\end{equation}
Selecting the basis functions as $\{k_x\hat{x}-k_y\hat{y},k_y\hat{x}+k_x\hat{y}\}$, we derive the representation matrices for $C_n^m$ and $C_2^{\prime}$:
\begin{equation}\label{eq:Dn}
\begin{split}
D(C_n^{m})&=
\begin{pmatrix}
\cos\frac{4m\pi}{n} & -\sin\frac{4m\pi}{n} \\
\sin\frac{4m\pi}{n} & \cos\frac{4m\pi}{n}
\end{pmatrix},\\
D(C_{2}^{\prime})&=
\begin{pmatrix}
1 & 0 \\ 
0 & -1
\end{pmatrix},
\end{split}
\end{equation}
where $m=0,1,\cdots,n-1$. Consequently, the characters of $C_n^m$ and $C_2^\prime$ are $\mbox{tr}D(C_n^{m})=2\cos(4m\pi/n)$ and $\mbox{tr}D(C_2^{\prime})=0$, respectively. Therefore, for $n=3$ or $n\ge{}5$, the two ESP states $k_x\hat{x}-k_y\hat{y}$ and $k_y\hat{x}+k_x\hat{y}$ serve as the basis functions of an irreducible representation ($E$ or $E_2$) of the group $D_n$. Note that the $D_4$ group is excluded here because the $C_4$ rotation only changes these two ESP states by a negative sign.

The effects of the Zeeman field and Rashba SOC on the four ESP states presented in Table~\ref{pwave} are examined below.

\subsubsection{Superconducting transition temperature}

\begin{figure*}[htb]
\subfigure[]{
\includegraphics[width=0.49\linewidth]{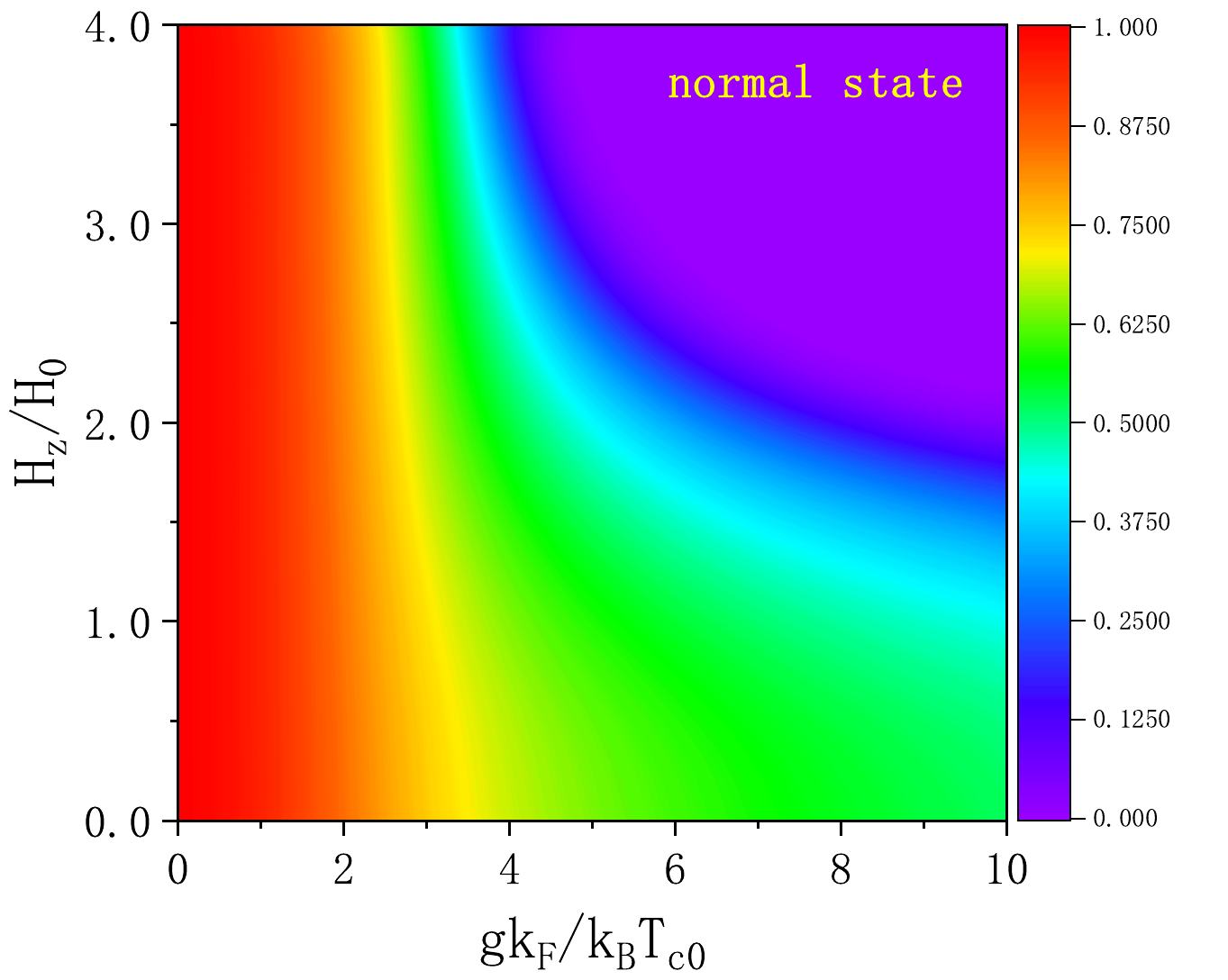}}
\subfigure[]{
\includegraphics[width=0.49\linewidth]{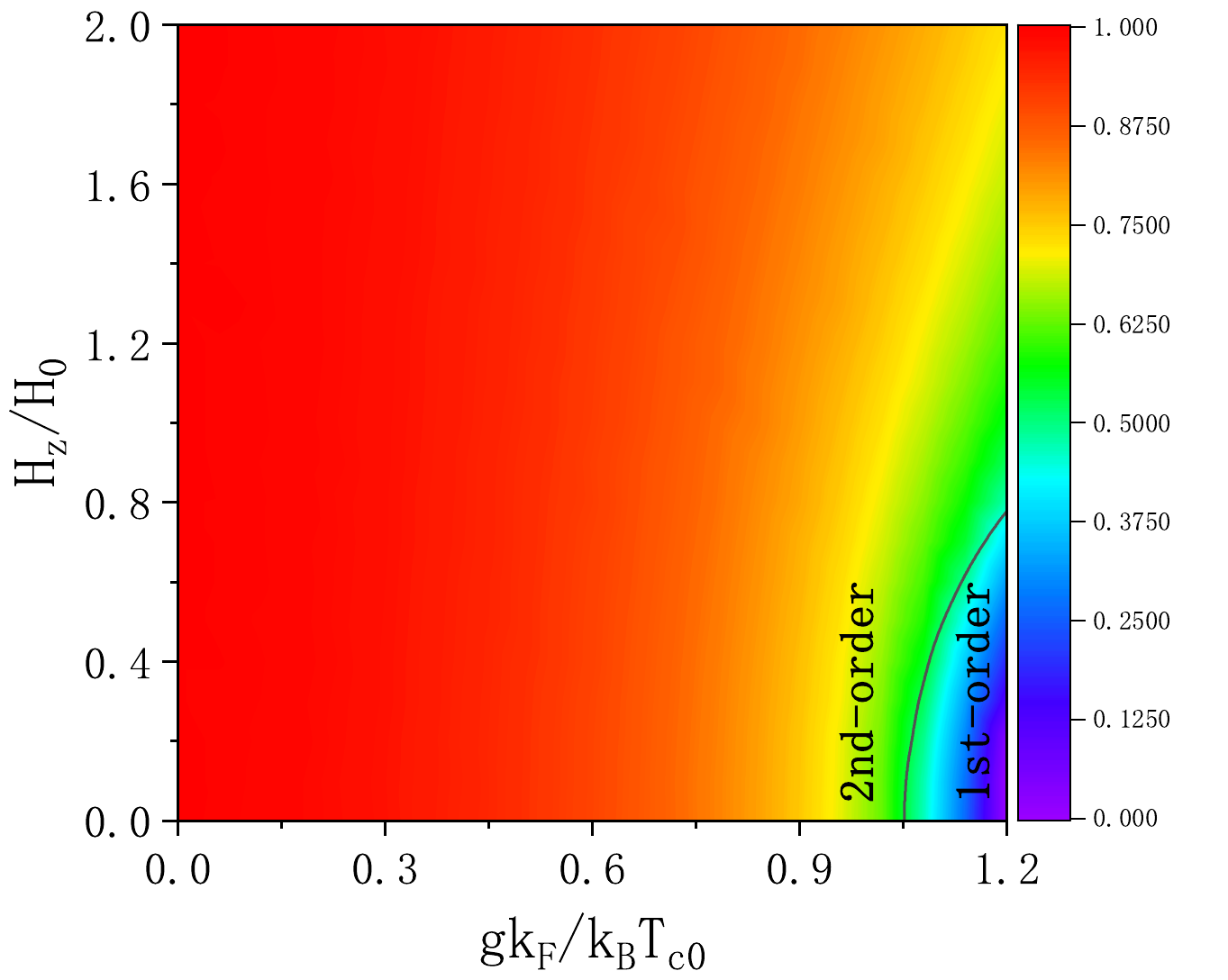}}
\subfigure[]{
\includegraphics[width=0.49\linewidth]{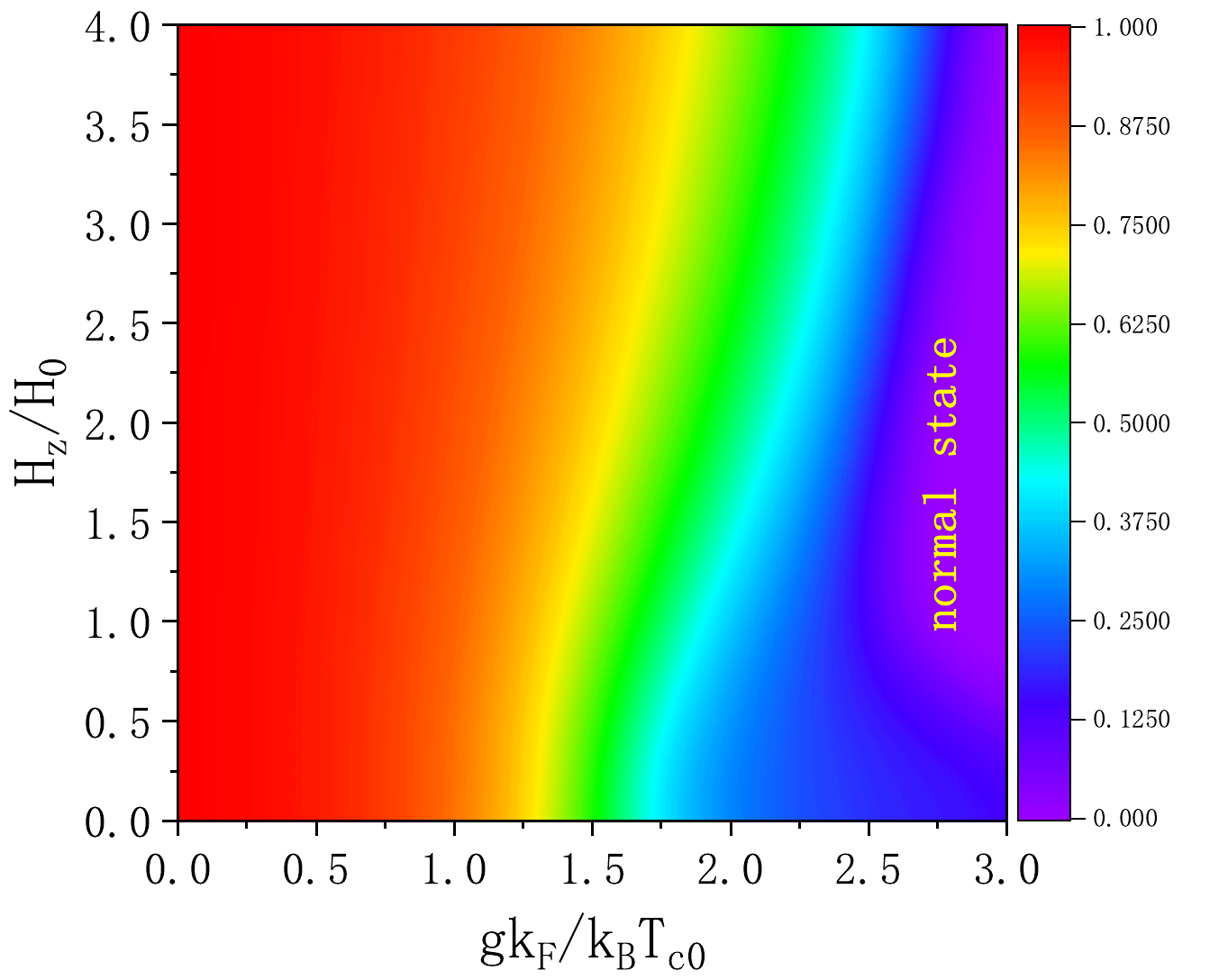}}
\caption{Superconducting transition temperature $T_c$, normalized to $T_{c0}$, as a function of the Rashba SOC $g$ and the Zeeman field $\mathbf{H}=H_z\hat{z}$. The four ESP states: (a) $k_x\hat{x}+k_y\hat{y}$, (b) $k_y\hat{x}-k_x\hat{y}$, (c) $k_x\hat{x}-k_y\hat{y}$, and $k_y\hat{x}+k_x\hat{y}$. The solid line in (b) marks the boundary between first-order and second-order phase transitions across $T_c$. For the pairing states in (a) and (c), regardless of how low $T_c$ is, the phase transition from superconducting to normal remains continuous.
}\label{fig:tc_esp_hz}
\end{figure*}

\begin{figure*}[htb]
\subfigure[]{
\includegraphics[width=0.49\linewidth]{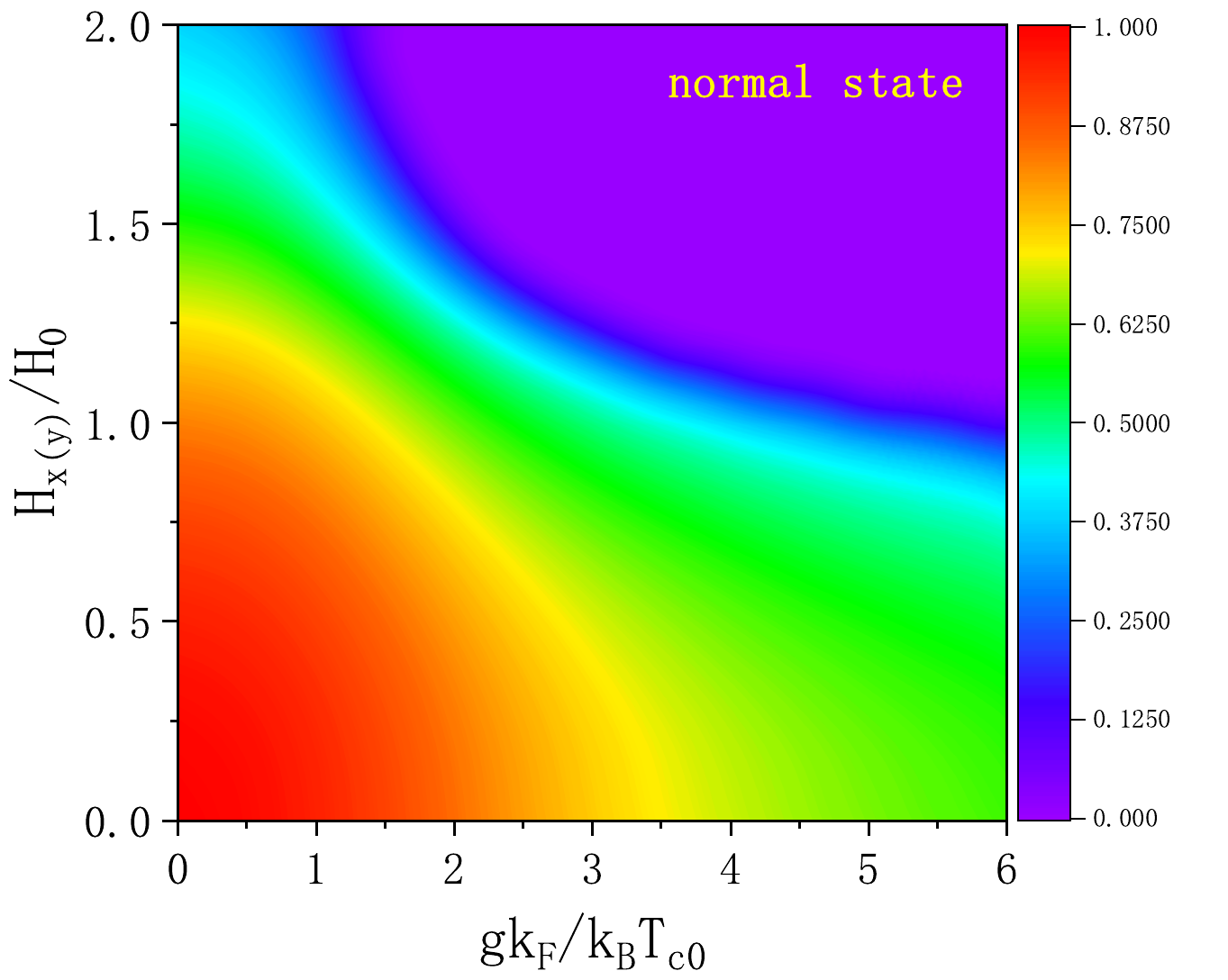}}
\subfigure[]{
\includegraphics[width=0.49\linewidth]{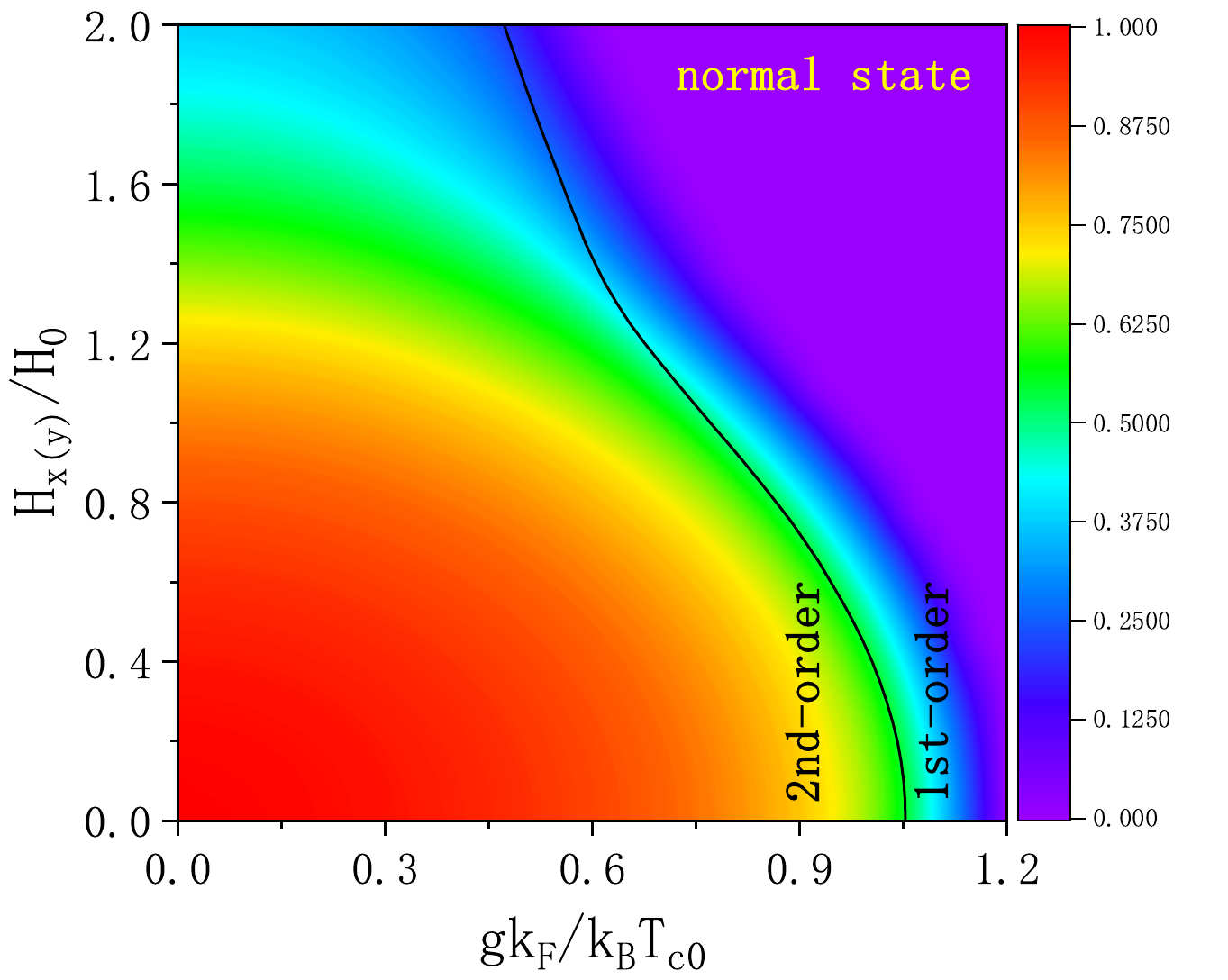}}
\subfigure[]{
\includegraphics[width=0.49\linewidth]{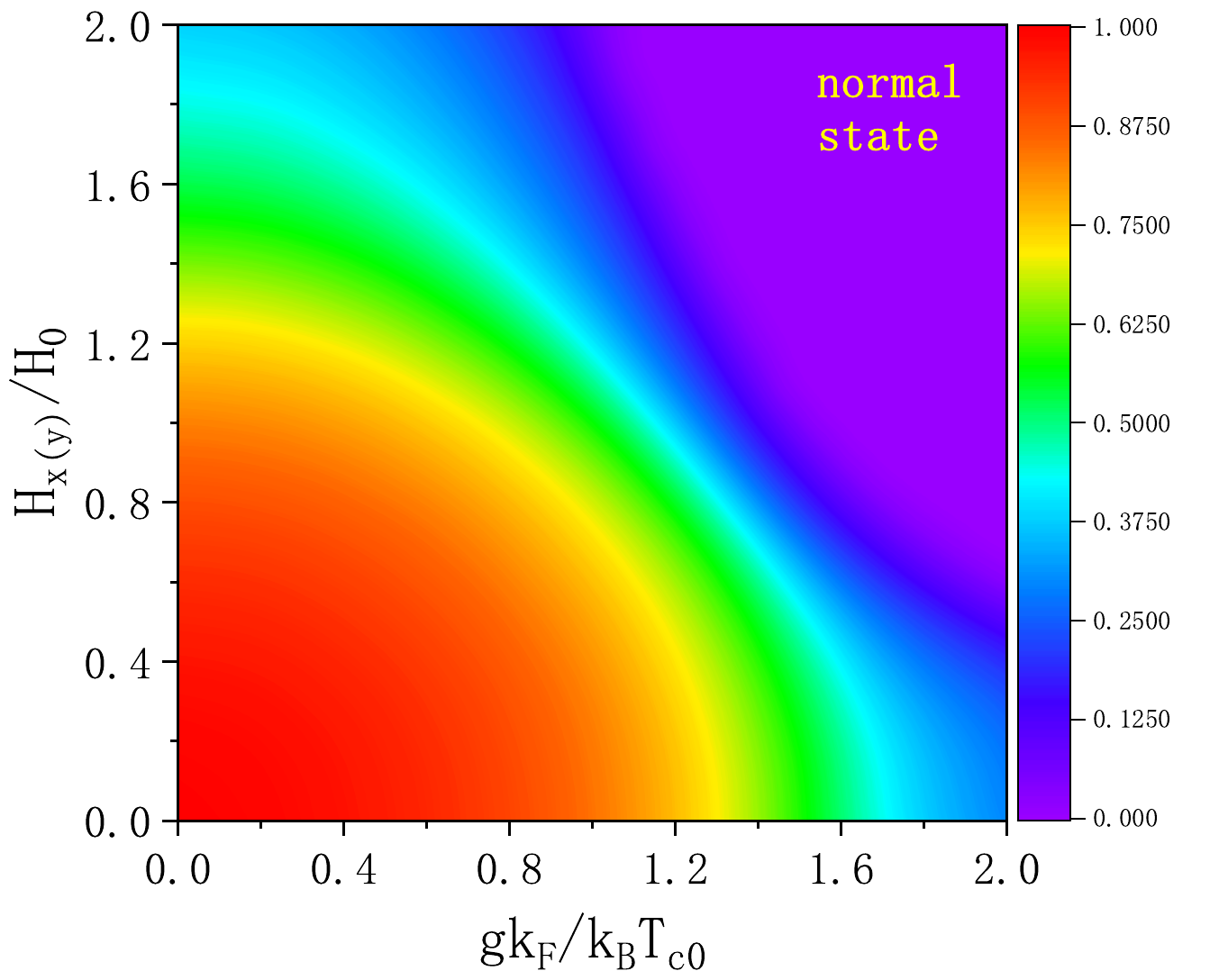}}
\subfigure[]{
\includegraphics[width=0.49\linewidth]{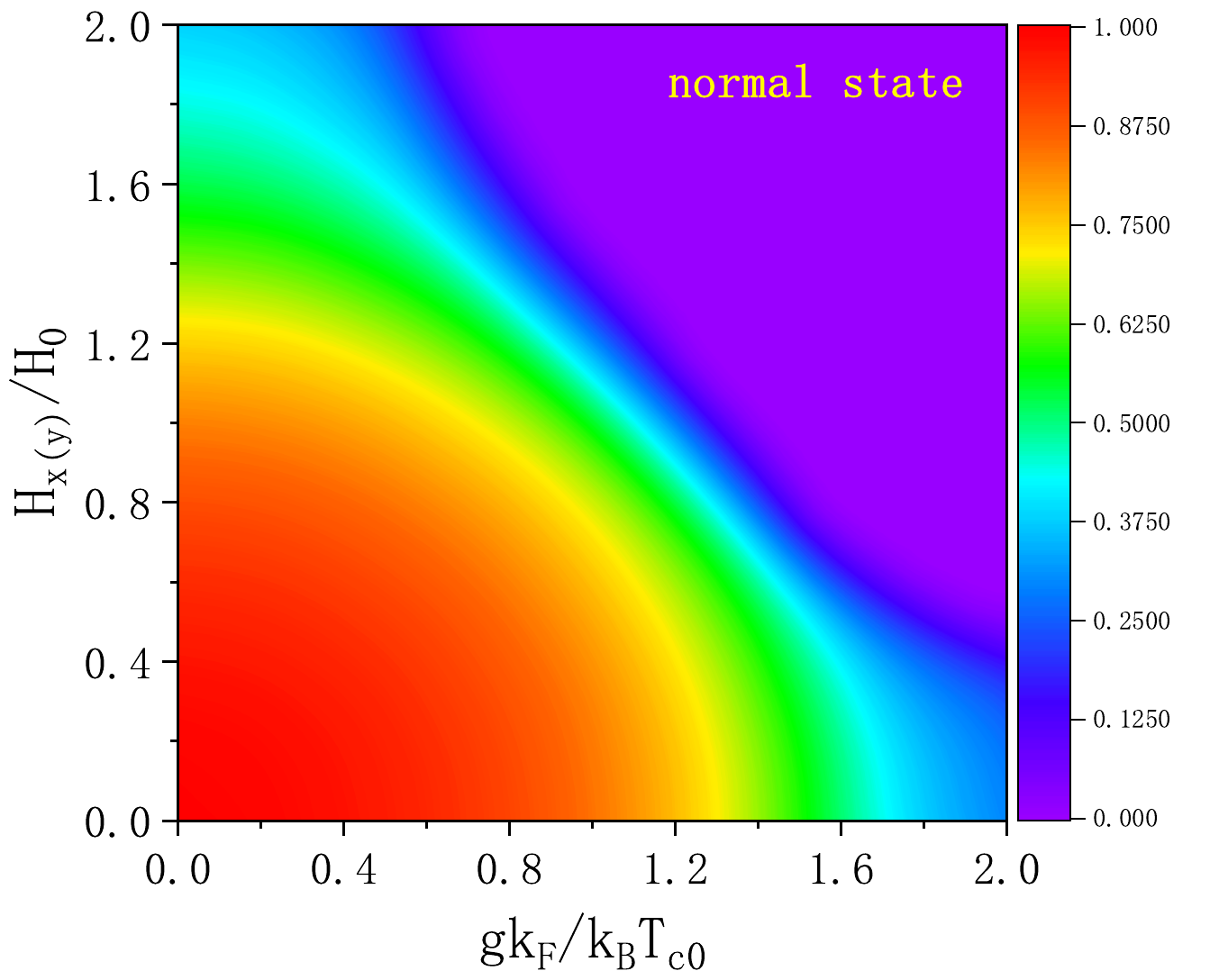}}
\caption{Superconducting transition temperature $T_c$ in units of $T_{c0}$ as a function of the Rashba SOC $g$ and the Zeeman field $\mathbf{H}=H_{x}\hat{x}$ (or $H_{y}\hat{y}$). The four ESP states: (a) $k_x\hat{x}+k_y\hat{y}$, (b) $k_y\hat{x}-k_x\hat{y}$, (c) $k_x\hat{x}-k_y\hat{y}$, and (d) $k_y\hat{x}+k_x\hat{y}$ pairing states. The solid line in (b) is the boundary between the first- and second-order phase transitions across $T_c$. For the pairing states in (a), (c), and (d), no matter how small $T_c$ is, the superconducting to normal state phase transition is always continuous. 
}\label{fig:tc_esp_hxy}
\end{figure*}

Analogous to the OSP states, the ESP states described by the d-vector in Eq.~\eqref{eq:dvec4} can be determined self-consistently using the pairing model of Eq.~\eqref{eq:Vp} and the gap equation outlined in Eq.~\eqref{gapeqwhole}. Figs. \ref{fig:tc_esp_hz} and \ref{fig:tc_esp_hxy} illustrate the numerical results for the superconducting transition temperature $T_c$, which depends on the Rashba SOC $g$ and the Zeeman field $\mathbf{H}$. The key features are discussed and analyzed as follows.

{\bf{}Effects of the Zeeman field:} In the absence of Rashba SOC, a parallel Zeeman field $\mathbf{H}=H_z\hat{z}$ does not alter the transition temperature $T_c$, i.e. $T_c$ remains at $T_{c0}$, as illustrated in Fig.~\ref{fig:tc_esp_hz}. This is because, under these conditions, $\mathbf{d}_{\mathbf{k}}\cdot\hat{\mathbf{H}}=0$ and $|\mathbf{d}_{\mathbf{k}}\times\hat{\mathbf{H}}|=\left|\mathbf{d}_{\mathbf{k}}\right|$ in Eq.~\eqref{eq:EtripltH}. Consequently, the only effect of the parallel Zeeman field is to shift the chemical potential by $\pm\mu_{B}H_z$, leaving the gap equation unaltered. On the other hand, a perpendicular Zeeman field, expressed as $\mathbf{H}=H_{\perp}\left(\cos\phi\hat{x}+\sin\phi\hat{y}\right)$, reduces $T_c$. For the four ESP states listed in Table~\ref{pwave}, they have the same gap equation when $g=0$, resulting in an identical expression for the superconducting $T_c$:
\begin{equation}\label{eq:Tc-ESP-g0}
\ln\left(\frac{T_c}{T_{c0}}\right)=\frac{1}{2}\left[\psi\left(\frac{1}{2}\right)-\mbox{Re}\,\psi\left(\frac{1}{2}+\frac{i\mu_BH_{\perp}}{2\pi k_BT_{c}}\right)\right].
\end{equation}
In the limit $k_B{}T_c\ll\mu_B{}H_{\perp}/2\pi$, this $T_c$ equation yields the solution:
\begin{equation}
\frac{T_c}{T_{c0}}=\sqrt{3}\mathrm{e}^{-\frac{5}{6}}\cdot \frac{H_0}{H_{\perp}}.
\end{equation}
For detailed calculations, refer to Appendix~\ref{app:tc_2nd}.

{\bf{}Rashba SOC effects:} Rashba SOC consistently reduces $T_c$ in all four ESP states, regardless of whether a Zeeman field is present or not. In particular, in the absence of a Zeeman field, the $T_c$ equation retains the same form as Eq.~\eqref{eq:Tc-p-H0}, with $\alpha$ assuming values $1/5, 1, 3/5$ for the pairing states $k_x\hat{x}+k_y\hat{y}$, $k_y\hat{x}-k_x\hat{y}$, $k_x\hat{x}-k_y\hat{y}$ (and $k_y\hat{x}+k_x\hat{y}$). It is also assumed that there is a continuous phase transition from the superconducting state to the normal state.

For the pairing state $k_y\hat{x}-k_x\hat{y}$ with $\alpha=1$ in Eq.~\eqref{eq:Tc-p-H0}, the $T_c$ equation can be numerically solved to reveal a continuous phase transition. Interestingly, for significantly large $g$, $T_c$ no longer behaves monotonically with $g$, implying a first-order phase transition when $T_c$ is sufficiently small. As $g$ increases even further, $T_c$ eventually vanishes.

In the remaining three ESP states, the transition from the superconducting phase to the normal phase is always continuous. When $k_BT_{c}\ll{}gk_F/2\pi$, the following $T_c$ relation holds:
\begin{equation}
\frac{T_c}{T_{c0}}=\left[\frac{\pi \mathrm{e}^{-\gamma}}{2}\cdot\frac{k_BT_{c0}}{gk_F}\right]^\frac{1}{4}
\end{equation}
for the pairing state $k_x\hat{x}+k_y\hat{y}$, and 
\begin{equation}
\frac{T_c}{T_{c0}}=\left[\frac{\pi \mathrm{e}^{-\gamma}}{2}\cdot\frac{k_BT_{c0}}{gk_F}\right]^\frac{3}{2}
\end{equation}
for the pairing states $k_x\hat{x}-k_y\hat{y}$ and $k_y\hat{x}+k_x\hat{y}$.

{\bf{}Combination of Zeeman field and Rashba SOC:} In the presence of both a Zeeman field and Rashba SOC, superconductivity can be completely suppressed in all four ESP states.
\begin{itemize}
\item{} For the four ESP states, an increase in Rashba SOC $g$ and the perpendicular Zeeman field $H_{\perp}$ results in a reduction of $T_c$.

\item{} When Rashba SOC is absent, the superconducting $T_c$ stays at $T_c$ with a parallel Zeeman field $H_{z}$. In contrast, when Rashba SOC is present, raising the parallel Zeeman field $H_z$ reduces $T_c$ in the pairing state $k_x\hat{x}+k_y\hat{y}$, but increases $T_c$ in the remaining three ESP states.

\item{} In the pairing state $k_y\hat{x}-k_x\hat{y}$, the phase transition from the superconducting state to the normal state becomes first order if the superconducting $T_c$ is sufficiently low, whereas in the other three states, the phase transition remains continuous.
\end{itemize}

\subsubsection{Quasiparticle excitation energy}

In the absence of Zeeman field and Rashba SOC, all four ESP states share the same quasiparticle energy dispersion:
\begin{equation}
E_{\mathbf{k}}=\sqrt{\xi_{\mathbf{k}}^2+\Delta^2\sin^2\theta_\mathbf{k}}.    
\end{equation}
Consequently, the energy gap exhibits two nodal points at $\theta_{\mathbf{k}}=0$ and $\pi$ on the Fermi surface.

In the following, we examine the influence of the Zeeman field and Rashba SOC on the excitation energy, with particular emphasis on their energy gap structure.

{\bf{}Zeeman field effect:}
We first consider a parallel Zeeman field $\mathbf{H}=H_z\hat{z}$. In this case, the quasiparticle energy dispersion given in Eq.~\eqref{eq:EtripltG} take a simpler form of
\begin{equation}\label{eq:espEkHz}
E_{\mathbf{k}\pm}=\sqrt{(\xi_{\mathbf{k}}\pm \mu_B H_z)^2 + \Delta^2\sin^2\theta_\mathbf{k}},
\end{equation} 
which resembles Eq.~\eqref{eq:ospEkHx} and leaves the quasiparticle energy gap unchanged. Therefore, there are always two nodal points at $\theta_{\mathbf{k}}=0$ and $\pi$ on the two Fermi surfaces that are separated by the Zeeman field.

On the other hand, when a perpendicular Zeeman field is applied, $\mathbf{H}=H_{\perp}\left(\cos\phi\hat{x}+\sin\phi\hat{y}\right)$, the d-vector in Eq.~\eqref{eq:dvec4} results in
\begin{equation}
|\mathbf{d}_{\mathbf{k}}\times\hat{\mathbf{H}}| = \Delta\sin\theta_{\mathbf{k}}|\sin(\phi_{\mathbf{k}}-\phi)|
\end{equation}
and 
\begin{equation} 
\mathbf{d}_{\mathbf{k}}\cdot\hat{\mathbf{H}} = \Delta\sin\theta_{\mathbf{k}}\cos(\phi_{\mathbf{k}}-\phi). 
\end{equation}
Therefore, as indicated by Eq.~\eqref{eq:EtripltH}, the positions of the nodes in the quasiparticle energy gap are given by
\begin{equation}
\sqrt{\xi_{\mathbf{k}}^2+\Delta^2\sin^2\theta_{\mathbf{k}}\cos^2(\phi_{\mathbf{k}}-\phi)} =\mu_{B}H_{\perp}    
\end{equation}
and
\begin{equation}
\sin\theta_{\mathbf{k}}\sin(\phi_{\mathbf{k}}-\phi)=0.    
\end{equation}

These lead to two possible solutions: (1) $\sin\theta_{\mathbf{k}}=0$ and $|\xi_{\mathbf{k}}|=\mu_B H_{\perp}$, and (2) $\sin(\phi_{\mathbf{k}}-\phi)=0$ and $\sqrt{\xi_{\mathbf{k}}^2 + \Delta^2 \sin^2 \theta_{\mathbf{k}}} = \mu_B H_{\perp}$ if $\mu_B H_{\perp} < \Delta$. This indicates that the two nodal points at $\theta_{\mathbf{k}}=0$ and $\pi$ evolve into two nodal lines on each of the separate Fermi surfaces. These lines are found in $\sqrt{\xi_{\mathbf{k}}^2 + \Delta^2 \sin^2 \theta_{\mathbf{k}}} = \mu_B H_{\perp}$ and $\phi_{\mathbf{k}}-\phi = 0, \pm \pi$.

{\bf{}Rashba SOC effect:} In the presence of Rashba SOC alone, with the d-vector as defined in Eq.~\eqref{eq:dvec4}, the energy dispersion described in Eq.~\eqref{eq:EtripltG} depends on the following components:
\begin{equation}
|\mathbf{d}_{\mathbf{k}}\times\hat{\mathbf{k}}| = \Delta\sin\theta_{\mathbf{k}}\sqrt{\cos^2\theta_{\mathbf{k}}+\sin^2\theta_{\mathbf{k}}\sin^2(\phi_{\mathbf{k}}-\varphi_{\mathbf{k}})}
\end{equation}
and 
\begin{equation}
\mathbf{d}_{\mathbf{k}}\cdot\hat{\mathbf{k}}  = \Delta\sin^2\theta_{\mathbf{k}}\cos(\phi_{\mathbf{k}}-\varphi_{\mathbf{k}}).
\end{equation}

In accordance with Eq.~\eqref{eq:EtripltG}, the quasiparticle energy gap closes if and only if $\sqrt{\xi_{\mathbf{k}}^2+\left|\mathbf{d}_{\mathbf{k}}\times\hat{\mathbf{k}}\right|^2} = g\left|\mathbf{k}\right|$ and $\mathbf{d}_{\mathbf{k}}\cdot\hat{\mathbf{k}}=0$. These criteria lead to:
\begin{equation}
\begin{split}
\xi_{\mathbf{k}}^2 + \Delta^2\sin^2\theta_{\mathbf{k}}\left[\cos^2\theta_{\mathbf{k}} + \sin^2\theta_{\mathbf{k}}\sin^2(\phi_{\mathbf{k}} - \varphi_{\mathbf{k}})\right] & = g^2\left|\mathbf{k}\right|^2,\\
\sin^2\theta_{\mathbf{k}}\cos(\phi_{\mathbf{k}} - \varphi_{\mathbf{k}}) & = 0.
\end{split}    
\end{equation}
Therefore, nodes in the energy gap appear in one of the following scenarios:
\begin{enumerate}
\item When $\sin\theta_{\mathbf{k}} = 0$ and $|\xi_{\mathbf{k}}| = g\left|\mathbf{k}\right|$: Nodal points occur at $\theta_{\mathbf{k}}=0$ and $\pi$, reminiscent of the scenario when $g=0$.
\item When $\cos(\phi_{\mathbf{k}} - \varphi_{\mathbf{k}}) = 0$ and $\xi_{\mathbf{k}}^2 + \Delta^2\sin^2\theta_{\mathbf{k}} = g^2\left|\mathbf{k}\right|^2$: For any given $\theta_{\mathbf{k}}$, the last equation will have a solution for $|\mathbf{k}|$ whenever $gk_{F} \ge \Delta\sin\theta_{\mathbf{k}}$. This leads to the formation of nodal lines or nodal surfaces, depending on the nature of $\cos(\phi_{\mathbf{k}}-\varphi_{\mathbf{k}})$. Explicitly, for the four ESP states listed in Table~\ref{pwave},
\begin{equation}
\cos(\phi_{\mathbf{k}}-\varphi_{\mathbf{k}}) = \begin{cases}
1, & k_x\hat{x}+k_y\hat{y},\\
0, & k_y\hat{x}-k_x\hat{y},\\
\cos2\varphi_{\mathbf{k}}, & k_x\hat{x}-k_y\hat{y},\\
\sin2\varphi_{\mathbf{k}}, & k_y\hat{x}+k_x\hat{y}.
\end{cases}
\end{equation}
As a result, Rashba SOC does not introduce additional gap nodes in the pairing state $k_x\hat{x}+k_y\hat{y}$. However, it causes the appearance of extra nodal surfaces in the pairing state $k_y\hat{x}-k_x\hat{y}$ and additional nodal lines in the pairing states $k_x\hat{x}-k_y\hat{y}$ and $k_y\hat{x}+k_x\hat{y}$.

The additional nodal surfaces are defined by $\sin\theta_{\mathbf{k}}\le{}gk_F/\Delta$,  whereas the extra nodal lines are characterized by $\sin\theta_{\mathbf{k}}\le{}gk_F/\Delta$ and $\varphi_{\mathbf{k}}=(2n+1)\pi/4$ for the pairing state $k_x\hat{x}-k_y\hat{y}$, and by $\sin\theta_{\mathbf{k}}\le{}gk_F/\Delta$ and $\varphi_{\mathbf{k}}=n\pi/2$ for the pairing state $k_y\hat{x}+k_x\hat{y}$, where $n=0,1,2,3$.
\end{enumerate}

\subsubsection{Discussions}
The structure of the gap nodes discussed previously agrees well with the characteristics observed in the ESP states $T_c$.

\begin{itemize}
\item{} The fact that the parallel Zeeman field does not modify the quasiparticle energy gap, combined with the perpendicular Zeeman field producing an identical nodal line structure across all four ESP states, both results in the same Zeeman field dependence on the superconducting transition temperature $T_c$.
\item{} The presence of Rashba SOC creates extra nodal surfaces in the pairing state $k_y\hat{x}-k_x\hat{y}$, causing a more rapid decline in $T_c$ relative to the other three ESP states and inducing a first-order phase transition. In contrast, it does not generate an additional gap node in the pairing state $k_x\hat{x}+k_y\hat{y}$, leading to a much slower reduction in $T_c$ with increasing $g$ compared to the other ESP states.
\end{itemize}

\section{Summary and Discussions}

In summary, we have studied the effects of an external Zeeman field and Rashba SOC on the superconducting transition temperature $T_c$ and quasiparticle excitations for $s$- and $p$- wave pairing superconductors. For physical relevance, Rashba SOC plays a crucial role in bulk superconductors that break spatial inversion symmetry. 

For the $s$-wave pairing superconductor, a phase diagram is illustrated in Fig.~\ref{fig:swave_phasediagram}. It is evident that the Zeeman field reduces the superconducting $T_c$, whereas Rashba SOC maintains $T_c$ when the Zeeman field is absent, and tends to restore $T_c$ to its inherent value $T_{c0}=T_{c}(g=0,\mathbf{H}=0)$ in the presence of the Zeeman field. The contrasting impacts of the Zeeman field and Rashba SOC on $T_c$ can be explained through the interband and intraband pairing scenario. Furthermore, as anticipated, a sufficiently strong Zeeman field induces an FFLO state before superconductivity is completely suppressed. In the phase diagram, the FFLO region expands with a sufficiently large Rashba SOC. Moreover, Bogoliubov Fermi surfaces arise from the combined effects of Rashba SOC and the Zeeman field, allowing both quasiparticles and quasiholes to have Fermi surfaces. The Bogoliubov Fermi surfaces are observed in both the FFLO phase and neighboring uniform superconducting states.

The $p$- wave pairing includes the OSP and ESP states listed in Table~\ref{pwave}.For the two OSP states $(k_x+ik_y)\hat{z}$ and $k_z\hat{z}$, both Rashba SOC and a Zeeman field aligned with the spin quantization axis reduce $T_c$, while a perpendicular Zeeman field does not affect $T_c$, as illustrated in Fig.~\ref{fig:tc_magsoc-dz}. For the four ESP states, both Rashba SOC and a perpendicular Zeeman field lower $T_c$, while a parallel Zeeman field leaves $T_c$ unchanged, as shown in Figs.~\ref{fig:tc_esp_hz} and \ref{fig:tc_esp_hxy}. Moreover, an external Zeeman field combined with Rashba SOC can greatly alter the nodal configuration of the quasiparticle energy gap. For instance, within the pairing state $(k_x+ik_y)\hat{z}$, a parallel Zeeman field can induce a nodal surface, as depicted in Fig.~\ref{fig:nodalS1}.

Finally, we discuss a few concerns related to the validation and potential applications of our theory.

\begin{itemize}
\item\emph{Mixture of $s$-wave and $p$-wave pairings:} When spatial inversion symmetry is broken, parity is no longer conserved, which permits the coexistence of $s$-wave and $p$-wave pairing under strong Rashba SOC~\cite{Frigeri2004,bauer2012non,Yip14,Samohkin15,Smidman17}. A mixture of this kind can qualitatively alter the structure of the gap node~\cite{NaIrO4}. However, the mixing ratio also depends on the pairing strength in each pairing channel characterized by $V_l$ in Eq.~\eqref{eq:Vl}.
Our analysis in this paper will still hold if the mixture is not particularly strong.

\item{
\emph{Feasibility of a uniform Zeeman field:}} 
In addition to the Zeeman effect considered in this paper, which concerns the electron spin degrees of freedom, a magnetic field also induces a diamagnetic current known as the orbital effect. In a type-I superconductor, a weak external magnetic field results in the complete expulsion of the magnetic field from the bulk. However, our theory is relevant for the intermediate state in type-I superconductors in a strong magnetic field and applies to type-II superconductors when $H_{c1}<H<H_{c2}$~\cite{deGennes,tinkham}.
Moreover, our theory can also be applied to systems where ferromagnetism and superconductivity coexist~\cite{HUXLEY2015} or hybridize~\cite{Liu2022} at the microscopic level. In such systems, it is possible to expand the calculations for the Zeeman field to encompass local magnetization.

\item\emph{Implications to K$_2$Cr$_3$As$_3$:} A novel family of superconductors, A$_2$Cr$_3$As$_3$ (A = Na, K, Rb, and Cs), has recently been discovered~\cite{Bao15,Tang15_1,Tang15_2,Mu18}, exhibiting a $T_c$ up to $8$ K. Density functional theory calculations indicate that the Cr-3d orbitals primarily influence the electronic states close to the Fermi level. At the Fermi level, there are three energy bands: two quasi-1D $\alpha$ and $\beta$ bands with flat Fermi surfaces and a 3D $\gamma$ band~\cite{Jiang15,Wu15}. 
Given that the 3D $\gamma$ band significantly contributes to the density of states around the Fermi level, our theoretical model assumes a spherical Fermi surface, which is relevant to these compounds. Additionally, the A$_2$Cr$_3$As$_3$ lattices possess the $P6m2$ space group and do not exhibit spatial inversion symmetry~\cite{Bao15}, which supports the inclusion of Rashba SOC in our model.

Of particular interest is the NMR experiment conducted on the K$_2$Cr$_3$As$_3$ single crystal~\cite{Triplet2021}, which indicates that the spin susceptibility, measured by the $^{75}$As Knight shift, remains unchanged below $T_c$ when the magnetic field $\mathbf{H}$ is applied within the $ab$ plane. However, it approaches zero as the temperature approaches zero when $\mathbf{H}$ is oriented along the $c$ axis. This clearly suggests that the compound is a spin-triplet pairing superconductor with its d-vector aligned along the $c$ axis, which agrees with previous theoretical predictions~\cite{ZHOU2017208}. Specifically, it points to an OSP state described by the d-vector $\mathbf{d}(\mathbf{k})=d_{z}(\mathbf{k})\hat{z}$ as studied in our paper. According to Table~\ref{pwave}, the superconducting ground state can be the pairing state $(k_{x}+ik_y)\hat{z}$ or $k_z\hat{z}$. Therefore, the results presented in this paper can be used in future experiments and theoretical research to differentiate between these two pairing states. For instance, as illustrated in Fig.~\ref{fig:tc_magsoc-dz}, when the Rashba SOC is sufficiently strong, specifically $gk_F/k_B{}T_{c0}\gtrsim{}1.5$, the pairing state $(k_{x}+ik_y)\hat{z}$ will have its superconductivity entirely suppressed, whereas it will persist in $k_z\hat{z}$. If the Rashba SOC strength can be accurately measured or calculated, distinguishing between these two pairing states becomes feasible.
\end{itemize}

\section*{Acknowledgement}
We would like to thank Prof. Guang-Han Cao for helpful discussions. This work is partially supported by National Key Research and Development Program of China (No. 2022YFA1403403), and National Natural Science Foundation of China (No. 12274441, 12034004).

\appendix

\section{Bogoliubov transformation matrices}\label{app:BTM}
The Hamiltonian governing a superconducting system can be written as follows:
\begin{equation}
H=\frac{1}{2}\sum_{\mathbf{k}}C^{\dagger}_{\mathbf{k}}H(\mathbf{k})C_{\mathbf{k}},
\end{equation} 
where $C_{\mathbf{k}}=(c_{\mathbf{k}\uparrow},
c_{\mathbf{k}\downarrow},
c_{-\mathbf{k}\uparrow}^{\dagger},
c_{-\mathbf{k}\downarrow}^{\dagger})^{\top}$, and
\begin{equation} \begin{aligned}\label{eq:ckhk}
H(\mathbf{k})&=\begin{pmatrix}
H_0(\mathbf{k})&\Delta(\mathbf{k})\\
\Delta^{\dagger}(\mathbf{k})&-[H_0(-\mathbf{k})]^\top
\end{pmatrix}.
\end{aligned}\end{equation} 
Introducing $
\Psi_{\mathbf{k}}=(\psi_{\mathbf{k}+},\psi_{\mathbf{k}-},\psi_{-\mathbf{k}+}^{\dagger},\psi_{-\mathbf{k}-}^{\dagger})^{\top}$, the Bogoliubov transformation can be abbreviated to $$C_{\mathbf{k}}=U_{\mathbf{k}}\Psi_{\mathbf{k}},$$where
\begin{equation}
U_{\mathbf{k}}=\begin{pmatrix}
u_{\mathbf{k}}&v_{\mathbf{k}}\\
v_{-\mathbf{k}}^*&u_{-\mathbf{k}}^*
\end{pmatrix}.
\label{Transm}
\end{equation} 
Then the Hamiltonian is diagonalized as follows, 
\begin{equation}
U_{\mathbf{k}}^{\dagger}H(\mathbf{k})U_{\mathbf{k}}=diag\begin{pmatrix}
E_{\mathbf{k}+}&E_{\mathbf{k}-}&-E_{-\mathbf{k}+}&-E_{-\mathbf{k}-}
\end{pmatrix}.
\end{equation} 
Note that $U_{\mathbf{k}}$ given in Eq.~\eqref{Transm} is a unitary matrix, i.e. $U_{\mathbf{k}}U_{\mathbf{k}}^\dagger=U_{\mathbf{k}}^\dagger U_{\mathbf{k}}=\mathbb{I}$. The substitution of Eq.~\eqref{Transm} leads to
\begin{equation}
\begin{aligned} u_{\mathbf{k}}u_{\mathbf{k}}^\dagger+v_{\mathbf{k}}v_{\mathbf{k}}^\dagger&=\mathbb{I} ,\\
u_{\mathbf{k}}^\dagger u_{\mathbf{k}}+v_{-\mathbf{k}}^\top v_{-\mathbf{k}}^*&=\mathbb{I}, \\
u_{\mathbf{k}}v_{-\mathbf{k}}^\top+v_{\mathbf{k}}u_{-\mathbf{k}}^\top&=0.
\end{aligned}
\end{equation}

\section{Intraband and interband pairing functions}\label{app:Theta}

In this appendix, we will explore the intraband and interband pairing functions expressed in the pseudo-spin basis as presented in Eq.~\eqref{eq:pair-pspin}. These functions rely on $\Theta_{\mathbf{k}}$ and $\Phi_{\mathbf{k}}$ defined in Eq.~\eqref{eq:Thetak}.

For an $s$-wave pairing state, according to Eq.~\eqref{eq:s-pseudo-spin} in the main text, we have the intraband and interband pairing functions as follows:
\begin{equation*}
\begin{split}
&\Delta_{++}=-\Delta_{--}=-\Delta{}\mathrm{e}^{-i\varphi_{\mathbf{k}}} \sin{\frac{\Theta_{\mathbf{k}}+\Theta_{-\mathbf{k}}}{2}},\\
&\Delta_{+-}=\Delta_{-+}=\Delta{}\mathrm{e}^{-i\varphi_{\mathbf{k}}} \cos{\frac{\Theta_{\mathbf{k}}+\Theta_{-\mathbf{k}}}{2}}.
\end{split}
\end{equation*} 

Without loss of generality, we consider $\mathbf{H}=H_z\hat{z}$ and have $\Phi_{\mathbf{k}}=\varphi_\mathbf{k}$, and
\begin{equation}
\begin{split}
\tan\Theta_{\mathbf{k}}&=\frac{g\lvert\mathbf{k}\lvert\sin \theta_{\mathbf{k}}}{\mu_BH_z+g\lvert\mathbf{k}\lvert\cos \theta_{\mathbf{k}}},\\
\tan\left(\Theta_{\mathbf{k}}+\Theta_{\mathbf{-k}}\right)&=\frac{2\mu_{B}H_z{}g\lvert\mathbf{k}\lvert\sin \theta_{\mathbf{k}}}{\mu_B^2{}H_z^2 - g^2\lvert\mathbf{k}\lvert^2},\\
\tan\left(\Theta_{\mathbf{k}}-\Theta_{\mathbf{-k}}\right)&=\frac{-g^2\lvert\mathbf{k}\rvert^2\sin 2\theta_{\mathbf{k}}}{\mu_B^2{}H_z^2 - g^2\lvert\mathbf{k}\lvert^2\cos2\theta_{\mathbf{k}}}.
\end{split}
\end{equation}
\begin{itemize}
\item{}If $g{}k_{F}/\mu_{B}H_z\to{}0$, both $\Theta_{\mathbf{k}}$ and $\Theta_{-\mathbf{k}}$ tend towards zero, resulting in $\Theta_{\mathbf{k}}\pm\Theta_{-\mathbf{k}}\to{}0$.
\item{}If $\mu_{B}H_z/g{}k_{F}\to{}0$, $\Theta_{\mathbf{k}}\to\theta_{\mathbf{k}}$ and $\Theta_{-\mathbf{k}}\to\theta_{-\mathbf{k}}=\pi-\theta_{\mathbf{k}}$, yielding $\Theta_{\mathbf{k}}+\Theta_{-\mathbf{k}}\to{}\pi$ and $\Theta_{\mathbf{k}}-\Theta_{-\mathbf{k}}\to{}2\theta_{\mathbf{k}}-\pi$.
\item{}As the ratio of $g{}k_{F}/\mu_{B}H_z$ increases from 0 to 1 and then to $+\infty$, the angle $\frac{\Theta_{\mathbf{k}}+\Theta_{\mathbf{-k}}}{2}$ increases monotonically from $0$ to $\pi/4$ and then to $\pi/2$; while the angle $\frac{\Theta_{\mathbf{k}}-\Theta_{\mathbf{-k}}}{2}$ monotonically decreases or increases from $0$ to $\theta_{\mathbf{k}}-\pi/2$, depending on $\theta_\mathbf{k}\in(0,\pi/2)$ or $\theta_\mathbf{k}\in(\pi/2,\pi)$.
\end{itemize}

\section{Gap equation and free energy: uniform $s$-wave pairing state}
For the $s$-wave pairing function specified in Eq.~\eqref{dels}, the gap equation \eqref{gapeqwhole} simplifies to the form \begin{subequations}\label{eq:sgap} \begin{equation}  
\Delta =- V_0\sum_{\mathbf{k}}\langle c_{-\mathbf{k}\downarrow}c_{\mathbf{k}\uparrow}\rangle,  
\end{equation}
which can be solved in a self-consistent way, given the pairing interaction $V_0$ and the energy cutoff $\Lambda=\hbar\omega_D$ are defined. Here, $\omega_{D}$ denotes the Debye frequency.

An analytical solution to the gap equation cannot be obtained in the presence of both the Zeeman field and Rashba SOC. Thus, we will approach the gap equation through numerical methods, which explicitly takes the form:
\begin{widetext}
\begin{equation}
\Delta=-V_0\sum_{\mathbf{k}}\left[(U_\mathbf{k})^*_{41}(U_\mathbf{k})_{11}f(E_{\mathbf{k}+})+(U_\mathbf{k})^*_{42}(U_\mathbf{k})_{12}f(E_{\mathbf{k}-})+(U_\mathbf{k})_{43}^*(U_\mathbf{k})_{13}f(-E_{-\mathbf{k}+})+(U_\mathbf{k})^*_{44}(U_\mathbf{k})_{14}f(-E_{-\mathbf{k}-})\right],    
\end{equation}    
\end{widetext}
\end{subequations}
where $f(E)$ is the Fermi distribution function and $U_{\mathbf{k}}$ is a $4\times{}4$ matrix specified by the Bogoliubov transformation in Eq.~\eqref{Transm}.

Furthermore, to validate the physical solution of the gap equation \eqref{eq:sgap}, which corresponds to the minimum of the free energy, we calculate the free energy difference between the superconducting and normal states as follows,
\begin{widetext}
\begin{equation}\label{free}
F_s-F_n=-\frac{1}{\beta}\sum_{\mathbf{k}} \ln\frac{(1+\mathrm{e}^{-\beta E_{\mathbf{k}+}})(1+\mathrm{e}^{-\beta E_{\mathbf{k}- }})}{(1+\mathrm{e}^{-\beta \xi_{\mathbf{k}+}})(1+\mathrm{e}^{-\beta \xi_{\mathbf{k}-}})}+\frac{\lvert\Delta\lvert^2}{V_0} +\sum_{\mathbf{k}}\left[\xi_{\mathbf{k}}-\frac{E_{-\mathbf{k}+}+E_{-\mathbf{k}-}}{2}\right],
\end{equation}
\end{widetext}
where $E_{\pm\mathbf{k}\pm}$ and $\xi_{\mathbf{k}\pm}$ are the eigenvalues of the Hamiltonian for the superconducting and normal states respectively. In principle, the gap function in Eq.~\eqref{gapeqwhole} can be seen as the saddle point of the free energy functional. The superconducting transition temperature $T_c$ is then determined by
\begin{equation}\label{eq:Tc}
F_s(T=T_c)-F_n(T=T_c)=0,    
\end{equation}
where $F_s-F_n$ is specified in Eq.~\eqref{free}.

\section{Bogoliubov transformation: $s$-wave pairing state when only Zeeman field or Rashba SOC is present}\label{app:BT}

\subsection{In a Zeeman field}
For an $s$-wave superconductor in a Zeeman field and without Rashba SOC, we explicitly have
\begin{equation} \begin{aligned}
H(\mathbf{k})&=\begin{pmatrix}
\xi_{\mathbf{k}+}&0&0&\Delta\\
0&\xi_{\mathbf{k}-}&-\Delta&0\\
0&-\Delta^*&-\xi_{\mathbf{k}+}&0\\
\Delta^*&0&0&-\xi_{\mathbf{k}-}
\end{pmatrix},
\end{aligned}\end{equation} 
where 
\begin{equation} \begin{aligned}
\xi_{\mathbf{k}\pm}=\xi_{\mathbf{k}}\pm\mu_BH_z.
\end{aligned}\end{equation} 
In this case, the Bogoliubov transformation matrix is
\begin{equation} \begin{aligned}
U_{\mathbf{k}}=\begin{pmatrix}
\frac{E_{\mathbf{k}}+\xi_{\mathbf{k}}}{[2E_{\mathbf{k}}(E_{\mathbf{k}}+\xi_{\mathbf{k}})]^{\frac{1}{2}}}&0&0&\frac{-\Delta}{[2E_{\mathbf{k}}(E_{\mathbf{k}}+\xi_{\mathbf{k}})]^{\frac{1}{2}}}\\		0&\frac{E_{\mathbf{k}}+\xi_{\mathbf{k}}}{[2E_{\mathbf{k}}(E_{\mathbf{k}}+\xi_{\mathbf{k}})]^{\frac{1}{2}}}&\frac{\Delta}{[2E_{\mathbf{k}}(E_{\mathbf{k}}+\xi_{\mathbf{k}})]^{\frac{1}{2}}}&0\\
0&\frac{-\Delta^*}{[2E_{\mathbf{k}}(E_{\mathbf{k}}+\xi_{\mathbf{k}})]^{\frac{1}{2}}}&\frac{E_{\mathbf{k}}+\xi_{\mathbf{k}}}{[2E_{\mathbf{k}}(E_{\mathbf{k}}+\xi_{\mathbf{k}})]^{\frac{1}{2}}}&0\\
\frac{\Delta^*}{[2E_{\mathbf{k}}(E_{\mathbf{k}}+\xi_{\mathbf{k}})]^{\frac{1}{2}}}&0&0&\frac{E_{\mathbf{k}}+\xi_{\mathbf{k}}}{[2E_{\mathbf{k}}(E_{\mathbf{k}}+\xi_{\mathbf{k}})]^{\frac{1}{2}}}
\end{pmatrix},
\end{aligned}\end{equation} 
which is exactly the same as the case without the Zeeman field.

\subsection{In the presence of Rashba SOC}
If only the finite Rashba SOC $g$ but no Zeeman field is present in an $s$-wave pairing superconductor, it is easy to verify that the gap equation is exactly the same as for a conventional $s$-wave superconductor for all values of $g$.

As shown in Fig.~\ref{soc}, the $\mathbf{k}$-space can be divided into three regions, and the corresponding transformation matrices are as follows
\begin{widetext}
\begin{equation}
u_{\mathbf{k}}=\left\{\begin{aligned}
&\begin{pmatrix}
a_{\mathbf{k}+}\cos{\frac{\theta_{\mathbf{k}}}{2}}& \sin{\frac{\theta_{\mathbf{k}}}{2}}\\
a_{\mathbf{k}+}\sin{\frac{\theta_{\mathbf{k}}}{2}}\mathrm{e}^{i\varphi_{\mathbf{k}}}&-\cos{\frac{\theta_{\mathbf{k}}}{2}}\mathrm{e}^{i\varphi_{\mathbf{k}}}
\end{pmatrix},\qquad &\mathbf{k}\in I_1\\
&\begin{pmatrix}
a_{\mathbf{k}+}\cos{\frac{\theta_{\mathbf{k}}}{2}}& a_{\mathbf{k}-}\sin{\frac{\theta_{\mathbf{k}}}{2}}\\
a_{\mathbf{k}+}\sin{\frac{\theta_{\mathbf{k}}}{2}}\mathrm{e}^{i\varphi_{\mathbf{k}}}&-a_{\mathbf{k}-}\cos{\frac{\theta_{\mathbf{k}}}{2}}\mathrm{e}^{i\varphi_{\mathbf{k}}}
\end{pmatrix},&\mathbf{k}\in I_2\\
&\begin{pmatrix}
\cos{\frac{\theta_{\mathbf{k}}}{2}}& a_{\mathbf{k}-}\sin{\frac{\theta_{\mathbf{k}}}{2}}\\
\sin{\frac{\theta_{\mathbf{k}}}{2}}\mathrm{e}^{i\varphi_{\mathbf{k}}}&-a_{\mathbf{k}-}\cos{\frac{\theta_{\mathbf{k}}}{2}}\mathrm{e}^{i\varphi_{\mathbf{k}}}
\end{pmatrix},&\mathbf{k}\in I_3
\end{aligned}\right., \quad
v_{\mathbf{k}}=\left\{
\begin{aligned}
&\begin{pmatrix}
b_{\mathbf{k}+}\cos{\frac{\theta_{\mathbf{k}}}{2}}\mathrm{e}^{-i\varphi_{\mathbf{k}}}&0\\
b_{\mathbf{k}+}\sin{\frac{\theta_{\mathbf{k}}}{2}}&0
\end{pmatrix},\quad &\mathbf{k}\in I_1\\
&\begin{pmatrix}
b_{\mathbf{k}+}\cos{\frac{\theta_{\mathbf{k}}}{2}}\mathrm{e}^{-i\varphi_{\mathbf{k}}}&-b_{\mathbf{k}-}\sin{\frac{\theta_{\mathbf{k}}}{2}}\mathrm{e}^{-i\varphi_{\mathbf{k}}}\\
b_{\mathbf{k}+}\sin{\frac{\theta_{\mathbf{k}}}{2}}&b_{\mathbf{k}-}\cos{\frac{\theta_{\mathbf{k}}}{2}}
\end{pmatrix}, &\mathbf{k}\in I_2\\
&\begin{pmatrix}
0&-b_{\mathbf{k}-}\sin{\frac{\theta_{\mathbf{k}}}{2}}\mathrm{e}^{-i\varphi_{\mathbf{k}}}\\
0&b_{\mathbf{k}-}\cos{\frac{\theta_{\mathbf{k}}}{2}}
\end{pmatrix},\qquad &\mathbf{k}\in I_3
\end{aligned}\right.,
\label{eq:bt_s_soc}
\end{equation}
where $a_{\mathbf{k}\pm}$ and $b_{\mathbf{k}\pm}$ are defined as
\begin{equation}
\begin{aligned}
a_{\mathbf{k}\pm}&=\frac{\xi_{\mathbf{k}\pm}+E_{\mathbf{k}\pm}}{[2E_{\mathbf{k}\pm}(\xi_{\mathbf{k}\pm}+E_{\mathbf{k}\pm})]^{\frac{1}{2}}},\quad
b_{\mathbf{k}\pm}&=\frac{\Delta}{[2E_{\mathbf{k}\pm}(\xi_{\mathbf{k}\pm}+E_{\mathbf{k}\pm})]^{\frac{1}{2}}}.
\end{aligned}
\end{equation} 
Substituting the above transformation matrices into Eq.~\eqref{eq:sgap}, we have the following equation,
\begin{equation}
\begin{aligned}
\Delta=&-V\left[\sum_{\mathbf{k}\in I_1}\langle c_{-\mathbf{k}\downarrow}c_{\mathbf{k}\uparrow}\rangle+\sum_{\mathbf{k}\in I_2}\langle c_{-\mathbf{k}\downarrow}c_{\mathbf{k}\uparrow}\rangle+\sum_{\mathbf{k}\in I_3}\langle c_{-\mathbf{k}\downarrow}c_{\mathbf{k}\uparrow}\rangle\right] \\
=&-V\Delta\left[\sum_{\mathbf{k}\in I_1,I_2}\frac{\cos^2{\frac{\theta_{\mathbf{k}}}{2}}}{2E_{\mathbf{k}+}}\left\langle\psi_{\mathbf{k}+}^{\dagger}\psi_{\mathbf{k}+}-\psi_{-\mathbf{k}+}\psi_{-\mathbf{k}+}^\dagger\right\rangle +\sum_{\mathbf{k}\in I_2,I_3}\frac{\sin^2{\frac{\theta_{\mathbf{k}}}{2}}}{2E_{\mathbf{k}+}}\left\langle\psi_{\mathbf{k}-}^{\dagger}\psi_{\mathbf{k}-}-\psi_{-\mathbf{k}-}\psi_{-\mathbf{k}-}^\dagger\right\rangle\right].
\end{aligned}
\end{equation}    
\end{widetext}
 
Since $\xi_{\mathbf{k}+}(\xi_{\mathbf{k}-})\in [-\Lambda,\Lambda]$, one can replace $\xi_{\mathbf{k}+}(\xi_{\mathbf{k}-})$ in the integrand by $\xi_{\mathbf{k}}$. This restores the gap equation to its conventional form:
\begin{equation} \begin{aligned}
1=N(0)V\int_{-\Lambda}^{\Lambda}d\xi_{\mathbf{k}}\frac{1}{2E_{\mathbf{k}}}\left[1-2f(E_{\mathbf{k}})\right].
\end{aligned}\end{equation}

\section{Finite total momentum $s$-wave pairing states}\label{free_ff}

\begin{figure*}[tb]
\centering
\subfigure[]{ \includegraphics[width=0.485\linewidth]{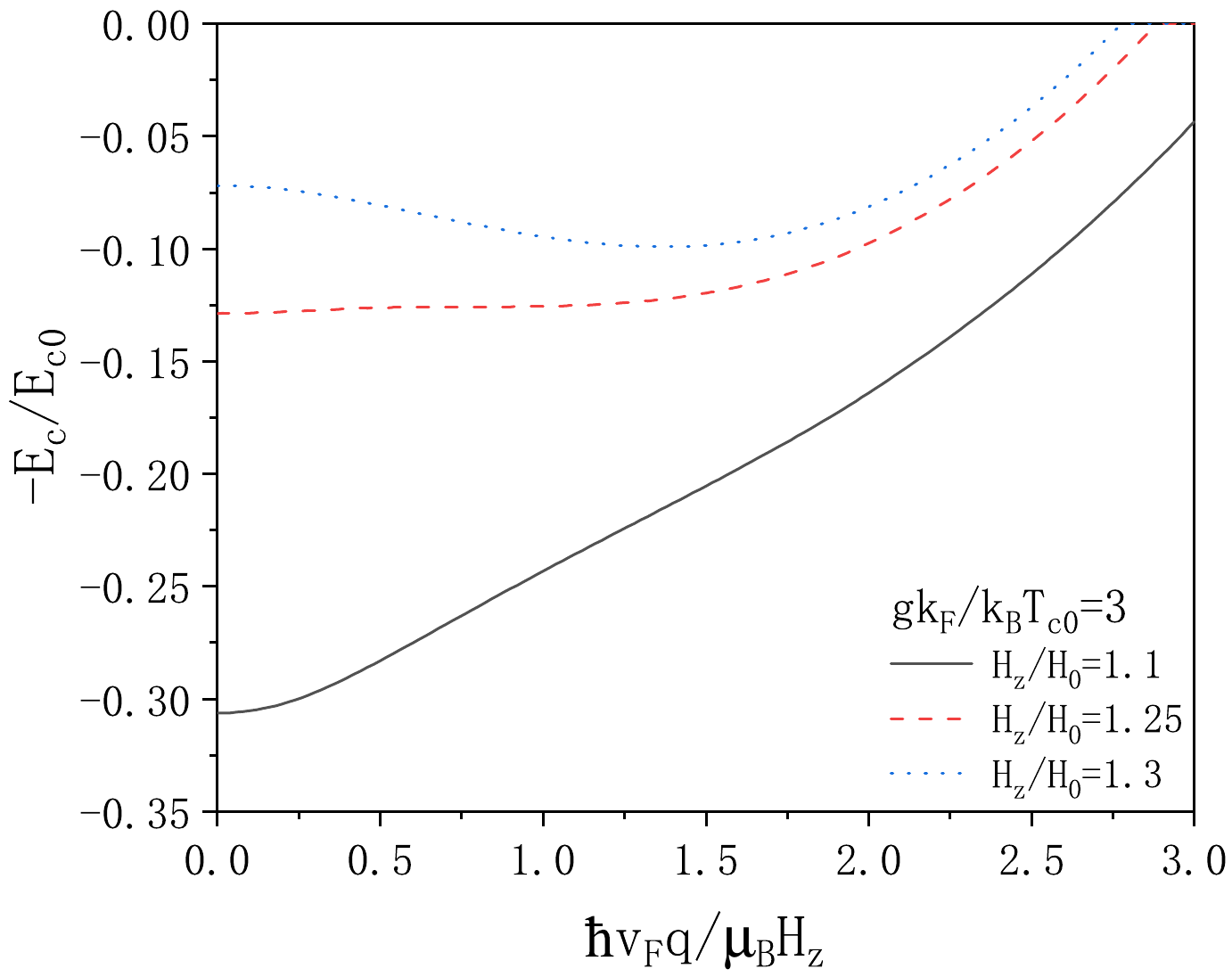} }
\subfigure[]{ \includegraphics[width=0.485\linewidth]{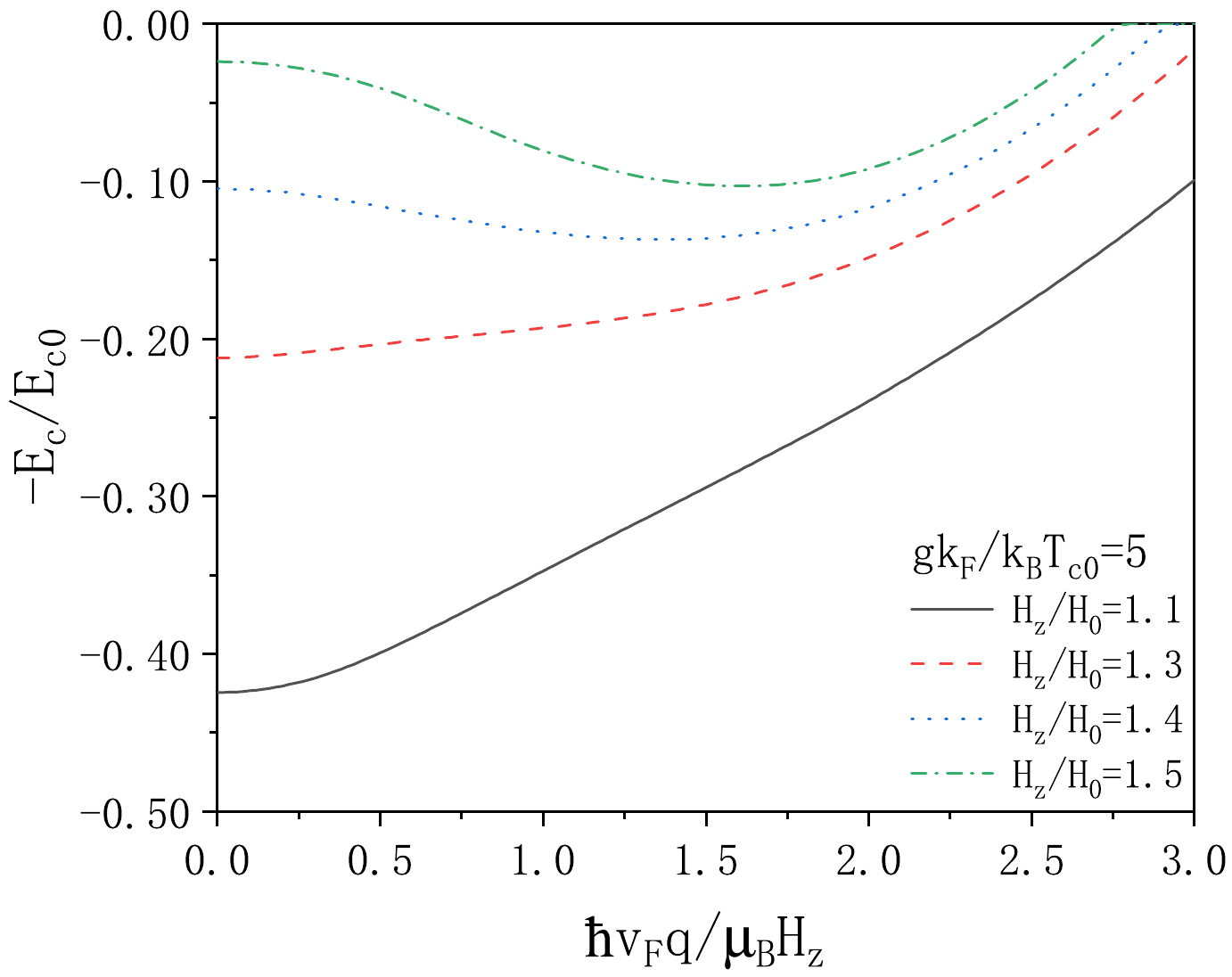} }
\caption{$s$-wave pairing FFLO state: The condensation energy $E_c$ is plotted versus $q=|\mathbf{q}|$. An FFLO ground state is identified when $E_c$ achieves its maximum at a nonzero $q$. In this state, the total momentum $\mathbf{q}$ points in the opposite direction to the Zeeman field $\mathbf{H}=H_z\hat{z}$, with $E_{c0}$ determined at $q=0$. We examine the $gk_F/k_BT_{c0}$ values of (a) $3$ and (b) $5$ as typical examples. }
\label{Fig:free_ff}
\end{figure*} 

\begin{figure}[tb]
\centering
\subfigure[]{ \includegraphics[width=0.97\linewidth]{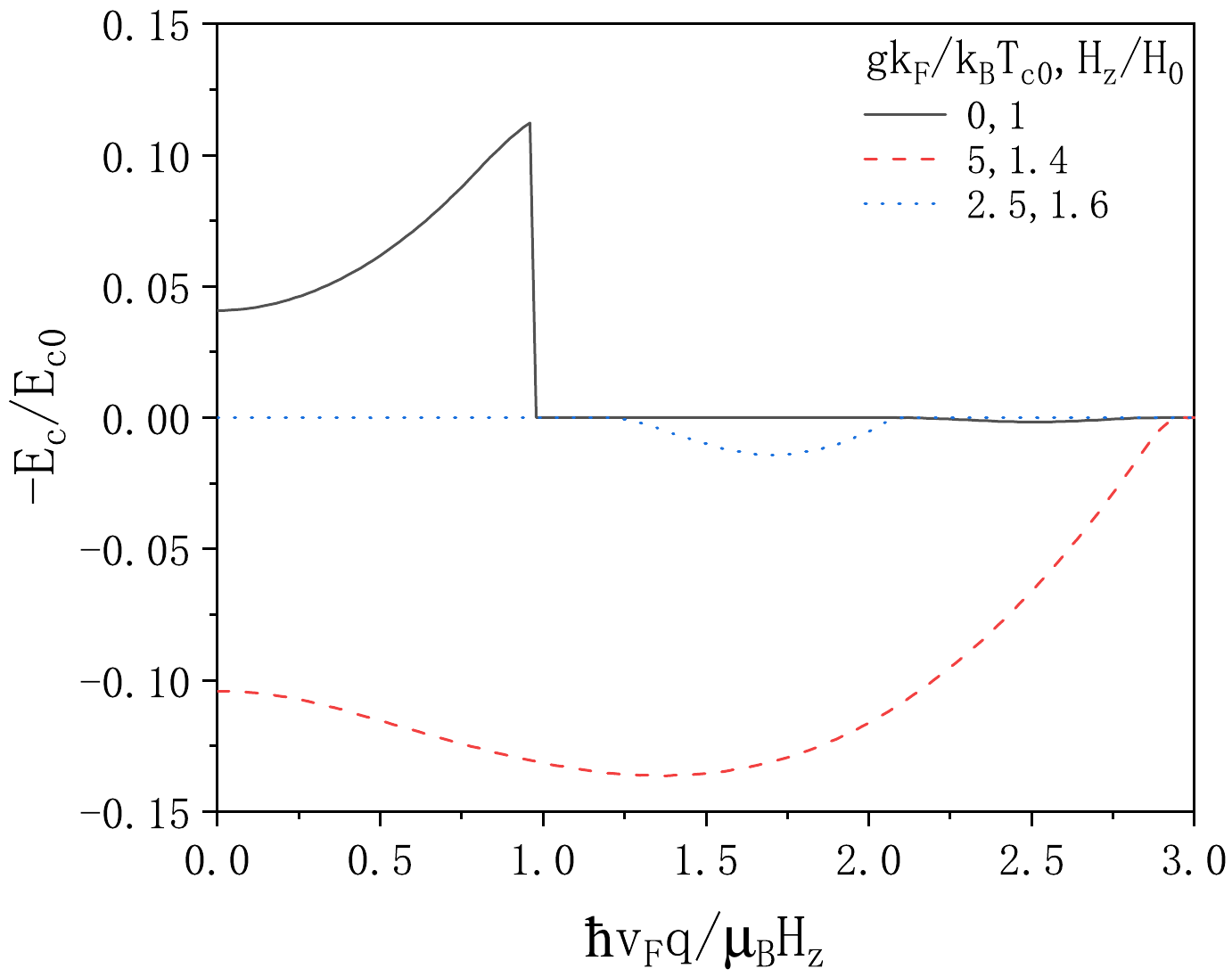} }
\caption{ $s$-wave pairing FFLO state: Condensation energy plotted versus the total momentum of a Cooper pair $q=|\mathbf{q}|$.\label{fig:FFLO-Ec} }
\end{figure} 

\begin{figure*}[tb]
\centering
\subfigure[]{ \includegraphics[width=0.485\linewidth]{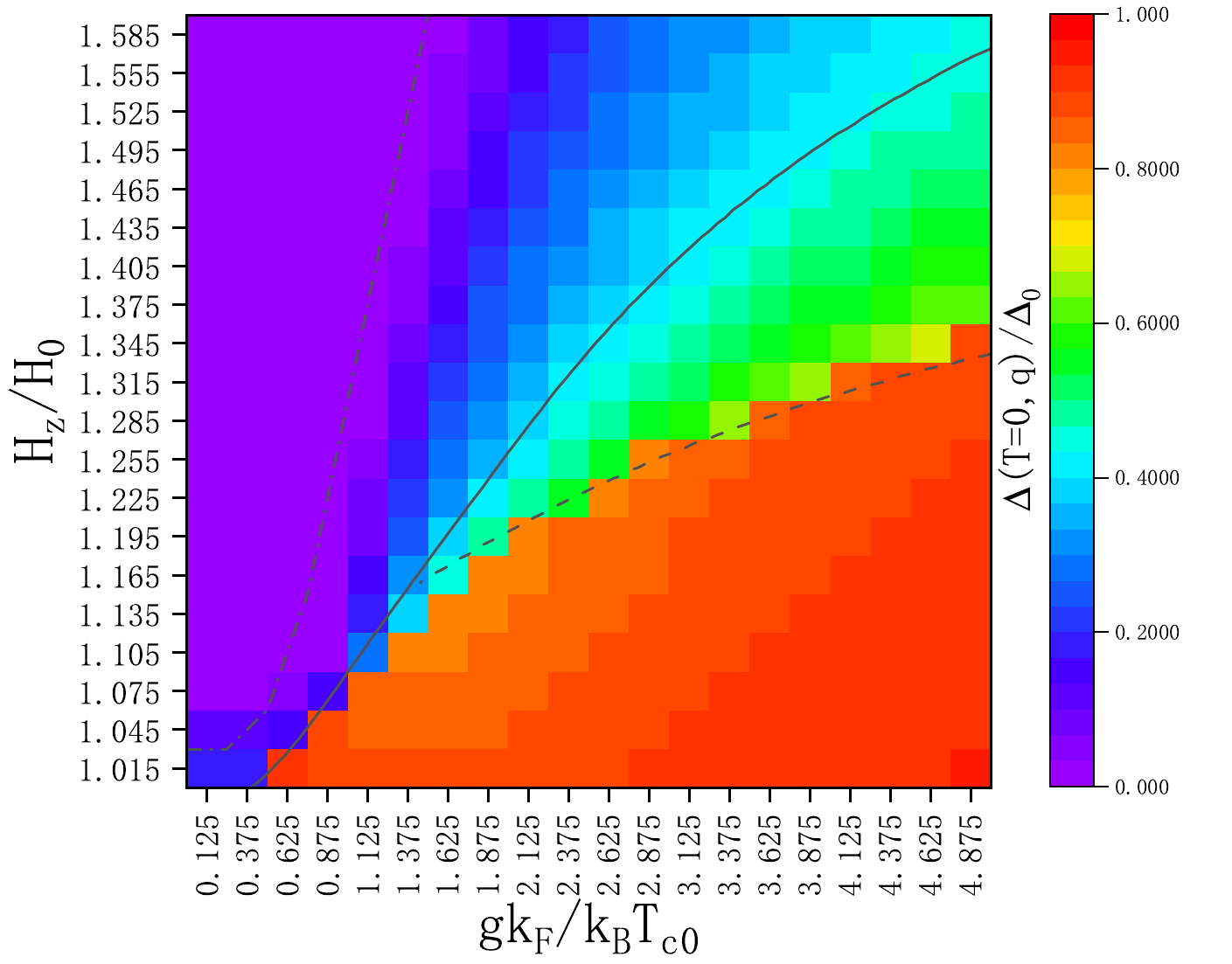} }
\subfigure[]{ \includegraphics[width=0.485\linewidth]{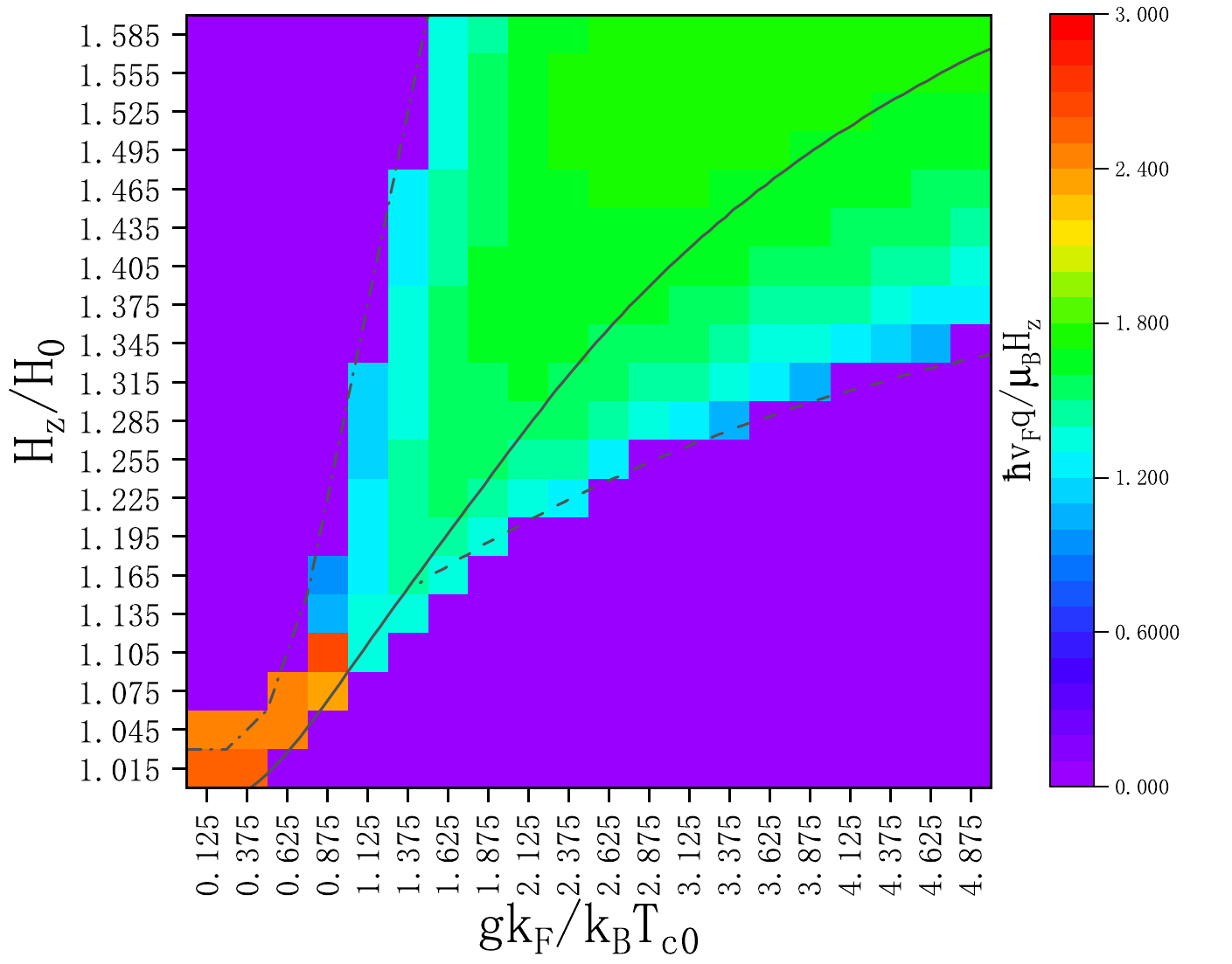} }
\caption{ $s$-wave pairing FFLO state: (a) Gap function $\Delta_{\mathbf{q}}(T=0)$ and (b) the magnitude of the total momentum $\mathbf{q}$ of a Cooper pair determined by the minimum of the free energy. The solid line is the boundary between the uniform $s$-wave pairing state and the normal state, the dashed line separates the FFLO state and the uniform $s$-wave state, and the dashed-dotted line represents the boundary between the FFLO state and the normal state.}\label{fig:FFLO-gap}
\end{figure*}

In addition to uniform superconductors, where Cooper pairs carry zero total momentum, we also examine finite total momentum $s$-wave pairing states for comparison. 

In an $s$-wave pairing state where each Cooper pair has a total momentum $\mathbf{q}$, the Hamiltonian presented in Eq.~\eqref{hamilt} undergoes the following modification,
\begin{widetext}
\begin{equation}
H=\sum_{\mathbf{k},\alpha,\beta}c^{\dagger}_{\mathbf{k}+\frac{\mathbf{q}}{2}\alpha}H^{\alpha\beta}_0\left(\mathbf{k}+\frac{\mathbf{q}}{2}\right) c_{\mathbf{k}+\frac{\mathbf{q}}{2}\beta} +\frac{1}{2}\sum_{\mathbf{k},\alpha,\beta} \left[c^{\dagger}_{\mathbf{k} +\frac{\mathbf{q}}{2}\alpha}\Delta_{\mathbf{q}}^{\alpha\beta}(\mathbf{k})c_{-\mathbf{k}+\frac{\mathbf{q}}{2}\beta}+h.c.\right],
\end{equation}    
\end{widetext}
where $\alpha, \beta$ are spin indices, $\Delta_{\mathbf{q}}$ is the $\mathbf{q}$-dependent $s$-wave pairing function, and $H_0^{\alpha\beta}(\mathbf{k}+\frac{\mathbf{q}}{2})$ can be obtained by replacing the wave vector $\mathbf{k}$ in Eq.~\eqref{hamilnormal} by $\mathbf{k}+\frac{\mathbf{q}}{2}$. 
Accordingly, the equation for the gap transforms into
\begin{equation}
\Delta_{\mathbf{q}}=-V\sum_{\mathbf{k}}\langle c_{-\mathbf{k}+\frac{\mathbf{q}}{2}\downarrow}c_{\mathbf{k}+\frac{\mathbf{q}}{2}\uparrow} \rangle. 
\end{equation}
Then the free energy is given by
\begin{widetext}
\begin{equation}\label{eq:free_energy_fflo}
F_{s}=-\frac{1}{\beta}\sum_{\mathbf{k}}\left[\ln(1+\mathrm{e}^{-\beta E_{\mathbf{k} + \frac{\mathbf{q}}{2},+}}) +\ln(1+\mathrm{e}^{-\beta E_{\mathbf{k}+\frac{\mathbf{q}}{2},-}})\right] + \sum_{\mathbf{k}}\left[\xi_{-\mathbf{k}+\frac{\mathbf{q}}{2}}-\frac{1}{2}(E_{-\mathbf{k}+\frac{\mathbf{q}}{2},+}+E_{-\mathbf{k}+\frac{\mathbf{q}}{2},-})\right]+\frac{|\Delta_{\mathbf{q}}|^2}{V}.
\end{equation}    
\end{widetext}
On the other hand, we have
\begin{equation}
F_{n}(T=0)= \sum_{ \xi_{\mathbf{k}\pm}<0 } \xi_{\mathbf{k}+}+\xi_{\mathbf{k}-}.
\end{equation}

Thus, the condensation energy is determined as $E_c = F_n(T=0) - F_s(T=0)$. By numerically solving the self-consistent gap equation, one can evaluate both the free energy and the resulting condensation energy. The condensation energy $E_c$ is depicted in Fig.~\ref{Fig:free_ff} and Fig.~\ref{fig:FFLO-Ec} as a function of $q = |\mathbf{q}|$. An FFLO ground state emerges when $E_c$ maximizes at a non-zero $q$, as shown in both Fig.~\ref{Fig:free_ff} and Fig.~\ref{fig:FFLO-Ec}. It is evident that with sufficiently large Zeeman field and Rashba SOC, an FFLO ground state is realized, as demonstrated in Fig.~\ref{fig:swave_phasediagram}. Moreover, the corresponding gap function $\Delta(T=0)$ and the value of $q$ are presented in Fig.~\ref{fig:FFLO-gap}.

\section{Spin quantization axes and pairing functions}\label{app:spin-quantization-axis}

Define single-spin eigenstates, $\left|\uparrow_{\alpha}\right\rangle$ and $\left|\downarrow_{\alpha}\right\rangle$ by $\sigma_{\alpha}\left|\uparrow_{\alpha}\right\rangle=+\left|\uparrow_{\alpha}\right\rangle$ and $\sigma_{\alpha}\left|\downarrow_{\alpha}\right\rangle=-\left|\downarrow_{\alpha}\right\rangle$ respectively, where $\alpha=x,z$. Then the transformation of the set of spin axes and its inverse are
\begin{equation}
\lvert\uparrow_{x}\rangle = \frac{\lvert\uparrow_{z}\rangle+\lvert\downarrow_{z}\rangle}{\sqrt{2}}, \,\,\,
\lvert\downarrow_{x}\rangle = \frac{\lvert\uparrow_{z}\rangle-\lvert\downarrow_{z}\rangle}{\sqrt{2}},
\end{equation}
and 
\begin{equation}
\lvert\uparrow_{z}\rangle = \frac{\lvert\uparrow_{x}\rangle+\lvert\downarrow_{x}\rangle}{\sqrt{2}}, \,\,\,
\lvert\downarrow_{z}\rangle = \frac{\lvert\uparrow_{x}\rangle-\lvert\downarrow_{x}\rangle}{\sqrt{2}}.
\end{equation}
The spin-triplet components of a pair of $S=1/2$ spins can be written as follows
\begin{widetext}
\begin{equation}
\left|\uparrow_{z}\uparrow_{z}\right\rangle = \frac{\left|\uparrow_{x}\uparrow_{x}\right\rangle+\left|\downarrow_{x}\downarrow_{x}\right\rangle}{2}+\frac{\left|\uparrow_{x}\downarrow_{x}\right\rangle+\left|\downarrow_{x}\uparrow_{x}\right\rangle}{2}, \,\,\, 
\left|\downarrow_{z}\downarrow_{z}\right\rangle = \frac{\left|\uparrow_{x}\uparrow_{x}\right\rangle+\left|\downarrow_{x}\downarrow_{x}\right\rangle}{2}-\frac{\left|\uparrow_{x}\downarrow_{x}\right\rangle+\left|\downarrow_{x}\uparrow_{x}\right\rangle}{2}, \,\,\, 
\left|\uparrow_{z}\downarrow_{z}\right\rangle+\left|\downarrow_{z}\uparrow_{z}\right\rangle = \left|\uparrow_{x}\uparrow_{x}\right\rangle -\left|\downarrow_{x}\downarrow_{x}\right\rangle.
\end{equation}    
Thus, a generic spin-triplet pairing state can be represented in these two spin quantization axes as follows:
\begin{equation}
\Delta_{\uparrow\uparrow} \left|\uparrow_{z}\uparrow_{z}\right\rangle + \Delta_{\downarrow\downarrow} \left|\downarrow_{z}\downarrow_{z}\right\rangle + \Delta_{0}\left(\left|\uparrow_{z}\downarrow_{z}\right\rangle+\left|\downarrow_{z}\uparrow_{z}\right\rangle \right) = \frac{\Delta_{\uparrow\uparrow}+\Delta_{\downarrow\downarrow}+2\Delta_0}{2} \left|\uparrow_{x}\uparrow_{x}\right\rangle + \frac{\Delta_{\uparrow\uparrow}+\Delta_{\downarrow\downarrow}-2\Delta_0}{2} \left|\downarrow_{x}\downarrow_{x}\right\rangle + \frac{\Delta_{\uparrow\uparrow}-\Delta_{\downarrow\downarrow}}{2} \left(\left|\uparrow_{x}\downarrow_{x}\right\rangle +\left|\downarrow_{x}\uparrow_{x}\right\rangle \right).
\end{equation}
\end{widetext}

\section{Bogoliubov transformation: $p$-wave pairing states in a single Zeeman field}\label{app:Bog_pwave_mag}

Consider a Zeeman field exerted along the direction labeled $(\theta,\phi)$, i.e.
$\mathbf{H}=|\mathbf{H}|(\sin\theta\cos\phi, \sin\theta\sin\phi,\cos\theta)$,
the Hamiltonian of a $p$-wave pairing state can be written as follows,
\begin{widetext}
\begin{equation} \begin{aligned}
H_{\mathbf{k}}= \begin{pmatrix}
\xi_\mathbf{k}+\mu_B|\mathbf{H}|\cos\theta&\mu_B|\mathbf{H}|\sin\theta \mathrm{e}^{-i\phi}&[-d^\prime_x(\mathbf{k})+id^\prime_y(\mathbf{k})]\mathrm{e}^{i\eta_{\mathbf{k}}}&d^\prime_z(\mathbf{k})\mathrm{e}^{i\eta_{\mathbf{k}}}\\
\mu_B|\mathbf{H}|\sin\theta \mathrm{e}^{i\phi}&\xi_\mathbf{k}-\mu_B|\mathbf{H}|\cos\theta_\mathbf{k}&d^\prime_z(\mathbf{k})\mathrm{e}^{i\eta_{\mathbf{k}}}&[d^\prime_x(\mathbf{k})+id^\prime_y(\mathbf{k})]\mathrm{e}^{i\eta_\mathbf{k}}\\
[-d^\prime_x(\mathbf{k})-id^\prime_y(\mathbf{k})]\mathrm{e}^{-i\eta_{\mathbf{k}}}&d^\prime_z(\mathbf{k})\mathrm{e}^{-i\eta_{\mathbf{k}}}&-\xi_\mathbf{k}-\mu_B|\mathbf{H}|\cos\theta&-\mu_B|\mathbf{H}|\sin\theta_\mathbf{k}\mathrm{e}^{i\phi}\\
d^\prime_z(\mathbf{k})\mathrm{e}^{-i\eta_{\mathbf{k}}}&[d^\prime_x(\mathbf{k})-id^\prime_y(\mathbf{k})]\mathrm{e}^{-i\eta_{\mathbf{k}}}&-\mu_B|\mathbf{H}|\sin\theta \mathrm{e}^{-i\phi}&-\xi_\mathbf{k}+\mu_B|\mathbf{H}|\cos\theta
\end{pmatrix},
\end{aligned}\end{equation}
where $\eta_{\mathbf{k}}$ is the overall phase of $\mathbf{d}(\mathbf{k})$ and $d^\prime_{x,y,z}(\mathbf{k})$ are real numbers.
\begin{equation}
    \begin{split}
\mathbf{d}(\mathbf{k})&=\mathbf{d}^\prime(\mathbf{k})\mathrm{e}^{i\eta_\mathbf{k}},\\
d_x(\mathbf{k})&=d_x^\prime (\mathbf{k})\mathrm{e}^{i\eta_\mathbf{k}},\\
d_y(\mathbf{k})&=d_y^\prime (\mathbf{k})\mathrm{e}^{i\eta_\mathbf{k}},\\
d_z(\mathbf{k})&=d_z^\prime (\mathbf{k})\mathrm{e}^{i\eta_\mathbf{k}}.
\label{eq:dkprime}
    \end{split}
\end{equation}
Using the transformation
\begin{equation} \begin{aligned}
&C_{\mathbf{k}}=U_{1}C'_{\mathbf{k}}\\
=&\begin{pmatrix}
\cos{\frac{\theta}{2}}&\sin{\frac{\theta}{2}}&0&0\\
\sin{\frac{\theta}{2}}\mathrm{e}^{i\phi}&-\cos{\frac{\theta}{2}}\mathrm{e}^{i\phi}&0&0\\
0&0&\cos{\frac{\theta}{2}}&\sin{\frac{\theta}{2}}\\
0&0&\sin{\frac{\theta}{2}}\mathrm{e}^{-i\phi}&-\cos{\frac{\theta}{2}}\mathrm{e}^{-i\phi}
\end{pmatrix}
\begin{pmatrix}
c_{\mathbf{k}+}\\
c_{\mathbf{k}-}\\
c_{-\mathbf{k}+}^\dagger\\
c_{-\mathbf{k}-}^\dagger
\end{pmatrix},
\end{aligned}\end{equation}
we rewrite the Hamiltonian in the basis of $C'_\mathbf{k}=\{c_{\mathbf{k}+},c_{\mathbf{k}-}, c_{-\mathbf{k}+}^\dagger, c_{-\mathbf{k}-}^\dagger\}^{T}$ as follows,
\begin{equation} \begin{aligned}
H^{\prime}_\mathbf{k} =
\begin{pmatrix}
\xi_\mathbf{k}+\mu_B|\mathbf{H}|&0&[-d_1(\mathbf{k})+id_2(\mathbf{k})]\mathrm{e}^{i\gamma_{\mathbf{k}}}&d_3(\mathbf{k})\mathrm{e}^{i\gamma_{\mathbf{k}}}\\
0&\xi_\mathbf{k}-\mu_B|\mathbf{H}|&d_3(\mathbf{k})\mathrm{e}^{i\gamma_{\mathbf{k}}}&[d_1(\mathbf{k})+id_2(\mathbf{k})]\mathrm{e}^{i\gamma_{\mathbf{k}}}\\
[-d_1(\mathbf{k})-id_2(\mathbf{k})]\mathrm{e}^{-i\gamma_{\mathbf{k}}}&d_3(\mathbf{k})\mathrm{e}^{-i\gamma_{\mathbf{k}}}&-\xi_\mathbf{k}-\mu_B|\mathbf{H}|&0\\
d_3(\mathbf{k})\mathrm{e}^{-i\gamma_{\mathbf{k}}}&[d_1(\mathbf{k})-id_2(\mathbf{k})]\mathrm{e}^{-i\gamma_{\mathbf{k}}}&0&-\xi_\mathbf{k}+\mu_B|\mathbf{H}|
\end{pmatrix},
\end{aligned}\end{equation}
where $d_{1,2,3}(\mathbf{k})$ are three components of $\mathbf{d}(\mathbf{k})$ in the local coordinates defined as follows,
\begin{equation}
    \begin{split}
\gamma_{\mathbf{k}}&=\eta_\mathbf{k}-\phi+\pi,\\
d_\alpha(\mathbf{k})&= \mathbf{d}^\prime(\mathbf{k})\cdot \hat{n}_\alpha,\,\,\,\alpha=1,2,3 ,\\
\hat{n}_1&=(-\cos\theta\cos\phi,-\cos\theta\sin\phi,\sin\theta) ,\\
\hat{n}_2&=(\sin\phi,-\cos\phi,0)  ,\\
\hat{n}_3&=\hat{\mathbf{H}}=(\sin\theta\cos\phi,\sin\theta\sin\phi,\cos\theta) ,\\
d^{2}_1(\mathbf{k})+d^{2}_2(\mathbf{k})&=|\mathbf{d}(\mathbf{k})\times \hat{\mathbf{H}}|^2.
\end{split}
\end{equation}
So we can diagonalize the matrix $H^{\prime}_{\mathbf{k}}=U_{1}^\dagger H_\mathbf{k} U_{1}$ to obtain the eigenvalues and corresponding eigenvectors of the Hamiltonian, resulting in
\begin{equation}
\begin{split}
E_{\mathbf{k}\pm}&=\sqrt{\left[\sqrt{\xi_\mathbf{k}^2+d_3^2(\mathbf{k})}\pm \mu_B|\mathbf{H}|\right]^2+d_{1}^2(\mathbf{k})+d_{2}^2(\mathbf{k})} = \sqrt{\left[\sqrt{\xi_\mathbf{k}^2+|\mathbf{d}(\mathbf{k})\cdot \hat{\mathbf{H}}|^2}\pm \mu_B|\mathbf{H}|\right]^2+|\mathbf{d}(\mathbf{k})\times \hat{\mathbf{H}}|^2}\\
&=\sqrt{\left[E_{\mathbf{k}\parallel}\pm \mu_B|\mathbf{H}|\right]^2+|\mathbf{d}(\mathbf{k})\times \hat{\mathbf{H}}|^2},
\end{split}
\end{equation}
where 
\begin{equation} \begin{aligned}
E_{\mathbf{k}\parallel}&=\sqrt{\xi_\mathbf{k}^2+|\mathbf{d}(\mathbf{k})\cdot \hat{\mathbf{H}}|^2}.
\end{aligned}\end{equation} 
Note that $E_{-\mathbf{k}\parallel}=E_{\mathbf{k}\parallel}$ and $E_{-\mathbf{k}\pm}=E_{\mathbf{k}\pm}$.
The transformation matrix is given by
\begin{equation}
\begin{split}
&U_{\mathbf{k}}=U_1U_{2\mathbf{k}}\\
&=\begin{pmatrix}
\cos{\frac{\theta}{2}}&\sin{\frac{\theta}{2}}&0&0\\
\sin{\frac{\theta}{2}}\mathrm{e}^{i\phi}&-\cos{\frac{\theta}{2}}\mathrm{e}^{i\phi}&0&0\\
0&0&\cos{\frac{\theta}{2}}&\sin{\frac{\theta}{2}}\\
0&0&\sin{\frac{\theta}{2}}\mathrm{e}^{-i\phi}&-\cos{\frac{\theta}{2}}\mathrm{e}^{-i\phi}
\end{pmatrix}\cdot\frac{1}{2\sqrt{E_{\mathbf{k}\parallel}}}\\
&\begin{pmatrix}
s_{\mathbf{k}3} \mathrm{e}^{i\gamma_{\mathbf{k}}}\sqrt{\frac{(E_{\mathbf{k}+}+E_{\mathbf{k}\parallel}+\mu_B|\mathbf{H}|)(E_{\mathbf{k}\parallel}+\xi_{\mathbf{k}})}{E_{\mathbf{k}+}}}&-s_{\mathbf{k}3}\mathrm{e}^{i\gamma_\mathbf{k}}\sqrt{\frac{(E_{\mathbf{k}-}-E_{\mathbf{k}\parallel}+\mu_B|\mathbf{H}|)(E_{\mathbf{k}\parallel}-\xi_{\mathbf{k}})}{E_{\mathbf{k}-}}}&s_{\mathbf{k}3}\mathrm{e}^{i\gamma_\mathbf{k}}\sqrt{\frac{(E_{\mathbf{k}+}-E_{\mathbf{k}\parallel}-\mu_B|\mathbf{H}|)(E_{\mathbf{k}\parallel}+\xi_{\mathbf{k}})}{E_{\mathbf{k}+}}}&-s_{\mathbf{k}3}\mathrm{e}^{i\gamma_\mathbf{k}}\sqrt{\frac{(E_{\mathbf{k}-}+E_{\mathbf{k}\parallel}-\mu_B|\mathbf{H}|)(E_{\mathbf{k}\parallel}-\xi_{\mathbf{k}})}{E_{\mathbf{k}-}}}\\
\mathrm{e}^{i\gamma_\mathbf{k}}\mathrm{e}^{i\Omega_{\mathbf{k}}}\sqrt{\frac{(E_{\mathbf{k}+}-E_{\mathbf{k}\parallel}-\mu_B|\mathbf{H}|)(E_{\mathbf{k}\parallel}-\xi_{\mathbf{k}})}{E_{\mathbf{k}+}}}&\mathrm{e}^{i\gamma_{\mathbf{k}}}\mathrm{e}^{i\Omega_\mathbf{k}}\sqrt{\frac{(E_{\mathbf{k}-}+E_{\mathbf{k}\parallel}-\mu_B|\mathbf{H}|)(E_{\mathbf{k}\parallel}+\xi_{\mathbf{k}})}{E_{\mathbf{k}-}}}&-\mathrm{e}^{i\gamma_\mathbf{k}}\mathrm{e}^{i\Omega_\mathbf{k}}\sqrt{\frac{(E_{\mathbf{k}+}+E_{\mathbf{k}\parallel}+\mu_B|\mathbf{H}|)(E_{\mathbf{k}\parallel}-\xi_{\mathbf{k}})}{E_{\mathbf{k}+}}}&-\mathrm{e}^{i\gamma_\mathbf{k}}\mathrm{e}^{i\Omega_\mathbf{k}}\sqrt{\frac{(E_{\mathbf{k}-}-E_{\mathbf{k}\parallel}+\mu_B|\mathbf{H}|)(E_{\mathbf{k}\parallel}+\xi_{\mathbf{k}})}{E_{\mathbf{k}-}}}\\
-s_{\mathbf{k}3}\mathrm{e}^{i\Omega_{\mathbf{k}}}\sqrt{\frac{(E_{\mathbf{k}+}-E_{\mathbf{k}\parallel}-\mu_B|\mathbf{H}|)(E_{\mathbf{k}\parallel}+\xi_{\mathbf{k}})}{E_{\mathbf{k}+}}}&s_{\mathbf{k}3}\mathrm{e}^{i\Omega_\mathbf{k}}\sqrt{\frac{(E_{\mathbf{k}-}+E_{\mathbf{k}\parallel}-\mu_B|\mathbf{H}|)(E_{\mathbf{k}\parallel}-\xi_{\mathbf{k}})}{E_{\mathbf{k}-}}}&s_{\mathbf{k}3}\mathrm{e}^{i\Omega_\mathbf{k}}\sqrt{\frac{(E_{\mathbf{k}+}+E_{\mathbf{k}\parallel}+\mu_B|\mathbf{H}|)(E_{\mathbf{k}\parallel}+\xi_{\mathbf{k}})}{E_{\mathbf{k}+}}}&-s_{\mathbf{k}3}\mathrm{e}^{i\Omega_\mathbf{k}}\sqrt{\frac{(E_{\mathbf{k}-}-E_{\mathbf{k}\parallel}+\mu_B|\mathbf{H}|)(E_{\mathbf{k}\parallel}-\xi_{\mathbf{k}})}{E_{\mathbf{k}-}}}\\
\sqrt{\frac{(E_{\mathbf{k}+}+E_{\mathbf{k}\parallel}+\mu_B|\mathbf{H}|)(E_{\mathbf{k}\parallel}-\xi_{\mathbf{k}})}{E_{\mathbf{k}+}}}&\sqrt{\frac{(E_{\mathbf{k}-}-E_{\mathbf{k}\parallel}+\mu_B|\mathbf{H}|)(E_{\mathbf{k}\parallel}+\xi_{\mathbf{k}})}{E_{\mathbf{k}-}}}&\sqrt{\frac{(E_{\mathbf{k}+}-E_{\mathbf{k}\parallel}-\mu_B|\mathbf{H}|)(E_{\mathbf{k}\parallel}-\xi_{\mathbf{k}})}{E_{\mathbf{k}+}}}&\sqrt{\frac{(E_{\mathbf{k}-}+E_{\mathbf{k}\parallel}-\mu_B|\mathbf{H}|)(E_{\mathbf{k}\parallel}+\xi_{\mathbf{k}})}{E_{\mathbf{k}-}}}
\end{pmatrix},
\end{split}
\end{equation}
where we have
\begin{equation}
\begin{split}
d_{12}(\mathbf{k})&=|\mathbf{d}(\mathbf{k})\times \hat{\mathbf{H}}|,\\
d_{12}(\mathbf{k})\mathrm{e}^{i\Omega_\mathbf{k}}&= d_1(\mathbf{k})+id_2(\mathbf{k}),\\
s_{\mathbf{k}3}&=\mbox{sgn} [d_3(\mathbf{k})].
\end{split}
\end{equation}

Defining 
\begin{equation} \begin{aligned}
a_{\mathbf{k}1\pm}=\sqrt{\frac{E_{\mathbf{k}\pm}+E_{\mathbf{k}\parallel}\pm \mu_B|\mathbf{H}|}{E_{\mathbf{k}\pm}}},\qquad  
a_{\mathbf{k}2\pm}=\sqrt{\frac{E_{\mathbf{k}\pm}-E_{\mathbf{k}\parallel}\mp \mu_B|\mathbf{H}|}{E_{\mathbf{k}\pm}}},\qquad b_{\mathbf{k}\pm}=\sqrt{\frac{E_{\mathbf{k}\parallel}\pm\xi_{\mathbf{k}}}{E_{\mathbf{k}\parallel}}},
\end{aligned}\end{equation} 
we can write transformation matrix in terms of $u_{\mathbf{k}}$ and $v_{\mathbf{k}}$ as follows,
\begin{equation}
\begin{split}
u_{\mathbf{k}} &=\frac{\mathrm{e}^{i \gamma_{\mathbf{k}}}}{2}  \begin{pmatrix}
s_{\mathbf{k}3} a_{\mathbf{k}1+} b_{\mathbf{k}+} \cos\frac{\theta}{2} + \mathrm{e}^{i \Omega_{\mathbf{k}}} a_{\mathbf{k}2+} b_{\mathbf{k}-} \sin\frac{\theta}{2} & 
\mathrm{e}^{i \Omega_{\mathbf{k}}} a_{\mathbf{k}1-} b_{\mathbf{k}+} \sin\frac{\theta}{2} - s_{\mathbf{k}3} a_{\mathbf{k}2-} b_{\mathbf{k}-} \cos\frac{\theta}{2} \\ 
\mathrm{e}^{i \phi} \left( s_{\mathbf{k}3} a_{\mathbf{k}1+} b_{\mathbf{k}+} \sin\frac{\theta}{2} - \mathrm{e}^{i \Omega_{\mathbf{k}}} a_{\mathbf{k}2+} b_{\mathbf{k}-} \cos\frac{\theta}{2} \right) & 
-\mathrm{e}^{i \phi} \left( \mathrm{e}^{i \Omega_{\mathbf{k}}} a_{\mathbf{k}1-} b_{\mathbf{k}+} \cos\frac{\theta}{2} + s_{\mathbf{k}3} a_{\mathbf{k}2-} b_{\mathbf{k}-} \sin\frac{\theta}{2} \right)
\end{pmatrix},\\
v_{\mathbf{k}} &=\frac{1}{2}  \begin{pmatrix}
a_{\mathbf{k}1+} b_{\mathbf{k}-} \sin\frac{\theta}{2} - \mathrm{e}^{- i \Omega_{\mathbf{k}}} s_{\mathbf{k}3} a_{\mathbf{k}2+} b_{\mathbf{k}+} \cos\frac{\theta}{2} & 
\mathrm{e}^{- i \Omega_{\mathbf{k}}} s_{\mathbf{k}3} a_{\mathbf{k}1-} b_{\mathbf{k}-} \cos\frac{\theta}{2} + a_{\mathbf{k}2-} b_{\mathbf{k}+} \sin\frac{\theta}{2} \\ 
-\mathrm{e}^{i \phi} \left(  a_{\mathbf{k}1+} b_{\mathbf{k}-} \cos\frac{\theta}{2} + \mathrm{e}^{- i \Omega_{\mathbf{k}}}  s_{\mathbf{k}3} a_{\mathbf{k}2+} b_{\mathbf{k}+} \sin\frac{\theta}{2} \right) & 
\mathrm{e}^{i \phi} \left( \mathrm{e}^{- i \Omega_{\mathbf{k}}} s_{\mathbf{k}3} a_{\mathbf{k}1-} b_{\mathbf{k}-} \sin\frac{\theta}{2} - a_{\mathbf{k}2-} b_{\mathbf{k}+} \cos\frac{\theta}{2}  \right)
\end{pmatrix},\\
v_{-\mathbf{k}}^* &= \frac{1}{2} \begin{pmatrix}
a_{\mathbf{k}1+} b_{\mathbf{k}-} \sin\frac{\theta}{2} - \mathrm{e}^{i \Omega_{\mathbf{k}}} s_{\mathbf{k}3} a_{\mathbf{k}2+} b_{\mathbf{k}+} \cos\frac{\theta}{2} & 
\mathrm{e}^{i \Omega_{\mathbf{k}}} s_{\mathbf{k}3} a_{\mathbf{k}1-} b_{\mathbf{k}-} \cos\frac{\theta}{2} + a_{\mathbf{k}2-} b_{\mathbf{k}+} \sin\frac{\theta}{2} \\ 
-\mathrm{e}^{-i \phi} \left( a_{\mathbf{k}1+} b_{\mathbf{k}-} \cos\frac{\theta}{2} + \mathrm{e}^{i \Omega_{\mathbf{k}}} s_{\mathbf{k}3} a_{\mathbf{k}2+} b_{\mathbf{k}+} \sin\frac{\theta}{2} \right) & 
\mathrm{e}^{-i \phi} \left( \mathrm{e}^{i \Omega_{\mathbf{k}}} s_{\mathbf{k}3} a_{\mathbf{k}1-} b_{\mathbf{k}-} \sin\frac{\theta}{2}  -a_{\mathbf{k}2-} b_{\mathbf{k}+} \cos\frac{\theta}{2} \right)
\end{pmatrix},\\
u_{-\mathbf{k}}^* &=-\frac{\mathrm{e}^{-i\gamma_{\mathbf{k}}}}{2} \begin{pmatrix}
s_{\mathbf{k}3} a_{\mathbf{k}1+} b_{\mathbf{k}+}  \cos\frac{\theta}{2} + \mathrm{e}^{- i \Omega_{\mathbf{k}}} a_{\mathbf{k}2+} b_{\mathbf{k}-} \sin\frac{\theta}{2} & \mathrm{e}^{- i \Omega_{\mathbf{k}}} a_{\mathbf{k}1-} b_{\mathbf{k}+} \sin\frac{\theta}{2}
-  s_{\mathbf{k}3} a_{\mathbf{k}2-} b_{\mathbf{k}-}  \cos\frac{\theta}{2}  \\ 
\mathrm{e}^{-i \phi} \left(  s_{\mathbf{k}3} a_{\mathbf{k}1+} b_{\mathbf{k}+}  \sin\frac{\theta}{2} - 
\mathrm{e}^{- i \Omega_{\mathbf{k}}} a_{\mathbf{k}2+} b_{\mathbf{k}-} \cos\frac{\theta}{2}  \right) & 
- \mathrm{e}^{-i \phi} \left( \mathrm{e}^{- i \Omega_{\mathbf{k}}} a_{\mathbf{k}1-} b_{\mathbf{k}+} \cos\frac{\theta}{2} + s_{\mathbf{k}3} a_{\mathbf{k}2-} b_{\mathbf{k}-}  \sin\frac{\theta}{2} \right)
\end{pmatrix}.
    \end{split}
\end{equation}

\section{Bogoliubov transformation: $p$-wave pairing states with only Rashba SOC}\label{app:Bog_pwave}

In the presence of only Rashba SOC, the Hamiltonian is given by
\begin{equation} \begin{aligned}
H_{\mathbf{k}}= \begin{pmatrix}
\xi_\mathbf{k}+g|\mathbf{k}|\cos\theta_\mathbf{k}&g|\mathbf{k}|\sin\theta_\mathbf{k}\mathrm{e}^{-i\varphi_\mathbf{k}}&[-d^\prime_x(\mathbf{k})+id^\prime_y(\mathbf{k})]\mathrm{e}^{i\eta_{\mathbf{k}}}&d^\prime_z(\mathbf{k})\mathrm{e}^{i\eta_{\mathbf{k}}}\\
g|\mathbf{k}|\sin\theta_\mathbf{k}\mathrm{e}^{i\varphi_\mathbf{k}}&\xi_\mathbf{k}-g|\mathbf{k}|\cos\theta_\mathbf{k}&d^\prime_z(\mathbf{k})\mathrm{e}^{i\eta_{\mathbf{k}}}&[d^\prime_x(\mathbf{k})+id^\prime_y(\mathbf{k})]\mathrm{e}^{i\eta_\mathbf{k}}\\
[-d^\prime_x(\mathbf{k})-id^\prime_y(\mathbf{k})]\mathrm{e}^{-i\eta_{\mathbf{k}}}&d^\prime_z(\mathbf{k})\mathrm{e}^{-i\eta_{\mathbf{k}}}&-\xi_\mathbf{k}+g|\mathbf{k}|\cos\theta_\mathbf{k}&g|\mathbf{k}|\sin\theta_\mathbf{k}\mathrm{e}^{i\varphi_\mathbf{k}}\\
d^\prime_z(\mathbf{k})\mathrm{e}^{-i\eta_{\mathbf{k}}}&[d^\prime_x(\mathbf{k})-id^\prime_y(\mathbf{k})]\mathrm{e}^{-i\eta_{\mathbf{k}}}&g|\mathbf{k}|\sin\theta_\mathbf{k}\mathrm{e}^{-i\varphi_\mathbf{k}}&-\xi_\mathbf{k}-g|\mathbf{k}|\cos\theta_\mathbf{k}
\end{pmatrix},   
\end{aligned}\end{equation} 
where $\eta_{\mathbf{k}}$  is the overall phase of $\mathbf{d}(\mathbf{k})$ and $d^\prime_{x,y,z}(\mathbf{k})$ are real numbers as defined in Eq.\eqref{eq:dkprime}.  Utilizing the transformation
\begin{equation} \begin{aligned}
C_{\mathbf{k}}&=U_{1\mathbf{k}}C'_{\mathbf{k}}=\begin{pmatrix}
\cos{\frac{\theta_\mathbf{k}}{2}}&\sin{\frac{\theta_\mathbf{k}}{2}}&0&0\\
\sin{\frac{\theta_\mathbf{k}}{2}}\mathrm{e}^{i\varphi_\mathbf{k}}&-\cos{\frac{\theta_\mathbf{k}}{2}}\mathrm{e}^{i\varphi_\mathbf{k}}&0&0\\
0&0&\cos{\frac{\theta_\mathbf{k}}{2}}&\sin{\frac{\theta_\mathbf{k}}{2}}\\
0&0&\sin{\frac{\theta_\mathbf{k}}{2}}\mathrm{e}^{-i\varphi_\mathbf{k}}&-\cos{\frac{\theta_\mathbf{k}}{2}}\mathrm{e}^{-i\varphi_\mathbf{k}}
\end{pmatrix}
\begin{pmatrix}
c_{\mathbf{k}+}\\
c_{\mathbf{k}-}\\
c_{-\mathbf{k}-}^\dagger\\
c_{-\mathbf{k}+}^\dagger
\end{pmatrix},
\end{aligned}\end{equation} 
we can rewrite the Hamiltonian in the basis of $C'_\mathbf{k}=\{c_{\mathbf{k}+},c_{\mathbf{k}-}, c_{-\mathbf{k}-}^\dagger, c_{-\mathbf{k}+}^\dagger\}^{T}$,
\begin{equation} \begin{aligned}
H^{\prime}_\mathbf{k} =
\begin{pmatrix}
\xi_\mathbf{k}+g|\mathbf{k}|&0&[-d_1(\mathbf{k})+id_2(\mathbf{k})]\mathrm{e}^{i\gamma_{\mathbf{k}}}&d_3(\mathbf{k})\mathrm{e}^{i\gamma_{\mathbf{k}}}\\
0&\xi_\mathbf{k}-g|\mathbf{k}|&d_3(\mathbf{k})\mathrm{e}^{i\gamma_{\mathbf{k}}}&[d_1(\mathbf{k})+id_2(\mathbf{k})]\mathrm{e}^{i\gamma_{\mathbf{k}}}\\
[-d_1(\mathbf{k})-id_2(\mathbf{k})]\mathrm{e}^{-i\gamma_{\mathbf{k}}}&d_3(\mathbf{k})\mathrm{e}^{-i\gamma_{\mathbf{k}}}&-\xi_\mathbf{k}+g|\mathbf{k}|&0\\
d_3(\mathbf{k})\mathrm{e}^{-i\gamma_{\mathbf{k}}}&[d_1(\mathbf{k})-id_2(\mathbf{k})]\mathrm{e}^{-i\gamma_{\mathbf{k}}}&0&-\xi_\mathbf{k}-g|\mathbf{k}|
\end{pmatrix},
\end{aligned}\end{equation} 
where $d_{1,2,3}(\mathbf{k})$ are three components of $\mathbf{d}(\mathbf{k})$ in the local coordinates defined as follows,
\begin{equation} \begin{aligned}
\label{eq:d123k}
\gamma_{\mathbf{k}}&=\eta_{\mathbf{k}}-\varphi_\mathbf{k}+\pi ,\\
d_\alpha(\mathbf{k})&= \mathbf{d}^\prime(\mathbf{k})\cdot \hat{n}_\alpha(\mathbf{k}),\,\,\,\alpha=1,2,3 ,\\
\hat{n}_1(\mathbf{k})&=(-\cos\theta_\mathbf{k}\cos\varphi_\mathbf{k},-\cos\theta_\mathbf{k}\sin\varphi_\mathbf{k},\sin\theta_\mathbf{k}), \\
\hat{n}_2(\mathbf{k})&=(\sin\varphi_\mathbf{k},-\cos\varphi_\mathbf{k},0),  \\
\hat{n}_3(\mathbf{k})&=\hat{\mathbf{k}}=(\sin\theta_\mathbf{k}\cos\varphi_\mathbf{k},\sin\theta_\mathbf{k}\sin\varphi_\mathbf{k},\cos\theta_\mathbf{k}), \\
d^{2}_1(\mathbf{k})+d^{2}_2(\mathbf{k})&=|\mathbf{d}(\mathbf{k})\times \hat{\mathbf{k}}|^2.
\end{aligned}\end{equation} 

So that, we can diagonalize the matrix $H^{\prime}_{\mathbf{k}}=U_{1\mathbf{k}}^\dagger H_\mathbf{k} U_{1\mathbf{k}}$ to obtain the eigvenvalues and corresponding eigenvectors of the Hamiltonian, resulting in
\begin{equation} \begin{aligned}
E_{\mathbf{k}\pm}&=\sqrt{\left[\sqrt{\xi_{\mathbf{k}}^2+d_1^{2}(\mathbf{k})+d_2^{2}(\mathbf{k})}\pm g|\mathbf{k}|\right]^2+d_3^{2}(\mathbf{k})}=\sqrt{\left[\sqrt{\xi_{\mathbf{k}}^2+|\mathbf{d}(\mathbf{k})\times \hat{\mathbf{k}}|^2}\pm g|\mathbf{k}|\right]^2+|\mathbf{d}(\mathbf{k})\cdot \hat{\mathbf{k}}|^2} \\
&=\sqrt{\left[E_{\mathbf{k}\perp}\pm g|\mathbf{k}|\right]^2+|\mathbf{d}(\mathbf{k})\cdot \hat{\mathbf{k}}|^2},
\end{aligned}\end{equation} 
where
\begin{equation} \begin{aligned}
E_{\mathbf{k}\perp}&= \sqrt{\xi_{\mathbf{k}}^2+|\mathbf{d}(\mathbf{k})\times \hat{\mathbf{k}}|^2}.
\end{aligned}\end{equation} 
Note that $E_{-\mathbf{k}\perp}=E_{\mathbf{k}\perp}$ and $E_{-\mathbf{k}\pm}=E_{\mathbf{k}\pm}$.
The transformation matrix is given by
\begin{equation} \begin{aligned}
&U_{\mathbf{k}}=U_{1\mathbf{k}}U_{2\mathbf{k}} \\
&=\begin{pmatrix}
\cos{\frac{\theta_\mathbf{k}}{2}}&\sin{\frac{\theta_\mathbf{k}}{2}}&0&0\\
\sin{\frac{\theta_\mathbf{k}}{2}}\mathrm{e}^{i\varphi_\mathbf{k}}&-\cos{\frac{\theta_\mathbf{k}}{2}}\mathrm{e}^{i\varphi_\mathbf{k}}&0&0\\
0&0&\cos{\frac{\theta_\mathbf{k}}{2}}&\sin{\frac{\theta_\mathbf{k}}{2}}\\
0&0&\sin{\frac{\theta_\mathbf{k}}{2}}\mathrm{e}^{-i\varphi_\mathbf{k}}&-\cos{\frac{\theta_\mathbf{k}}{2}}\mathrm{e}^{-i\varphi_\mathbf{k}}
\end{pmatrix}\cdot\frac{1}{2\sqrt{E_{\mathbf{k}\perp}}} \\
&\begin{pmatrix}
\mathrm{e}^{i\gamma_{\mathbf{k}}}s_{\mathbf{k}3}\sqrt{\frac{(E_{\mathbf{k}+}+E_{\mathbf{k}\perp}+g|\mathbf{k}|)(E_{\mathbf{k}\perp}+\xi_{\mathbf{k}})}{E_{\mathbf{k}+}}}&\mathrm{e}^{i\gamma_{\mathbf{k}}}s_{\mathbf{k}3}\sqrt{\frac{(E_{\mathbf{k}-}-E_{\mathbf{k}\perp}+g|\mathbf{k}|)(E_{\mathbf{k}\perp}-\xi_{\mathbf{k}})}{E_{\mathbf{k}-}}}&-\mathrm{e}^{i\gamma_{\mathbf{k}}}s_{\mathbf{k}3}\sqrt{\frac{(E_{\mathbf{k}+}-E_{\mathbf{k}\perp}-g|\mathbf{k}|)(E_{\mathbf{k}\perp}+\xi_{\mathbf{k}})}{E_{\mathbf{k}+}}}&-\mathrm{e}^{i\gamma_{\mathbf{k}}}s_{\mathbf{k}3}\sqrt{\frac{(E_{\mathbf{k}-}+E_{\mathbf{k}\perp}-g|\mathbf{k}|)(E_{\mathbf{k}\perp}-\xi_{\mathbf{k}})}{E_{\mathbf{k}-}}}\\
-\mathrm{e}^{i\gamma_{\mathbf{k}}}\mathrm{e}^{i\Omega_{\mathbf{k}}}\sqrt{\frac{(E_{\mathbf{k}+}-E_{\mathbf{k}\perp}-g|\mathbf{k}|)(E_{\mathbf{k}\perp}-\xi_{\mathbf{k}})}{E_{\mathbf{k}+}}}&\mathrm{e}^{i\gamma_{\mathbf{k}}}\mathrm{e}^{i\Omega_{\mathbf{k}}}\sqrt{\frac{(E_{\mathbf{k}-}+E_{\mathbf{k}\perp}-g|\mathbf{k}|)(E_{\mathbf{k}\perp}+\xi_{\mathbf{k}})}{E_{\mathbf{k}-}}}&-\mathrm{e}^{i\gamma_{\mathbf{k}}}\mathrm{e}^{i\Omega_{\mathbf{k}}}\sqrt{\frac{(E_{\mathbf{k}+}+E_{\mathbf{k}\perp}+g|\mathbf{k}|)(E_{\mathbf{k}\perp}-\xi_{\mathbf{k}})}{E_{\mathbf{k}+}}}&\mathrm{e}^{i\gamma_{\mathbf{k}}}\mathrm{e}^{i\Omega_{\mathbf{k}}}\sqrt{\frac{(E_{\mathbf{k}-}-E_{\mathbf{k}\perp}+g|\mathbf{k}|)(E_{\mathbf{k}\perp}+\xi_{\mathbf{k}})}{E_{\mathbf{k}-}}}\\
-\mathrm{e}^{i\Omega_{\mathbf{k}}}s_{\mathbf{k}3}\sqrt{\frac{(E_{\mathbf{k}+}+E_{\mathbf{k}\perp}+g|\mathbf{k}|)(E_{\mathbf{k}\perp}-\xi_{\mathbf{k}})}{E_{\mathbf{k}+}}}&\mathrm{e}^{i\Omega_{\mathbf{k}}}s_{\mathbf{k}3}\sqrt{\frac{(E_{\mathbf{k}-}-E_{\mathbf{k}\perp}+g|\mathbf{k}|)(E_{\mathbf{k}\perp}+\xi_{\mathbf{k}})}{E_{\mathbf{k}-}}}&\mathrm{e}^{i\Omega_{\mathbf{k}}}s_{\mathbf{k}3}\sqrt{\frac{(E_{\mathbf{k}+}-E_{\mathbf{k}\perp}-g|\mathbf{k}|)(E_{\mathbf{k}\perp}-\xi_{\mathbf{k}})}{E_{\mathbf{k}+}}}&-\mathrm{e}^{i\Omega_{\mathbf{k}}}s_{\mathbf{k}3}\sqrt{\frac{(E_{\mathbf{k}-}+E_{\mathbf{k}\perp}-g|\mathbf{k}|)(E_{\mathbf{k}\perp}+\xi_{\mathbf{k}})}{E_{\mathbf{k}-}}}\\
\sqrt{\frac{(E_{\mathbf{k}+}-E_{\mathbf{k}\perp}-g|\mathbf{k}|)(E_{\mathbf{k}\perp}+\xi_{\mathbf{k}})}{E_{\mathbf{k}+}}}&\sqrt{\frac{(E_{\mathbf{k}-}+E_{\mathbf{k}\perp}-g|\mathbf{k}|)(E_{\mathbf{k}\perp}-\xi_{\mathbf{k}})}{E_{\mathbf{k}-}}}&\sqrt{\frac{(E_{\mathbf{k}+}+E_{\mathbf{k}\perp}+g|\mathbf{k}|)(E_{\mathbf{k}\perp}+\xi_{\mathbf{k}})}{E_{\mathbf{k}+}}}&\sqrt{\frac{(E_{\mathbf{k}-}-E_{\mathbf{k}\perp}+g|\mathbf{k}|)(E_{\mathbf{k}\perp}-\xi_{\mathbf{k}})}{E_{\mathbf{k}-}}}
\end{pmatrix},
\label{eq:Bog_pwavesoc}
\end{aligned}\end{equation} 
where we have
\begin{equation} \begin{aligned}
d_{12}(\mathbf{k})&=|\mathbf{d}(\mathbf{k})\times \hat{\mathbf{k}}|, \\
d_{12}(\mathbf{k})\mathrm{e}^{i\Omega_{\mathbf{k}}}&= d_1(\mathbf{k})+id_2(\mathbf{k}), \\
s_{\mathbf{k}3}&= \mbox{sgn}[d_3(\mathbf{k})].
\end{aligned}\end{equation} 

Defining 
\begin{equation} \begin{aligned}
a_{\mathbf{k}1\pm}=\sqrt{\frac{E_{\mathbf{k}\pm}+E_{\mathbf{k}\perp}\pm g|\mathbf{k}|}{E_{\mathbf{k}\pm}}},\qquad  
a_{\mathbf{k}2\pm}=\sqrt{\frac{E_{\mathbf{k}\pm}-E_{\mathbf{k}\perp}\mp g|\mathbf{k}|}{E_{\mathbf{k}\pm}}},\qquad b_{\mathbf{k}\pm}=\sqrt{\frac{E_{\mathbf{k}\perp}\pm\xi_{\mathbf{k}}}{E_{\mathbf{k}\perp}}},
\end{aligned}\end{equation} 
we can write transformation matrix in terms of $u_{\mathbf{k}}$ and $v_{\mathbf{k}}$ as follows,
\begin{equation} \begin{aligned}
u_{\mathbf{k}}
&=\frac{\mathrm{e}^{i\gamma_{\mathbf{k}}}}{2}\begin{pmatrix}
\cos\frac{\theta_{\mathbf{k}}}{2}s_{\mathbf{k}3}a_{\mathbf{k}1+}b_{\mathbf{k}+}-\mathrm{e}^{i\Omega_{\mathbf{k}}}\sin\frac{\theta_{\mathbf{k}}}{2}a_{\mathbf{k}2+}b_{\mathbf{k}-}&\cos\frac{\theta_{\mathbf{k}}}{2}s_{\mathbf{k}3}a_{\mathbf{k}2-}b_{\mathbf{k}-}+\mathrm{e}^{i\Omega_{\mathbf{k}}}\sin\frac{\theta_{\mathbf{k}}}{2}a_{\mathbf{k}1-}b_{\mathbf{k}+}\\
\mathrm{e}^{i\varphi_\mathbf{k}}(\sin\frac{\theta_{\mathbf{k}}}{2}s_{\mathbf{k}3}a_{\mathbf{k}1+}b_{\mathbf{k}+}+\mathrm{e}^{i\Omega_{\mathbf{k}}}\cos\frac{\theta_{\mathbf{k}}}{2}a_{\mathbf{k}2+}b_{\mathbf{k}-})&\mathrm{e}^{i\varphi_\mathbf{k}}(\sin\frac{\theta_{\mathbf{k}}}{2}s_{\mathbf{k}3}a_{\mathbf{k}2-}b_{\mathbf{k}-}-\mathrm{e}^{i\Omega_{\mathbf{k}}}\cos\frac{\theta_{\mathbf{k}}}{2}a_{\mathbf{k}1-}b_{\mathbf{k}+})
\end{pmatrix}, \\
v_{\mathbf{k}}&=\frac{ 1 }{2}\begin{pmatrix}
\cos{\frac{\theta_{\mathbf{k}}}{2}}a_{\mathbf{k}2+}b_{\mathbf{k}+}+\mathrm{e}^{i\Omega_{\mathbf{k}}}\sin{\frac{\theta_{\mathbf{k}}}{2}}s_{\mathbf{k}3}a_{\mathbf{k}1+}b_{\mathbf{k}-}&\cos{\frac{\theta_{\mathbf{k}}}{2}}a_{\mathbf{k}1-}b_{\mathbf{k}-}- 
 \mathrm{e}^{i\Omega_{\mathbf{k}}}\sin{\frac{\theta_{\mathbf{k}}}{2}}s_{\mathbf{k}3}a_{\mathbf{k}2-}b_{\mathbf{k}+}\\
\mathrm{e}^{i\varphi_\mathbf{k}}(\sin{\frac{\theta_{\mathbf{k}}}{2}}a_{\mathbf{k}2+}b_{\mathbf{k}+}-\mathrm{e}^{i\Omega_{\mathbf{k}}}\cos{\frac{\theta_{\mathbf{k}}}{2}}s_{\mathbf{k}3}a_{\mathbf{k}1+}b_{\mathbf{k}-})&\mathrm{e}^{i\varphi_\mathbf{k}}( \sin{\frac{\theta_{\mathbf{k}}}{2}}a_{\mathbf{k}1-}b_{\mathbf{k}-}+ \mathrm{e}^{i\Omega_{\mathbf{k}}}\cos{\frac{\theta_{\mathbf{k}}}{2}}s_{\mathbf{k}3}a_{\mathbf{k}2-}b_{\mathbf{k}+})
\end{pmatrix},
\end{aligned}\end{equation} 
and
\begin{equation} \begin{aligned}
v_{-\mathbf{k}}^*&=\frac{1}{2}\begin{pmatrix}
-\mathrm{e}^{i\Omega_{\mathbf{k}}}\cos{\frac{\theta_{\mathbf{k}}}{2}}s_{\mathbf{k}3}a_{\mathbf{k}1+}b_{\mathbf{k}-}+\sin{\frac{\theta_{\mathbf{k}}}{2}}a_{\mathbf{k}2+}b_{\mathbf{k}+}&\mathrm{e}^{i\Omega_{\mathbf{k}}}\cos{\frac{\theta_{\mathbf{k}}}{2}}s_{\mathbf{k}3}a_{\mathbf{k}2-}b_{\mathbf{k}+}+\sin{\frac{\theta_{\mathbf{k}}}{2}}a_{\mathbf{k}1-}b_{\mathbf{k}-}\\
\mathrm{e}^{-i\varphi_\mathbf{k}}(-\mathrm{e}^{i\Omega_{\mathbf{k}}}\sin{\frac{\theta_{\mathbf{k}}}{2}}s_{\mathbf{k}3}a_{\mathbf{k}1+}b_{\mathbf{k}-}-\cos{\frac{\theta_{\mathbf{k}}}{2}}a_{\mathbf{k}2+}b_{\mathbf{k}+})&\mathrm{e}^{-i\varphi_\mathbf{k}}(\mathrm{e}^{i\Omega_{\mathbf{k}}}\sin{\frac{\theta_{\mathbf{k}}}{2}}s_{\mathbf{k}3}a_{\mathbf{k}2-}b_{\mathbf{k}+}-\cos{\frac{\theta_{\mathbf{k}}}{2}}a_{\mathbf{k}1-}b_{\mathbf{k}-})
\end{pmatrix}, \\
u_{-\mathbf{k}}^*&=\frac{ \mathrm{e}^{-i\gamma_{\mathbf{k}}} }{2}\begin{pmatrix}
-\mathrm{e}^{i\Omega_{\mathbf{k}}}\cos{\frac{\theta_{\mathbf{k}}}{2}}a_{\mathbf{k}2+}b_{\mathbf{k}-}-\sin{\frac{\theta_{\mathbf{k}}}{2}}s_{\mathbf{k}3}a_{\mathbf{k}1+}b_{\mathbf{k}+}&\mathrm{e}^{i\Omega_{\mathbf{k}}}\cos{\frac{\theta_{\mathbf{k}}}{2}}a_{\mathbf{k}1-}b_{\mathbf{k}+}-\sin{\frac{\theta_{\mathbf{k}}}{2}}s_{\mathbf{k}3}a_{\mathbf{k}2-}b_{\mathbf{k}-}\\
\mathrm{e}^{-i\varphi_\mathbf{k}}(-\mathrm{e}^{i\Omega_{\mathbf{k}}}\sin{\frac{\theta_{\mathbf{k}}}{2}}a_{\mathbf{k}2+}b_{\mathbf{k}-}+\cos{\frac{\theta_{\mathbf{k}}}{2}}s_{\mathbf{k}3}a_{\mathbf{k}1+}b_{\mathbf{k}+})&\mathrm{e}^{-i\varphi_\mathbf{k}}(\mathrm{e}^{i\Omega_{\mathbf{k}}}\sin{\frac{\theta_{\mathbf{k}}}{2}}a_{\mathbf{k}1-}b_{\mathbf{k}+}+\cos{\frac{\theta_{\mathbf{k}}}{2}}s_{\mathbf{k}3}a_{\mathbf{k}2-}b_{\mathbf{k}-})
\end{pmatrix}.
\end{aligned}\end{equation} 
\end{widetext}

\section{Continuous phase transitions in $p$-wave pairing states: the superconducting $T_c$}
\label{app:tc_2nd}
In this appendix, we will explore potential continuous superconducting phase transitions for each $p$-wave pairing state presented in Table~\ref{pwave} and identify the corresponding $T_c$. To keep things simple, we will examine either the Zeeman field or Rashba SOC individually and concentrate on the continuous phase transition where the superconducting gap function $\Delta(T)$ approaches zero at $T=T_c-0^{+}$. For these straightforward scenarios, analytical solutions are attainable.

\subsection{ESP states: The effects of a sole perpendicular Zeeman field}

First, we consider only the effect of the perpendicular Zeeman field. We begin, without loss of generality, with the $k_x\hat{x}+k_y\hat{y}$ state and the Zeeman field aligned along the $\hat{x}$ axis. Here, the nonzero components of the pairing function are $\Delta_{\uparrow\uparrow}(\mathbf{k})=-\Delta_{\downarrow\downarrow}(\mathbf{k})^{\ast}$. Thus, the self-consistent gap equation is given by
\begin{widetext}
\begin{equation}
\begin{aligned}
-\sqrt{\frac{8\pi}{3}}\Delta Y_{1,-1}(\hat{\mathbf{k}})&=-\sum_{\mathbf{k}'}4\pi V_1 \sum_{m=0,\pm1}Y_{1,m}(\hat{\mathbf{k}})Y^*_{1,m}(\hat{\mathbf{k}}')\{(U_{\mathbf{k}'})_{31}^*(U_{\mathbf{k}'})_{11}f(E_{\mathbf{k}'+})+(U_{\mathbf{k}'})_{32}^*(U_{\mathbf{k}'})_{12}f(E_{\mathbf{k}'-})\\
&\quad+(U_{\mathbf{k}'})_{33}^*(U_{\mathbf{k}'})_{13}[1-f(E_{-\mathbf{k}'+})]+(U_{\mathbf{k}'})_{34}^*(U_{\mathbf{k}'})_{14}[1-f(E_{-\mathbf{k}'-})]\},
\end{aligned}
\end{equation}
which can be simplified as 
\begin{equation}
\begin{aligned}
\Delta=\sum_{\mathbf{k}}\frac{3}{2}V_1\sin\theta_{\mathbf{k}}\mathrm{e}^{i\varphi_{\mathbf{k}}}\{(U_\mathbf{k})_{31}^*(U_\mathbf{k})_{11}f(E_{\mathbf{k}+})+(U_\mathbf{k})_{32}^*(U_\mathbf{k})_{12}f(E_{\mathbf{k}-})+(U_\mathbf{k})_{33}^*(U_\mathbf{k})_{13}[1-f(E_{-\mathbf{k}+})]+(U_\mathbf{k})_{34}^*(U_\mathbf{k})_{14}[1-f(E_{-\mathbf{k}-})]\}.
\end{aligned}
\end{equation}
Substituting the Bogoliubov transformation matrices given in Appendix \ref{app:Bog_pwave_mag} into the above equation, we have
\begin{equation}
\begin{aligned}
\Delta=&\sum_{\mathbf{k}}\frac{3V_1}{2}\sin\theta_\mathbf{k}\mathrm{e}^{i\varphi_{\mathbf{k}}}\cdot\frac{1}{4}\sum_{ s=\pm }\left[\frac{d_3(E_{\mathbf{k}\parallel}+s \mu_BH_x)}{E_{\mathbf{k}s }E_{\mathbf{k}\parallel}}+\frac{id_2}{ E_{\mathbf{k}s }}\right]\tanh{\frac{\beta E_{\mathbf{k}s}}{2}}\\
=&\sum_{ s=\pm }\frac{3N(0)V_1}{4}\int_0^{\hbar\omega_D}d\xi_{\mathbf{k}}\int\frac{d\Omega_\mathbf{k}}{4\pi}\sin\theta_\mathbf{k}\mathrm{e}^{i\varphi_{\mathbf{k}}}\cdot\left[\frac{d_3(E_{\mathbf{k}\parallel}+s  \mu_BH_x)}{E_{\mathbf{k}s }E_{\mathbf{k}\parallel}}+\frac{id_2}{E_{\mathbf{k}s}}\right]\tanh{\frac{\beta E_{\mathbf{k}s}}{2}}.
\end{aligned}
\end{equation}
It is easy to verify that the self-consistent gap equations for the other three ESP states take the same form as above.

To obtain the superconducting $T_c$ of a continuous phase transition, we take the limit $\Delta\to{}0$ and put $E_{\mathbf{k}\parallel}$ and $E_{\mathbf{k}\pm}$, and $d_2$ and $d_3$ given in Appendix~\ref{app:Bog_pwave_mag} into the above gap equation. The resulting $T_c$ equation is the following
\begin{equation}
\begin{aligned}
1&=\sum_{s=\pm}\frac{3N(0)V_1}{4}\int_0^{\hbar\omega_D}d\xi_{\mathbf{k}}\int\frac{d\Omega_\mathbf{k}}{4\pi}\left[\frac{\sin^2\theta_{\mathbf{k}}\cos^2\varphi_{\mathbf{k}}}{\xi_{\mathbf{k}}}+\frac{\sin^2\theta_{\mathbf{k}}\sin^2\varphi_{\mathbf{k}}}{\xi_{\mathbf{k}}+s \mu_BH_x}\right]\tanh{\frac{\beta_c (\xi_{\mathbf{k}}+s\mu_BH_x)}{2}}\\
&=N(0)V_1\ln\left[\left(\frac{2e^\gamma}{\pi}\right)\left(\frac{\hbar\omega_D}{k_BT_c}\right)\right]+\frac{N(0)V_1}{4}\int_0^{\hbar\omega_D}d\xi_{\mathbf{k}}\frac{\mu_BH_x}{\xi_{\mathbf{k}}}\left[\frac{\tanh{\frac{\beta_c (\xi_{\mathbf{k}}+\mu_BH_x)}{2}}}{\xi_{\mathbf{k}}+ \mu_BH_x}-\frac{\tanh{\frac{\beta_c (\xi_{\mathbf{k}}-\mu_BH_x)}{2}}}{\xi_{\mathbf{k}}- \mu_BH_x}\right].\\
\end{aligned}
\end{equation}
Here $\gamma$ is the Euler constant, and the second term in the last line exists only when $H_{x}\neq{}0$. Using the integral
\begin{equation}
\int_0^{\hbar\omega_D}d\xi_{\mathbf{k}}\frac{\mu_BH_x}{\xi_{\mathbf{k}}}\left[\frac{\tanh{\frac{\beta_c (\xi_{\mathbf{k}}+\mu_BH_x)}{2}}}{\xi_{\mathbf{k}}+ \mu_BH_x}-\frac{\tanh{\frac{\beta_c (\xi_{\mathbf{k}}-\mu_BH_x)}{2}}}{\xi_{\mathbf{k}}- \mu_BH_x}\right] = 
2\left[\psi\left(\frac{1}{2}\right)-\mbox{Re}\psi\left(\frac{1}{2}+\frac{i\mu_BH_x}{2\pi k_BT_{c}}\right)\right],
\end{equation}
\end{widetext}
and comparing the $T_c$ solution at $H_{x}=0$:
\begin{equation}
N(0)V_1\ln\left[\left(\frac{2e^\gamma}{\pi}\right)\left(\frac{\hbar\omega_D}{k_BT_{c0}}\right)\right]=1
\end{equation}
we obtain the analytical $T_c$ equation as follows,
\begin{equation*}
\begin{aligned}
\ln\left(\frac{T_c}{T_{c0}}\right)&=\frac{1}{2}\left[\psi\left(\frac{1}{2}\right)-\mbox{Re}\,\psi\left(\frac{1}{2}+\frac{i\mu_BH_x}{2\pi k_BT_{c}}\right)\right].
\end{aligned}
\end{equation*}
It is nothing but Eq.~\eqref{eq:Tc-ESP-g0} in the main text.

When $\mu_BH_x\gg{}2\pi k_BT_{c}$, using
\begin{align*}
\mbox{Re}\psi\left(\frac{1}{2}+\frac{i\mu_BH_x}{2\pi k_BT_{c}}\right)&\approx \ln\frac{\mu_BH_x}{2\pi k_BT_{c}},
\end{align*}
we have
\begin{equation}
\begin{aligned}
\ln\left(\frac{T_c}{T_{c0}}\right)&\approx \frac{1}{2}\left[-2\ln2-\gamma -\ln\frac{\mu_BH_x}{2\pi k_BT_{c}}\right].
\end{aligned}
\end{equation}
This yields an analytical solution as follows:
\begin{equation*}
\begin{aligned}
\frac{T_c}{T_{c0}}&=\frac{\pi \mathrm{e}^{-\gamma}}{2}\cdot \frac{ k_BT_{c0}}{\mu_BH_x},
\end{aligned}
\end{equation*}
or
\begin{equation*}
\begin{aligned}
\frac{T_c}{T_{c0}}&=\sqrt{3}\mathrm{e}^{-\frac{5}{6}}\cdot \frac{ H_0}{H_x},
\end{aligned}
\end{equation*}
where the solutions $$k_BT_{c0}=\hbar\omega_D\left( \frac{2\mathrm{e}^{\gamma}}{\pi } \right) \mathrm{e}^{-\frac{1}{N(0)V_1}}$$ and       $$\Delta_0=\hbar\omega_D{}\mathrm{e}^{\frac{5}{6}} \mathrm{e}^{-\frac{1}{N(0)V_1}},$$ and the relation $\mu_BH_0=\Delta_0/\sqrt{3}\mu_B$ (see Table~\ref{tab:notation}) have been used.

\subsection{ESP states: The effects of only Rashba SOC}
Then we consider the four ESP states with only Rashba SOC. Substituting the Bogoliubov transformation matrices given in Appendix \ref{app:Bog_pwave} into the gap equation, we have
\begin{widetext}
\begin{equation}
\begin{aligned}
&(U_\mathbf{k})_{31}^*(U_\mathbf{k})_{11}f(E_{\mathbf{k}+})+(U_\mathbf{k})_{32}^*(U_\mathbf{k})_{12}f(E_{\mathbf{k}-})+(U_\mathbf{k})_{33}^*(U_\mathbf{k})_{13}[1-f(E_{-\mathbf{k}+})]+(U_\mathbf{k})_{34}^*(U_\mathbf{k})_{14}[1-f(E_{-\mathbf{k}-})]\\
=&\sum_{s=\pm}\frac{\mathrm{e}^{i\gamma_\mathbf{k}}}{4}\left[\frac{d_3\sin\theta_{\mathbf{k}}}{E_{\mathbf{k}s}}-\frac{d_1\cos\theta_{\mathbf{k}}}{E_{\mathbf{k}\perp}}\cdot\frac{E_{\mathbf{k}\perp}+s gk_F}{E_{\mathbf{k}s}}+\frac{id_2}{E_{\mathbf{k}\perp}}\cdot\frac{E_{\mathbf{k}\perp}+s  gk_F}{E_{\mathbf{k}s}}\right](-1)\tanh\frac{\beta E_{\mathbf{k}s}}{2}.
\end{aligned}
\end{equation}
\end{widetext}
Thus, the analytical $T_c$ equations can be found following the same steps as in the previous subsection, resulting in the equation $T_c$ given in Eq.~\eqref{eq:Tc-p-H0}, where $\alpha=1/5$ for the pairing state $k_x\hat{x}+k_y\hat{y}$, $\alpha=1$ for the pairing state $k_y\hat{x}-k_x\hat{y}$ and $\alpha=3/5$ for the pairing states $k_x\hat{x}-k_y\hat{y}$ and $k_y\hat{x}+k_x\hat{y}$.

(1) For the $k_x\hat{x}+k_y\hat{y}$ pairing state, we have
\begin{equation}
\begin{aligned}
\ln\left(\frac{T_c}{T_{c0}}\right)&=\frac{1}{5}\left[\psi\left(\frac{1}{2}\right)-\mbox{Re}\psi\left(\frac{1}{2}+\frac{igk_F}{2\pi k_BT_{c}}\right)\right],
\end{aligned}
\end{equation}
which gives rise to
\begin{equation*}
\begin{aligned}
\frac{T_c}{T_{c0}}&=\left[\frac{\pi \mathrm{e}^{-\gamma}}{2}\cdot\frac{k_BT_{c0}}{gk_F}\right]^\frac{1}{4},
\end{aligned}
\end{equation*}
when $k_BT_{c}\ll{}gk_F/2\pi$.

\begin{figure}[tb]
\centering
\includegraphics[width=0.98\linewidth]{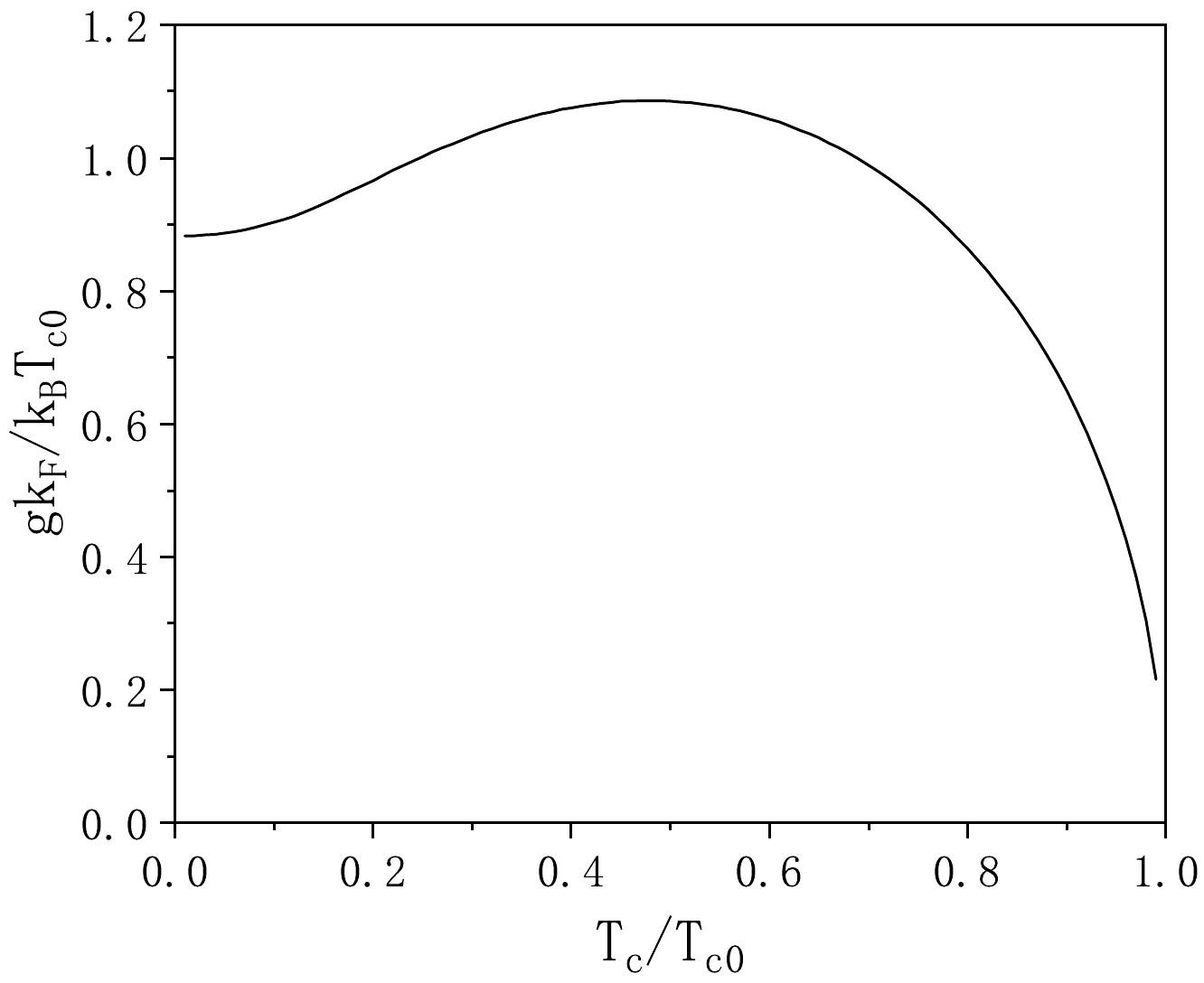}
\caption{The $k_y\hat{x} - k_x\hat{y}$ pairing state: The numerical solution to Eq.~\eqref{eq:Tc-p-H0} with $\alpha=1$ reveals non-monotonic behavior, indicating a first-order phase transition at a small $T_c$.}
\label{fig:kyx-kxy_conti_tc}
\end{figure}

(2) For the $k_y\hat{x}-k_x\hat{y}$ pairing state, the $T_c$ equation for a second-order phase transition is
\begin{equation}\label{eq:Tc-numeric}
\begin{aligned}
\ln\left(\frac{T_c}{T_{c0}}\right)&=\psi\left(\frac{1}{2}\right)-\mbox{Re}\psi\left(\frac{1}{2}+\frac{igk_F}{2\pi k_BT_{c}}\right).
\end{aligned}
\end{equation}
The numerical solution to the above equation is plotted in Fig.~\ref{fig:kyx-kxy_conti_tc}. It turns out that $T_c$ is not a single-value function of $g$, indicating a first-order phase transition for small $T_c$.

(3) For the $k_x\hat{x}-k_y\hat{y}$ and $k_y\hat{x}+k_x\hat{y}$ pairing states, the $T_c$ equation is
\begin{equation}
\begin{aligned}
\ln\left(\frac{T_c}{T_{c0}}\right)&=\frac{3}{5}\left[\psi\left(\frac{1}{2}\right)-\mbox{Re}\psi\left(\frac{1}{2}+\frac{igk_F}{2\pi k_BT_{c}}\right)\right].
\end{aligned}
\end{equation}
This gives a solution
\begin{equation*}
\begin{aligned}
\frac{T_c}{T_{c0}}&=\left[\frac{\pi \mathrm{e}^{-\gamma}}{2}\cdot\frac{k_BT_{c0}}{gk_F}\right]^\frac{3}{2},
\end{aligned}
\end{equation*}
when $k_BT_{c}\ll{}gk_F/2\pi$.

\subsection{OSP states: The effects of only Rashba SOC}

For OSP states, the self-consistent gap equation is
\begin{equation}
\Delta_{\uparrow\downarrow}(\mathbf{k})=-\sum_{\mathbf{k}'}V(\mathbf{k},\mathbf{k}')\langle c_{-\mathbf{k}'\downarrow} c_{\mathbf{k}' \uparrow} \rangle.
\end{equation}
Putting the Bogoliubov transformation matrices in Appendix~\ref{app:Bog_pwave} into the gap equation leads to 
\begin{widetext}
\begin{equation}
\begin{aligned}
\langle c_{-\mathbf{k}\downarrow} c_{\mathbf{k} \uparrow} \rangle
&=(U_\mathbf{k})_{41}^*(U_\mathbf{k})_{11}f(E_{\mathbf{k}+})+(U_\mathbf{k})_{42}^*(U_\mathbf{k})_{12}f(E_{\mathbf{k}-})+(U_\mathbf{k})_{43}^*(U_\mathbf{k})_{13}[1-f(E_{-\mathbf{k}+})]+(U_\mathbf{k})_{44}^*(U_\mathbf{k})_{14}[1-f(E_{-\mathbf{k}-})]\\
&=\sum_{\pm} -\frac{ \mathrm{e}^{ i\eta_{\mathbf{k}} } }{4}\left[ \frac{d_3}{ E_{\mathbf{k}\pm} }\left( \cos\theta_{\mathbf{k}} \pm \frac{ \xi_{\mathbf{k} } }{ 
E_{\mathbf{k}\perp}} \right) +\frac{d_1\sin{\theta_{\mathbf{k}}}}{E_{\mathbf{k}\perp}} \cdot \frac{E_{\mathbf{k}\perp} \pm gk_F }{ E_{\mathbf{k}\pm} } \right] \tanh{ \frac{\beta E_{\mathbf{k}\pm} }{2} }.
\end{aligned}
\end{equation}
\end{widetext} 
Analogous to previous subsections, we obtain the $T_c$ equation identical to Eq.~\eqref{eq:Tc-p-H0}, with $\alpha=2/5$ and $4/5$ for pairing states $k_z\hat{z}$ and $(k_x+ik_y)\hat{z}$, respectively.

For the pairing state $k_z\hat{z}$, where the superconducting phase transition across $T_c$ remains continuous if only Rashba SOC is present, the above equation yields a solution
\begin{equation*}
\frac{T_c}{T_{c0}}=\left[\frac{ \mathrm{e}^{-\gamma}}{4}\cdot \frac{2\pi k_BT_{c0}}{gk_F}\right]^{\frac{2}{3}},
\end{equation*}
when $k_BT_{c}\ll gk_F/2\pi$.

\bibliographystyle{apsrev4-2}
\bibliography{mag-soc}

\begin{thebibliography}{45}%
\makeatletter
\providecommand \@ifxundefined [1]{%
 \@ifx{#1\undefined}
}%
\providecommand \@ifnum [1]{%
 \ifnum #1\expandafter \@firstoftwo
 \else \expandafter \@secondoftwo
 \fi
}%
\providecommand \@ifx [1]{%
 \ifx #1\expandafter \@firstoftwo
 \else \expandafter \@secondoftwo
 \fi
}%
\providecommand \natexlab [1]{#1}%
\providecommand \enquote  [1]{``#1''}%
\providecommand \bibnamefont  [1]{#1}%
\providecommand \bibfnamefont [1]{#1}%
\providecommand \citenamefont [1]{#1}%
\providecommand \href@noop [0]{\@secondoftwo}%
\providecommand \href [0]{\begingroup \@sanitize@url \@href}%
\providecommand \@href[1]{\@@startlink{#1}\@@href}%
\providecommand \@@href[1]{\endgroup#1\@@endlink}%
\providecommand \@sanitize@url [0]{\catcode `\\12\catcode `\$12\catcode
  `\&12\catcode `\#12\catcode `\^12\catcode `\_12\catcode `\%12\relax}%
\providecommand \@@startlink[1]{}%
\providecommand \@@endlink[0]{}%
\providecommand \url  [0]{\begingroup\@sanitize@url \@url }%
\providecommand \@url [1]{\endgroup\@href {#1}{\urlprefix }}%
\providecommand \urlprefix  [0]{URL }%
\providecommand \Eprint [0]{\href }%
\providecommand \doibase [0]{https://doi.org/}%
\providecommand \selectlanguage [0]{\@gobble}%
\providecommand \bibinfo  [0]{\@secondoftwo}%
\providecommand \bibfield  [0]{\@secondoftwo}%
\providecommand \translation [1]{[#1]}%
\providecommand \BibitemOpen [0]{}%
\providecommand \bibitemStop [0]{}%
\providecommand \bibitemNoStop [0]{.\EOS\space}%
\providecommand \EOS [0]{\spacefactor3000\relax}%
\providecommand \BibitemShut  [1]{\csname bibitem#1\endcsname}%
\let\auto@bib@innerbib\@empty
\bibitem [{\citenamefont {Tinkham}(2004)}]{tinkham}%
  \BibitemOpen
  \bibfield  {author} {\bibinfo {author} {\bibfnamefont {M.}~\bibnamefont
  {Tinkham}},\ }\href@noop {} {\emph {\bibinfo {title} {Introduction to
  Superconductivity}}},\ \bibinfo {edition} {2nd}\ ed.\ (\bibinfo  {publisher}
  {Dover Publications},\ \bibinfo {year} {2004})\BibitemShut {NoStop}%
\bibitem [{\citenamefont {Bardeen}\ \emph {et~al.}(1957)\citenamefont
  {Bardeen}, \citenamefont {Cooper},\ and\ \citenamefont
  {Schrieffer}}]{BCS1957}%
  \BibitemOpen
  \bibfield  {author} {\bibinfo {author} {\bibfnamefont {J.}~\bibnamefont
  {Bardeen}}, \bibinfo {author} {\bibfnamefont {L.~N.}\ \bibnamefont
  {Cooper}},\ and\ \bibinfo {author} {\bibfnamefont {J.~R.}\ \bibnamefont
  {Schrieffer}},\ }\href {https://doi.org/10.1103/PhysRev.108.1175} {\bibfield
  {journal} {\bibinfo  {journal} {Phys. Rev.}\ }\textbf {\bibinfo {volume}
  {108}},\ \bibinfo {pages} {1175} (\bibinfo {year} {1957})}\BibitemShut
  {NoStop}%
\bibitem [{\citenamefont {Leggett}(1975)}]{Leggett1975}%
  \BibitemOpen
  \bibfield  {author} {\bibinfo {author} {\bibfnamefont {A.~J.}\ \bibnamefont
  {Leggett}},\ }\href {https://doi.org/10.1103/RevModPhys.47.331} {\bibfield
  {journal} {\bibinfo  {journal} {Rev. Mod. Phys.}\ }\textbf {\bibinfo {volume}
  {47}},\ \bibinfo {pages} {331} (\bibinfo {year} {1975})}\BibitemShut
  {NoStop}%
\bibitem [{\citenamefont {Anderson}\ and\ \citenamefont
  {Morel}(1961)}]{AndersonandMorel1961}%
  \BibitemOpen
  \bibfield  {author} {\bibinfo {author} {\bibfnamefont {P.~W.}\ \bibnamefont
  {Anderson}}\ and\ \bibinfo {author} {\bibfnamefont {P.}~\bibnamefont
  {Morel}},\ }\href {https://doi.org/10.1103/PhysRev.123.1911} {\bibfield
  {journal} {\bibinfo  {journal} {Phys. Rev.}\ }\textbf {\bibinfo {volume}
  {123}},\ \bibinfo {pages} {1911} (\bibinfo {year} {1961})}\BibitemShut
  {NoStop}%
\bibitem [{\citenamefont {Anderson}\ and\ \citenamefont
  {Brinkman}(1973)}]{AndersonandBrinkman1973}%
  \BibitemOpen
  \bibfield  {author} {\bibinfo {author} {\bibfnamefont {P.~W.}\ \bibnamefont
  {Anderson}}\ and\ \bibinfo {author} {\bibfnamefont {W.~F.}\ \bibnamefont
  {Brinkman}},\ }\href {https://doi.org/10.1103/PhysRevLett.30.1108} {\bibfield
   {journal} {\bibinfo  {journal} {Phys. Rev. Lett.}\ }\textbf {\bibinfo
  {volume} {30}},\ \bibinfo {pages} {1108} (\bibinfo {year}
  {1973})}\BibitemShut {NoStop}%
\bibitem [{\citenamefont {Balian}\ and\ \citenamefont
  {Werthamer}(1963)}]{Balian1963}%
  \BibitemOpen
  \bibfield  {author} {\bibinfo {author} {\bibfnamefont {R.}~\bibnamefont
  {Balian}}\ and\ \bibinfo {author} {\bibfnamefont {N.~R.}\ \bibnamefont
  {Werthamer}},\ }\href {https://doi.org/10.1103/PhysRev.131.1553} {\bibfield
  {journal} {\bibinfo  {journal} {Phys. Rev.}\ }\textbf {\bibinfo {volume}
  {131}},\ \bibinfo {pages} {1553} (\bibinfo {year} {1963})}\BibitemShut
  {NoStop}%
\bibitem [{\citenamefont {Dresselhaus}(1955)}]{Dresselhaus}%
  \BibitemOpen
  \bibfield  {author} {\bibinfo {author} {\bibfnamefont {G.}~\bibnamefont
  {Dresselhaus}},\ }\href {https://doi.org/10.1103/PhysRev.100.580} {\bibfield
  {journal} {\bibinfo  {journal} {Phys. Rev.}\ }\textbf {\bibinfo {volume}
  {100}},\ \bibinfo {pages} {580} (\bibinfo {year} {1955})}\BibitemShut
  {NoStop}%
\bibitem [{\citenamefont {Rashba}(1960)}]{Rashiba1960}%
  \BibitemOpen
  \bibfield  {author} {\bibinfo {author} {\bibfnamefont {E.~I.}\ \bibnamefont
  {Rashba}},\ }\href@noop {} {\bibfield  {journal} {\bibinfo  {journal} {Sov.
  Phys. Solid. State}\ }\textbf {\bibinfo {volume} {2}},\ \bibinfo {pages}
  {1109} (\bibinfo {year} {1960})}\BibitemShut {NoStop}%
\bibitem [{\citenamefont {Dresselhaus}\ \emph {et~al.}(2007)\citenamefont
  {Dresselhaus}, \citenamefont {Dresselhaus},\ and\ \citenamefont
  {Jorio}}]{dresselhaus2007group}%
  \BibitemOpen
  \bibfield  {author} {\bibinfo {author} {\bibfnamefont {M.~S.}\ \bibnamefont
  {Dresselhaus}}, \bibinfo {author} {\bibfnamefont {G.}~\bibnamefont
  {Dresselhaus}},\ and\ \bibinfo {author} {\bibfnamefont {A.}~\bibnamefont
  {Jorio}},\ }\href@noop {} {\emph {\bibinfo {title} {Group theory: application
  to the physics of condensed matter}}}\ (\bibinfo  {publisher} {Springer
  Science \& Business Media},\ \bibinfo {year} {2007})\BibitemShut {NoStop}%
\bibitem [{\citenamefont {Bauer}\ and\ \citenamefont
  {Sigrist}(2012)}]{bauer2012non}%
  \BibitemOpen
  \bibfield  {author} {\bibinfo {author} {\bibfnamefont {E.}~\bibnamefont
  {Bauer}}\ and\ \bibinfo {author} {\bibfnamefont {M.}~\bibnamefont
  {Sigrist}},\ }\href@noop {} {\emph {\bibinfo {title} {Non-Centrosymmetric
  Superconductors: Introduction and Overview}}},\ Lecture Notes in Physics\
  (\bibinfo  {publisher} {Springer Berlin Heidelberg},\ \bibinfo {year}
  {2012})\BibitemShut {NoStop}%
\bibitem [{\citenamefont {Yip}(2014)}]{Yip14}%
  \BibitemOpen
  \bibfield  {author} {\bibinfo {author} {\bibfnamefont {S.}~\bibnamefont
  {Yip}},\ }\href
  {https://doi.org/https://doi.org/10.1146/annurev-conmatphys-031113-133912}
  {\bibfield  {journal} {\bibinfo  {journal} {Annual Review of Condensed Matter
  Physics}\ }\textbf {\bibinfo {volume} {5}},\ \bibinfo {pages} {15} (\bibinfo
  {year} {2014})}\BibitemShut {NoStop}%
\bibitem [{\citenamefont {Samokhin}(2015)}]{Samohkin15}%
  \BibitemOpen
  \bibfield  {author} {\bibinfo {author} {\bibfnamefont {K.}~\bibnamefont
  {Samokhin}},\ }\href
  {https://doi.org/https://doi.org/10.1016/j.aop.2015.04.024} {\bibfield
  {journal} {\bibinfo  {journal} {Annals of Physics}\ }\textbf {\bibinfo
  {volume} {359}},\ \bibinfo {pages} {385} (\bibinfo {year}
  {2015})}\BibitemShut {NoStop}%
\bibitem [{\citenamefont {Smidman}\ \emph {et~al.}(2017)\citenamefont
  {Smidman}, \citenamefont {Salamon}, \citenamefont {Yuan},\ and\ \citenamefont
  {Agterberg}}]{Smidman17}%
  \BibitemOpen
  \bibfield  {author} {\bibinfo {author} {\bibfnamefont {M.}~\bibnamefont
  {Smidman}}, \bibinfo {author} {\bibfnamefont {M.~B.}\ \bibnamefont
  {Salamon}}, \bibinfo {author} {\bibfnamefont {H.~Q.}\ \bibnamefont {Yuan}},\
  and\ \bibinfo {author} {\bibfnamefont {D.~F.}\ \bibnamefont {Agterberg}},\
  }\href {https://doi.org/10.1088/1361-6633/80/3/036501} {\bibfield  {journal}
  {\bibinfo  {journal} {Reports on Progress in Physics}\ }\textbf {\bibinfo
  {volume} {80}},\ \bibinfo {pages} {036501} (\bibinfo {year}
  {2017})}\BibitemShut {NoStop}%
\bibitem [{\citenamefont {Reif}(1957)}]{Reif1957}%
  \BibitemOpen
  \bibfield  {author} {\bibinfo {author} {\bibfnamefont {F.}~\bibnamefont
  {Reif}},\ }\href {https://doi.org/10.1103/PhysRev.106.208} {\bibfield
  {journal} {\bibinfo  {journal} {Phys. Rev.}\ }\textbf {\bibinfo {volume}
  {106}},\ \bibinfo {pages} {208} (\bibinfo {year} {1957})}\BibitemShut
  {NoStop}%
\bibitem [{\citenamefont {Androes}\ and\ \citenamefont
  {Knight}(1959)}]{AndroesKnight1959}%
  \BibitemOpen
  \bibfield  {author} {\bibinfo {author} {\bibfnamefont {G.~M.}\ \bibnamefont
  {Androes}}\ and\ \bibinfo {author} {\bibfnamefont {W.~D.}\ \bibnamefont
  {Knight}},\ }\href {https://doi.org/10.1103/PhysRevLett.2.386} {\bibfield
  {journal} {\bibinfo  {journal} {Phys. Rev. Lett.}\ }\textbf {\bibinfo
  {volume} {2}},\ \bibinfo {pages} {386} (\bibinfo {year} {1959})}\BibitemShut
  {NoStop}%
\bibitem [{\citenamefont {Ferrell}(1959)}]{Ferrell1959}%
  \BibitemOpen
  \bibfield  {author} {\bibinfo {author} {\bibfnamefont {R.~A.}\ \bibnamefont
  {Ferrell}},\ }\href {https://doi.org/10.1103/PhysRevLett.3.262} {\bibfield
  {journal} {\bibinfo  {journal} {Phys. Rev. Lett.}\ }\textbf {\bibinfo
  {volume} {3}},\ \bibinfo {pages} {262} (\bibinfo {year} {1959})}\BibitemShut
  {NoStop}%
\bibitem [{\citenamefont {Martin}\ and\ \citenamefont
  {Kadanoff}(1959)}]{Kadanoff1959}%
  \BibitemOpen
  \bibfield  {author} {\bibinfo {author} {\bibfnamefont {P.~C.}\ \bibnamefont
  {Martin}}\ and\ \bibinfo {author} {\bibfnamefont {L.~P.}\ \bibnamefont
  {Kadanoff}},\ }\href {https://doi.org/10.1103/PhysRevLett.3.322} {\bibfield
  {journal} {\bibinfo  {journal} {Phys. Rev. Lett.}\ }\textbf {\bibinfo
  {volume} {3}},\ \bibinfo {pages} {322} (\bibinfo {year} {1959})}\BibitemShut
  {NoStop}%
\bibitem [{\citenamefont {Schrieffer}(1959)}]{Schrieffer1959}%
  \BibitemOpen
  \bibfield  {author} {\bibinfo {author} {\bibfnamefont {J.~R.}\ \bibnamefont
  {Schrieffer}},\ }\href {https://doi.org/10.1103/PhysRevLett.3.323} {\bibfield
   {journal} {\bibinfo  {journal} {Phys. Rev. Lett.}\ }\textbf {\bibinfo
  {volume} {3}},\ \bibinfo {pages} {323} (\bibinfo {year} {1959})}\BibitemShut
  {NoStop}%
\bibitem [{\citenamefont {Anderson}(1959{\natexlab{a}})}]{AndersonKnight1959}%
  \BibitemOpen
  \bibfield  {author} {\bibinfo {author} {\bibfnamefont {P.~W.}\ \bibnamefont
  {Anderson}},\ }\href {https://doi.org/10.1103/PhysRevLett.3.325} {\bibfield
  {journal} {\bibinfo  {journal} {Phys. Rev. Lett.}\ }\textbf {\bibinfo
  {volume} {3}},\ \bibinfo {pages} {325} (\bibinfo {year}
  {1959}{\natexlab{a}})}\BibitemShut {NoStop}%
\bibitem [{\citenamefont
  {Anderson}(1959{\natexlab{b}})}]{Andersontheoryofdirty1959}%
  \BibitemOpen
  \bibfield  {author} {\bibinfo {author} {\bibfnamefont {P.}~\bibnamefont
  {Anderson}},\ }\href
  {https://doi.org/https://doi.org/10.1016/0022-3697(59)90036-8} {\bibfield
  {journal} {\bibinfo  {journal} {Journal of Physics and Chemistry of Solids}\
  }\textbf {\bibinfo {volume} {11}},\ \bibinfo {pages} {26} (\bibinfo {year}
  {1959}{\natexlab{b}})}\BibitemShut {NoStop}%
\bibitem [{\citenamefont {Appel}(1965)}]{Appel1965}%
  \BibitemOpen
  \bibfield  {author} {\bibinfo {author} {\bibfnamefont {J.}~\bibnamefont
  {Appel}},\ }\href {https://doi.org/10.1103/PhysRev.139.A1536} {\bibfield
  {journal} {\bibinfo  {journal} {Phys. Rev.}\ }\textbf {\bibinfo {volume}
  {139}},\ \bibinfo {pages} {A1536} (\bibinfo {year} {1965})}\BibitemShut
  {NoStop}%
\bibitem [{\citenamefont {Shiba}(1976)}]{shiba1976effect}%
  \BibitemOpen
  \bibfield  {author} {\bibinfo {author} {\bibfnamefont {H.}~\bibnamefont
  {Shiba}},\ }\href@noop {} {\bibfield  {journal} {\bibinfo  {journal} {Journal
  of Low Temperature Physics}\ }\textbf {\bibinfo {volume} {22}},\ \bibinfo
  {pages} {105} (\bibinfo {year} {1976})}\BibitemShut {NoStop}%
\bibitem [{\citenamefont {Zhogolev}\ and\ \citenamefont
  {Glasser}(1972)}]{zhogolev1972magnetic}%
  \BibitemOpen
  \bibfield  {author} {\bibinfo {author} {\bibfnamefont {D.}~\bibnamefont
  {Zhogolev}}\ and\ \bibinfo {author} {\bibfnamefont {M.}~\bibnamefont
  {Glasser}},\ }\href@noop {} {\bibfield  {journal} {\bibinfo  {journal}
  {Journal of Low Temperature Physics}\ }\textbf {\bibinfo {volume} {9}},\
  \bibinfo {pages} {347} (\bibinfo {year} {1972})}\BibitemShut {NoStop}%
\bibitem [{\citenamefont {Frigeri}\ \emph
  {et~al.}(2004{\natexlab{a}})\citenamefont {Frigeri}, \citenamefont
  {Agterberg}, \citenamefont {Koga},\ and\ \citenamefont
  {Sigrist}}]{Frigeri2004}%
  \BibitemOpen
  \bibfield  {author} {\bibinfo {author} {\bibfnamefont {P.~A.}\ \bibnamefont
  {Frigeri}}, \bibinfo {author} {\bibfnamefont {D.~F.}\ \bibnamefont
  {Agterberg}}, \bibinfo {author} {\bibfnamefont {A.}~\bibnamefont {Koga}},\
  and\ \bibinfo {author} {\bibfnamefont {M.}~\bibnamefont {Sigrist}},\ }\href
  {https://doi.org/10.1103/PhysRevLett.92.097001} {\bibfield  {journal}
  {\bibinfo  {journal} {Phys. Rev. Lett.}\ }\textbf {\bibinfo {volume} {92}},\
  \bibinfo {pages} {097001} (\bibinfo {year} {2004}{\natexlab{a}})}\BibitemShut
  {NoStop}%
\bibitem [{\citenamefont {Frigeri}\ \emph
  {et~al.}(2004{\natexlab{b}})\citenamefont {Frigeri}, \citenamefont
  {Agterberg},\ and\ \citenamefont {Sigrist}}]{Frigeri_2004}%
  \BibitemOpen
  \bibfield  {author} {\bibinfo {author} {\bibfnamefont {P.~A.}\ \bibnamefont
  {Frigeri}}, \bibinfo {author} {\bibfnamefont {D.~F.}\ \bibnamefont
  {Agterberg}},\ and\ \bibinfo {author} {\bibfnamefont {M.}~\bibnamefont
  {Sigrist}},\ }\href {https://doi.org/10.1088/1367-2630/6/1/115} {\bibfield
  {journal} {\bibinfo  {journal} {New Journal of Physics}\ }\textbf {\bibinfo
  {volume} {6}},\ \bibinfo {pages} {115} (\bibinfo {year}
  {2004}{\natexlab{b}})}\BibitemShut {NoStop}%
\bibitem [{\citenamefont {Samokhin}(2007)}]{Samokhin2007}%
  \BibitemOpen
  \bibfield  {author} {\bibinfo {author} {\bibfnamefont {K.~V.}\ \bibnamefont
  {Samokhin}},\ }\href {https://doi.org/10.1103/PhysRevB.76.094516} {\bibfield
  {journal} {\bibinfo  {journal} {Phys. Rev. B}\ }\textbf {\bibinfo {volume}
  {76}},\ \bibinfo {pages} {094516} (\bibinfo {year} {2007})}\BibitemShut
  {NoStop}%
\bibitem [{\citenamefont {Edelstein}(2008)}]{Edelstein2008}%
  \BibitemOpen
  \bibfield  {author} {\bibinfo {author} {\bibfnamefont {V.~M.}\ \bibnamefont
  {Edelstein}},\ }\href {https://doi.org/10.1103/PhysRevB.78.094514} {\bibfield
   {journal} {\bibinfo  {journal} {Phys. Rev. B}\ }\textbf {\bibinfo {volume}
  {78}},\ \bibinfo {pages} {094514} (\bibinfo {year} {2008})}\BibitemShut
  {NoStop}%
\bibitem [{\citenamefont {Sigrist}\ and\ \citenamefont
  {Ueda}(1991)}]{Sigrist1991}%
  \BibitemOpen
  \bibfield  {author} {\bibinfo {author} {\bibfnamefont {M.}~\bibnamefont
  {Sigrist}}\ and\ \bibinfo {author} {\bibfnamefont {K.}~\bibnamefont {Ueda}},\
  }\href {https://doi.org/10.1103/RevModPhys.63.239} {\bibfield  {journal}
  {\bibinfo  {journal} {Rev. Mod. Phys.}\ }\textbf {\bibinfo {volume} {63}},\
  \bibinfo {pages} {239} (\bibinfo {year} {1991})}\BibitemShut {NoStop}%
\bibitem [{\citenamefont {Clogston}(1962)}]{Clogston1962}%
  \BibitemOpen
  \bibfield  {author} {\bibinfo {author} {\bibfnamefont {A.~M.}\ \bibnamefont
  {Clogston}},\ }\href {https://doi.org/10.1103/PhysRevLett.9.266} {\bibfield
  {journal} {\bibinfo  {journal} {Phys. Rev. Lett.}\ }\textbf {\bibinfo
  {volume} {9}},\ \bibinfo {pages} {266} (\bibinfo {year} {1962})}\BibitemShut
  {NoStop}%
\bibitem [{\citenamefont {Maki}\ and\ \citenamefont
  {Tsuneto}(1964)}]{Maki1964}%
  \BibitemOpen
  \bibfield  {author} {\bibinfo {author} {\bibfnamefont {K.}~\bibnamefont
  {Maki}}\ and\ \bibinfo {author} {\bibfnamefont {T.}~\bibnamefont {Tsuneto}},\
  }\href {https://doi.org/10.1143/PTP.31.945} {\bibfield  {journal} {\bibinfo
  {journal} {Progress of Theoretical Physics}\ }\textbf {\bibinfo {volume}
  {31}},\ \bibinfo {pages} {945} (\bibinfo {year} {1964})}\BibitemShut
  {NoStop}%
\bibitem [{\citenamefont {Fulde}\ and\ \citenamefont
  {Ferrell}(1964)}]{FF_state}%
  \BibitemOpen
  \bibfield  {author} {\bibinfo {author} {\bibfnamefont {P.}~\bibnamefont
  {Fulde}}\ and\ \bibinfo {author} {\bibfnamefont {R.~A.}\ \bibnamefont
  {Ferrell}},\ }\href {https://doi.org/10.1103/PhysRev.135.A550} {\bibfield
  {journal} {\bibinfo  {journal} {Phys. Rev.}\ }\textbf {\bibinfo {volume}
  {135}},\ \bibinfo {pages} {A550} (\bibinfo {year} {1964})}\BibitemShut
  {NoStop}%
\bibitem [{\citenamefont {{Larkin}}\ and\ \citenamefont
  {{Ovchinnikov}}(1969)}]{LO_state}%
  \BibitemOpen
  \bibfield  {author} {\bibinfo {author} {\bibfnamefont {A.~I.}\ \bibnamefont
  {{Larkin}}}\ and\ \bibinfo {author} {\bibfnamefont {Y.~N.}\ \bibnamefont
  {{Ovchinnikov}}},\ }\href
  {https://ui.adsabs.harvard.edu/abs/1969JETP...28.1200L} {\bibfield  {journal}
  {\bibinfo  {journal} {Sov. Phys. JETP}\ }\textbf {\bibinfo {volume} {28}},\
  \bibinfo {pages} {1200} (\bibinfo {year} {1969})}\BibitemShut {NoStop}%
\bibitem [{\citenamefont {Agterberg}\ \emph {et~al.}(2020)\citenamefont
  {Agterberg}, \citenamefont {Davis}, \citenamefont {Edkins}, \citenamefont
  {Fradkin}, \citenamefont {Van~Harlingen}, \citenamefont {Kivelson},
  \citenamefont {Lee}, \citenamefont {Radzihovsky}, \citenamefont {Tranquada},\
  and\ \citenamefont {Wang}}]{PDW}%
  \BibitemOpen
  \bibfield  {author} {\bibinfo {author} {\bibfnamefont {D.~F.}\ \bibnamefont
  {Agterberg}}, \bibinfo {author} {\bibfnamefont {J.~S.}\ \bibnamefont
  {Davis}}, \bibinfo {author} {\bibfnamefont {S.~D.}\ \bibnamefont {Edkins}},
  \bibinfo {author} {\bibfnamefont {E.}~\bibnamefont {Fradkin}}, \bibinfo
  {author} {\bibfnamefont {D.~J.}\ \bibnamefont {Van~Harlingen}}, \bibinfo
  {author} {\bibfnamefont {S.~A.}\ \bibnamefont {Kivelson}}, \bibinfo {author}
  {\bibfnamefont {P.~A.}\ \bibnamefont {Lee}}, \bibinfo {author} {\bibfnamefont
  {L.}~\bibnamefont {Radzihovsky}}, \bibinfo {author} {\bibfnamefont {J.~M.}\
  \bibnamefont {Tranquada}},\ and\ \bibinfo {author} {\bibfnamefont
  {Y.}~\bibnamefont {Wang}},\ }\href
  {https://doi.org/10.1146/annurev-conmatphys-031119-050711} {\bibfield
  {journal} {\bibinfo  {journal} {Annual Review of Condensed Matter Physics}\
  }\textbf {\bibinfo {volume} {11}},\ \bibinfo {pages} {231} (\bibinfo {year}
  {2020})}\BibitemShut {NoStop}%
\bibitem [{\citenamefont {Zhou}\ \emph {et~al.}(2008)\citenamefont {Zhou},
  \citenamefont {Lee}, \citenamefont {Ng},\ and\ \citenamefont
  {Zhang}}]{NaIrO4}%
  \BibitemOpen
  \bibfield  {author} {\bibinfo {author} {\bibfnamefont {Y.}~\bibnamefont
  {Zhou}}, \bibinfo {author} {\bibfnamefont {P.~A.}\ \bibnamefont {Lee}},
  \bibinfo {author} {\bibfnamefont {T.-K.}\ \bibnamefont {Ng}},\ and\ \bibinfo
  {author} {\bibfnamefont {F.-C.}\ \bibnamefont {Zhang}},\ }\href
  {https://doi.org/10.1103/PhysRevLett.101.197201} {\bibfield  {journal}
  {\bibinfo  {journal} {Phys. Rev. Lett.}\ }\textbf {\bibinfo {volume} {101}},\
  \bibinfo {pages} {197201} (\bibinfo {year} {2008})}\BibitemShut {NoStop}%
\bibitem [{\citenamefont {De~Gennes}(1999)}]{deGennes}%
  \BibitemOpen
  \bibfield  {author} {\bibinfo {author} {\bibfnamefont {P.}~\bibnamefont
  {De~Gennes}},\ }\href {https://doi.org/10.1201/9780429497032} {\emph
  {\bibinfo {title} {Superconductivity Of Metals And Alloys}}}\ (\bibinfo
  {publisher} {CRC Press},\ \bibinfo {year} {1999})\BibitemShut {NoStop}%
\bibitem [{\citenamefont {Huxley}(2015)}]{HUXLEY2015}%
  \BibitemOpen
  \bibfield  {author} {\bibinfo {author} {\bibfnamefont {A.~D.}\ \bibnamefont
  {Huxley}},\ }\href
  {https://doi.org/https://doi.org/10.1016/j.physc.2015.02.026} {\bibfield
  {journal} {\bibinfo  {journal} {Physica C: Superconductivity and its
  Applications}\ }\textbf {\bibinfo {volume} {514}},\ \bibinfo {pages} {368}
  (\bibinfo {year} {2015})},\ \bibinfo {note} {superconducting Materials:
  Conventional, Unconventional and Undetermined}\BibitemShut {NoStop}%
\bibitem [{\citenamefont {Liu}\ \emph {et~al.}(2021)\citenamefont {Liu},
  \citenamefont {Liu},\ and\ \citenamefont {Cao}}]{Liu2022}%
  \BibitemOpen
  \bibfield  {author} {\bibinfo {author} {\bibfnamefont {Y.-B.}\ \bibnamefont
  {Liu}}, \bibinfo {author} {\bibfnamefont {Y.}~\bibnamefont {Liu}},\ and\
  \bibinfo {author} {\bibfnamefont {G.-H.}\ \bibnamefont {Cao}},\ }\href
  {https://doi.org/10.1088/1361-648X/ac3cf2} {\bibfield  {journal} {\bibinfo
  {journal} {Journal of Physics: Condensed Matter}\ }\textbf {\bibinfo {volume}
  {34}},\ \bibinfo {pages} {093001} (\bibinfo {year} {2021})}\BibitemShut
  {NoStop}%
\bibitem [{\citenamefont {Bao}\ \emph {et~al.}(2015)\citenamefont {Bao},
  \citenamefont {Liu}, \citenamefont {Ma}, \citenamefont {Meng}, \citenamefont
  {Tang}, \citenamefont {Sun}, \citenamefont {Zhai}, \citenamefont {Jiang},
  \citenamefont {Bai}, \citenamefont {Feng}, \citenamefont {Xu},\ and\
  \citenamefont {Cao}}]{Bao15}%
  \BibitemOpen
  \bibfield  {author} {\bibinfo {author} {\bibfnamefont {J.-K.}\ \bibnamefont
  {Bao}}, \bibinfo {author} {\bibfnamefont {J.-Y.}\ \bibnamefont {Liu}},
  \bibinfo {author} {\bibfnamefont {C.-W.}\ \bibnamefont {Ma}}, \bibinfo
  {author} {\bibfnamefont {Z.-H.}\ \bibnamefont {Meng}}, \bibinfo {author}
  {\bibfnamefont {Z.-T.}\ \bibnamefont {Tang}}, \bibinfo {author}
  {\bibfnamefont {Y.-L.}\ \bibnamefont {Sun}}, \bibinfo {author} {\bibfnamefont
  {H.-F.}\ \bibnamefont {Zhai}}, \bibinfo {author} {\bibfnamefont
  {H.}~\bibnamefont {Jiang}}, \bibinfo {author} {\bibfnamefont
  {H.}~\bibnamefont {Bai}}, \bibinfo {author} {\bibfnamefont {C.-M.}\
  \bibnamefont {Feng}}, \bibinfo {author} {\bibfnamefont {Z.-A.}\ \bibnamefont
  {Xu}},\ and\ \bibinfo {author} {\bibfnamefont {G.-H.}\ \bibnamefont {Cao}},\
  }\href {https://doi.org/10.1103/PhysRevX.5.011013} {\bibfield  {journal}
  {\bibinfo  {journal} {Phys. Rev. X}\ }\textbf {\bibinfo {volume} {5}},\
  \bibinfo {pages} {011013} (\bibinfo {year} {2015})}\BibitemShut {NoStop}%
\bibitem [{\citenamefont {Tang}\ \emph
  {et~al.}(2015{\natexlab{a}})\citenamefont {Tang}, \citenamefont {Bao},
  \citenamefont {Liu}, \citenamefont {Sun}, \citenamefont {Ablimit},
  \citenamefont {Zhai}, \citenamefont {Jiang}, \citenamefont {Feng},
  \citenamefont {Xu},\ and\ \citenamefont {Cao}}]{Tang15_1}%
  \BibitemOpen
  \bibfield  {author} {\bibinfo {author} {\bibfnamefont {Z.-T.}\ \bibnamefont
  {Tang}}, \bibinfo {author} {\bibfnamefont {J.-K.}\ \bibnamefont {Bao}},
  \bibinfo {author} {\bibfnamefont {Y.}~\bibnamefont {Liu}}, \bibinfo {author}
  {\bibfnamefont {Y.-L.}\ \bibnamefont {Sun}}, \bibinfo {author} {\bibfnamefont
  {A.}~\bibnamefont {Ablimit}}, \bibinfo {author} {\bibfnamefont {H.-F.}\
  \bibnamefont {Zhai}}, \bibinfo {author} {\bibfnamefont {H.}~\bibnamefont
  {Jiang}}, \bibinfo {author} {\bibfnamefont {C.-M.}\ \bibnamefont {Feng}},
  \bibinfo {author} {\bibfnamefont {Z.-A.}\ \bibnamefont {Xu}},\ and\ \bibinfo
  {author} {\bibfnamefont {G.-H.}\ \bibnamefont {Cao}},\ }\href
  {https://doi.org/10.1103/PhysRevB.91.020506} {\bibfield  {journal} {\bibinfo
  {journal} {Phys. Rev. B}\ }\textbf {\bibinfo {volume} {91}},\ \bibinfo
  {pages} {020506} (\bibinfo {year} {2015}{\natexlab{a}})}\BibitemShut
  {NoStop}%
\bibitem [{\citenamefont {Tang}\ \emph
  {et~al.}(2015{\natexlab{b}})\citenamefont {Tang}, \citenamefont {Bao},
  \citenamefont {Wang}, \citenamefont {Bai}, \citenamefont {Jiang},
  \citenamefont {Liu}, \citenamefont {Zhai}, \citenamefont {Feng},
  \citenamefont {Xu},\ and\ \citenamefont {Cao}}]{Tang15_2}%
  \BibitemOpen
  \bibfield  {author} {\bibinfo {author} {\bibfnamefont {Z.-T.}\ \bibnamefont
  {Tang}}, \bibinfo {author} {\bibfnamefont {J.-K.}\ \bibnamefont {Bao}},
  \bibinfo {author} {\bibfnamefont {Z.}~\bibnamefont {Wang}}, \bibinfo {author}
  {\bibfnamefont {H.}~\bibnamefont {Bai}}, \bibinfo {author} {\bibfnamefont
  {H.}~\bibnamefont {Jiang}}, \bibinfo {author} {\bibfnamefont
  {Y.}~\bibnamefont {Liu}}, \bibinfo {author} {\bibfnamefont {H.-F.}\
  \bibnamefont {Zhai}}, \bibinfo {author} {\bibfnamefont {C.-M.}\ \bibnamefont
  {Feng}}, \bibinfo {author} {\bibfnamefont {Z.-A.}\ \bibnamefont {Xu}},\ and\
  \bibinfo {author} {\bibfnamefont {G.-H.}\ \bibnamefont {Cao}},\ }\href
  {https://doi.org/10.1007/s40843-015-0021-x} {\bibfield  {journal} {\bibinfo
  {journal} {Science China Materials}\ }\textbf {\bibinfo {volume} {58}},\
  \bibinfo {pages} {16} (\bibinfo {year} {2015}{\natexlab{b}})}\BibitemShut
  {NoStop}%
\bibitem [{\citenamefont {Mu}\ \emph {et~al.}(2018)\citenamefont {Mu},
  \citenamefont {Ruan}, \citenamefont {Pan}, \citenamefont {Liu}, \citenamefont
  {Yu}, \citenamefont {Zhao}, \citenamefont {Chen},\ and\ \citenamefont
  {Ren}}]{Mu18}%
  \BibitemOpen
  \bibfield  {author} {\bibinfo {author} {\bibfnamefont {Q.-G.}\ \bibnamefont
  {Mu}}, \bibinfo {author} {\bibfnamefont {B.-B.}\ \bibnamefont {Ruan}},
  \bibinfo {author} {\bibfnamefont {B.-J.}\ \bibnamefont {Pan}}, \bibinfo
  {author} {\bibfnamefont {T.}~\bibnamefont {Liu}}, \bibinfo {author}
  {\bibfnamefont {J.}~\bibnamefont {Yu}}, \bibinfo {author} {\bibfnamefont
  {K.}~\bibnamefont {Zhao}}, \bibinfo {author} {\bibfnamefont {G.-F.}\
  \bibnamefont {Chen}},\ and\ \bibinfo {author} {\bibfnamefont {Z.-A.}\
  \bibnamefont {Ren}},\ }\href
  {https://doi.org/10.1103/PhysRevMaterials.2.034803} {\bibfield  {journal}
  {\bibinfo  {journal} {Phys. Rev. Mater.}\ }\textbf {\bibinfo {volume} {2}},\
  \bibinfo {pages} {034803} (\bibinfo {year} {2018})}\BibitemShut {NoStop}%
\bibitem [{\citenamefont {Jiang}\ \emph {et~al.}(2015)\citenamefont {Jiang},
  \citenamefont {Cao},\ and\ \citenamefont {Cao}}]{Jiang15}%
  \BibitemOpen
  \bibfield  {author} {\bibinfo {author} {\bibfnamefont {H.}~\bibnamefont
  {Jiang}}, \bibinfo {author} {\bibfnamefont {G.}~\bibnamefont {Cao}},\ and\
  \bibinfo {author} {\bibfnamefont {C.}~\bibnamefont {Cao}},\ }\href
  {https://doi.org/10.1038/srep16054} {\bibfield  {journal} {\bibinfo
  {journal} {Scientific Reports}\ }\textbf {\bibinfo {volume} {5}},\ \bibinfo
  {pages} {16054} (\bibinfo {year} {2015})}\BibitemShut {NoStop}%
\bibitem [{\citenamefont {Wu}\ \emph {et~al.}(2015)\citenamefont {Wu},
  \citenamefont {Le}, \citenamefont {Yuan}, \citenamefont {Fan},\ and\
  \citenamefont {Hu}}]{Wu15}%
  \BibitemOpen
  \bibfield  {author} {\bibinfo {author} {\bibfnamefont {X.-X.}\ \bibnamefont
  {Wu}}, \bibinfo {author} {\bibfnamefont {C.-C.}\ \bibnamefont {Le}}, \bibinfo
  {author} {\bibfnamefont {J.}~\bibnamefont {Yuan}}, \bibinfo {author}
  {\bibfnamefont {H.}~\bibnamefont {Fan}},\ and\ \bibinfo {author}
  {\bibfnamefont {J.-P.}\ \bibnamefont {Hu}},\ }\href
  {https://doi.org/10.1088/0256-307X/32/5/057401} {\bibfield  {journal}
  {\bibinfo  {journal} {Chinese Physics Letters}\ }\textbf {\bibinfo {volume}
  {32}},\ \bibinfo {pages} {057401} (\bibinfo {year} {2015})}\BibitemShut
  {NoStop}%
\bibitem [{\citenamefont {Yang}\ \emph {et~al.}(2021)\citenamefont {Yang},
  \citenamefont {Luo}, \citenamefont {Yi}, \citenamefont {Shi}, \citenamefont
  {Zhou},\ and\ \citenamefont {qing Zheng}}]{Triplet2021}%
  \BibitemOpen
  \bibfield  {author} {\bibinfo {author} {\bibfnamefont {J.}~\bibnamefont
  {Yang}}, \bibinfo {author} {\bibfnamefont {J.}~\bibnamefont {Luo}}, \bibinfo
  {author} {\bibfnamefont {C.}~\bibnamefont {Yi}}, \bibinfo {author}
  {\bibfnamefont {Y.}~\bibnamefont {Shi}}, \bibinfo {author} {\bibfnamefont
  {Y.}~\bibnamefont {Zhou}},\ and\ \bibinfo {author} {\bibfnamefont
  {G.}~\bibnamefont {qing Zheng}},\ }\href
  {https://doi.org/10.1126/sciadv.abl4432} {\bibfield  {journal} {\bibinfo
  {journal} {Science Advances}\ }\textbf {\bibinfo {volume} {7}},\ \bibinfo
  {pages} {eabl4432} (\bibinfo {year} {2021})}\BibitemShut {NoStop}%
\bibitem [{\citenamefont {Zhou}\ \emph {et~al.}(2017)\citenamefont {Zhou},
  \citenamefont {Cao},\ and\ \citenamefont {Zhang}}]{ZHOU2017208}%
  \BibitemOpen
  \bibfield  {author} {\bibinfo {author} {\bibfnamefont {Y.}~\bibnamefont
  {Zhou}}, \bibinfo {author} {\bibfnamefont {C.}~\bibnamefont {Cao}},\ and\
  \bibinfo {author} {\bibfnamefont {F.-C.}\ \bibnamefont {Zhang}},\ }\href
  {https://doi.org/https://doi.org/10.1016/j.scib.2017.01.011} {\bibfield
  {journal} {\bibinfo  {journal} {Science Bulletin}\ }\textbf {\bibinfo
  {volume} {62}},\ \bibinfo {pages} {208} (\bibinfo {year} {2017})}\BibitemShut
  {NoStop}%
\end{thebibliography}%
	
\end{document}